\documentclass[journal,10pt,twocolumn,twoside]{IEEEtran}

\usepackage{mathtools}
\usepackage{empheq}
\usepackage{algpseudocode, algorithm}
\usepackage{algorithmicx}
\usepackage[caption=false]{subfig}
\usepackage{url}
\usepackage{amssymb}
\usepackage{amsthm}
\usepackage{amsmath}
\usepackage{changes}
\usepackage{cite}
\usepackage{multirow}
\usepackage{bm}
\usepackage{booktabs}
\usepackage{hyperref}
\usepackage{balance}
\usepackage{xcolor}
\usepackage{makecell}
\usepackage{amsbsy}
\usepackage{float}
\usepackage{color}
\usepackage{graphicx}
\usepackage{epstopdf}
\usepackage{colortbl}
\usepackage{hhline}
\usepackage{soul}

\definecolor{kugray5}{RGB}{224,224,224}

\usepackage[normalem]{ulem}
\newcommand\rsout{\bgroup\markoverwith
	{\textcolor{red}{\rule[0.5ex]{2pt}{0.8pt}}}\ULon}

\usepackage{array,ragged2e}
\newcolumntype{P}[1]{>{\RaggedRight\arraybackslash}p{#1}}



\makeatletter
\newcommand{\ALOOP}[1]{\ALC@it\algorithmicloop\ #1%
	\begin{ALC@loop}}
	\newcommand{\ENDALOOP}{\end{ALC@loop}\ALC@it\algorithmicendloop}

\makeatother

\usepackage{etoolbox}
\let\mybibitem\bibitem
\renewcommand{\bibitem}[1]{%
	\ifstrequal{#1}{nature}
	{\color{blue}\mybibitem{#1}}
	{\color{black}\mybibitem{#1}}%
}

\graphicspath{ {Figures/} }

\DeclareCaptionLabelSeparator{periodspace}{.\quad}

\renewcommand{\figurename}{Fig.}
\newcommand\numberthis{\addtocounter{equation}{1}\tag{\theequation}}

\newcommand{\norm}[1]{\left\lVert#1\right\rVert} 
\newcommand{\eqn}[1]{\begin{align}#1\end{align}} 
\newcommand{\abs}[1]{\left|#1\right|} 
\newcommand{\nb}{\numberthis}

\newcommand{\mean}[1]{\mathbb{E} \left\{#1\right\}}

\newcommand{\mR}{\textbf{\textit{R}}}
\newcommand{\mH}{\textbf{\textit{H}}} 

\newcommand{\mA}{\textbf{\textit{A}}}
\newcommand{\mW}{\textbf{\textit{W}}}

\newcommand{\mI}{\textbf{\textit{I}}}
\newcommand{\mF}{\textbf{\textit{F}}}

\newcommand{\mY}{\textbf{\textit{Y}}}
\newcommand{\mS}{\textbf{\textit{S}}}
\newcommand{\mN}{\textbf{\textit{N}}}

\newcommand{\setC}{\mathbb{C}} 

\newcommand{\setA}{\mathcal{A}}

\newcommand{\quan}[1]{\mathcal{Q}\left(#1\right)} 

\newcommand{\vg}{\textbf{\textit{g}}}
\newcommand{\va}{\textbf{\textit{a}}}

\newcommand{\vs}{\textbf{\textit{s}}}
\newcommand{\vx}{\textbf{\textit{x}}}
\newcommand{\vy}{\textbf{\textit{y}}}

\newcommand{\vv}{\textbf{\textit{v}}}
\newcommand{\vn}{\textbf{\textit{n}}}

\newcommand{\vz}{\textbf{\textit{z}}} 
\newcommand{\vh}{\textbf{\textit{h}}} 

\newcommand{\vq}{\textbf{\textit{q}}}
\newcommand{\vb}{\textbf{\textit{b}}}

\newcommand{\vf}{\textbf{\textit{f}}}
\newcommand{\sinr}{\text{SINR}_k}
\newcommand{\pdbf}{(\mathcal{P}_{\text{DBF}})}
\newcommand{\pabf}{(\mathcal{P}_{\text{ABF}})}
\newcommand{\phbf}{(\mathcal{P}_{\text{HBF}})}

\newcommand{\sopt}{\hat{\vs}_{\text{opt}}}
\newcommand{\sZF}{\hat{\vs}_{\text{ZF}}}
\newcommand{\sMMSE}{\hat{\vs}_{\text{MMSE}}}

\newcommand{\sML}{\hat{\vs}_{\text{ML}}}

\DeclareMathOperator*{\argmax}{arg\,max}

\begin{document}

\title{Intelligent Radio Signal Processing: A Survey}

\author{Quoc-Viet Pham, 
Nhan Thanh Nguyen, 
Thien Huynh-The, 
\\Long Bao Le,
Kyungchun Lee, 
and Won-Joo Hwang
\thanks{Quoc-Viet Pham is with the Korean Southeast Center for the 4th Industrial Revolution Leader Education, Pusan National University, Busan 46241, Korea (e-mail: vietpq@pusan.ac.kr).}

\thanks{Nhan Thanh Nguyen and Kyungchun Lee are with the Department of Electrical and Information Engineering, Seoul National University of Science and Technology, Seoul 01811, Republic of Korea (e-mail: \{nhan.nguyen, kclee\}@seoultech.ac.kr).}

\thanks{Thien Huynh-The is with the ICT Convergence Research Center, Kumoh National Institute of Technology, Gyeongsangbuk-do 39177, Republic of Korea (e-mail: thienht@kumoh.ac.kr).}


\thanks{Long Bao Le is with the INRS, University of Quebec, Montreal, Quebec, Canada (e-mail: le@emt.inrs.ca).}

\thanks{Won-Joo Hwang is with the Department of Biomedical Convergence Engineering, Pusan National University, Yangsan 50612, Republic of Korea (e-mail: wjhwang@pusan.ac.kr).}


\IEEEcompsocitemizethanks{This work was supported by a National Research Foundation of Korea (NRF) Grant funded by the Korean Government (MSIT) under Grants NRF-2019R1C1C1006143, NRF-2019R1I1A3A01060518, and NRF-2019R1F1A1061934, and in part by Pusan National University Research Grant, 2020. This work was also supported by Institute of Information \& communications Technology Planning \& Evaluation(IITP) grant funded by the Korea government (MSIT) (No. 2020-0-01450, Artificial Intelligence Convergence Research Center [Pusan National University]). This work was also supported by the MSIT (Ministry of Science and ICT), Korea, under the Grand Information Technology Research Center support program(IITP-2021-2016-0-00318) supervised by the IITP (Institute for Information \& communications Technology Planning \& Evaluation).
}
}


\IEEEtitleabstractindextext{
\begin{abstract}
Intelligent signal processing for wireless communications is a vital task in modern wireless systems, but it faces new challenges because of network heterogeneity, diverse service requirements, a massive number of connections, and various radio characteristics.
Owing to recent advancements in big data and computing technologies, \textit{artificial intelligence} (AI) has become a useful tool for \textit{radio signal processing} and has enabled the realization of \textit{intelligent radio signal processing}. 
This survey covers four intelligent signal processing topics for the wireless physical layer, including modulation classification, signal detection, beamforming, and channel estimation. In particular, each theme is presented in a dedicated section, starting with the most fundamental principles, followed by a review of up-to-date studies and a summary.
To provide the necessary background, we first present a brief overview of AI techniques such as machine learning, deep learning, and federated learning.
Finally, we highlight a number of research challenges and future directions in the area of intelligent radio signal processing. 
We expect this survey to be a good source of information for anyone interested in intelligent radio signal processing, and the perspectives we provide therein will stimulate many more novel ideas and contributions in the future. 
\end{abstract}

\begin{IEEEkeywords}
Artificial intelligence, beamforming, channel estimation, deep learning, federated learning, machine learning, modulation classification, radio frequency, signal processing.
\end{IEEEkeywords}}

\maketitle
\IEEEdisplaynontitleabstractindextext
\IEEEpeerreviewmaketitle

\section{Introduction}
\label{Sec:Introduction}
Radio signal processing plays a vital role in the engineering of all generations of wireless networks. With the emergence of many advanced wireless technologies 
and massive connectivity, processing radio signals in an efficient and intelligent way presents both challenges and opportunities. 
Additionally, next-generation wireless systems are likely to rely not only on the sub-6 GHz, but also on the mmWave and THz frequency bands, and non-radio frequencies (RFs) such as the visible and optical bands \cite{saad2020vision}. Furthermore, the use of massive multiple input and multiple output (massive MIMO) in fifth generation (5G) wireless systems and beyond demands sophisticated radio signal processing schemes. 
Radio signals were conventionally processed by mathematical model-based algorithms.  
Despite promising results, these conventional methods have various shortcomings including high complexity as well as poor scalability, online implementation, and adaptivity to dynamic environments. 
Recent advancements in computing hardware and big data processing have rendered AI a useful tool for radio signal processing, thereby realizing the term \textit{intelligent signal processing}. Undoubtedly, AI is expected to play a key role in solving many complex problems that are neither tractably nor efficiently overcome by conventional model-based approaches. 

\subsection{Intelligent Signal Processing: An Overview}
The past three years have witnessed growing interest in the application of AI to wireless signal processing. A good example is the IEEE initiative (https://mlc.committees.comsoc.org/) to promote the use of AI for physical layer signal processing, e.g., modulation recognition (also known as modulation classification), channel estimation, signal detection, channel encoding and decoding, localization, and beamforming. Various AI-based algorithms and deep learning (DL)-based models have been proposed as alternatives to the present model-based approaches. 
An unspoken consensus is that model- and AI-based approaches have different particularities but complementary capabilities, i.e., AI is not a universal solution and should be used for tasks that cannot be efficiently attempted by conventional approaches. 
For instance, the globally optimal solution for signal processing problems can be obtained via existing model-based mechanisms such as optimal signal detection \cite{AMC-WenWei2000Likelihood} and optimal beamforming \cite{huang2009rank}. In general, AI-based algorithms cannot outperform optimal model-based schemes if they are used to solve the same problem, but they have the potential for real-time signal processing. 
Moreover, several scenarios exist in which AI may significantly improve radio signal processing over conventional model-based approaches. In the following, we briefly discuss these scenarios along with representative examples.

\subsubsection{Algorithmic Approximation} A common limitation preventing algorithms from finding the optimal solution is the difficulty of real-time executions; therefore, 
they are impractical for real-time implementation.
Several approaches, e.g., heuristics, metaheuristics, and problem decomposition, have been proposed to optimize the tradeoff between computational complexity and performance. However, the real-time implementation of the underlying algorithms is quite challenging.  
For this case, the use of AI techniques appears to be a promising solution. In particular, the data generation and training phases can be executed offline while the system operates in real time by using the trained model. 
For instance, Huynh \emph{et al.} \cite{AMC-Huynh2020MCNet} proposed a DL architecture for automatic modulation classification (AMC), namely MCNet, which was 93.59\% accurate at a signal-to-noise ratio (SNR) of 20 dB with an inference time of only 0.095~ms.

\subsubsection{Unknown Model and Nonlinearities} Many physical phenomena 
cannot be accurately modeled. Therefore, conventional model-based algorithms usually fail to obtain efficient solutions.
For instance, fiber nonlinearities (e.g., signal distortion and self-phase modulation) in optical systems together with the adoption of coherent communication render model-based methods ineffective for network optimization \cite{Musumeci2019AnOverview}. 
To mitigate the nonlinearities and perform signal detection, AI techniques (e.g., an end-to-end learning approach \cite{karanov2018end}) can be utilized with very low bit error rates (BER).
The end-to-end learning approach \cite{o2017introduction} has found many applications in scenarios in which the channel model is unknown or well-established mathematical models are unavailable. Another application that involves the use of DL to address hardware nonlinearities in MIMO systems (e.g., hardware impairments) was presented \cite{demir2019channel}. These researchers proposed two DL-based estimators to exploit the nonlinear characteristics with the aim of improving the estimation performance. Nonlinearity was also observed in MIMO systems with low-bit analog-to-digital converters. In an attempt to mitigate this nonlinear effect, Nguyen \emph{et al.} \cite{nguyen2020neural} proposed a DNN model to jointly optimize the channel estimation and training signal. The model outperformed the linear channel estimator in various practical settings.

\subsubsection{Algorithm Acceleration} Another direction intelligent signal processing has been taking is to use AI to facilitate and accelerate existing algorithms. This approach differs markedly from the two scenarios discussed above in that an existing model-based algorithm is completely replaced by an AI-based algorithm, i.e., an end-to-end learning paradigm. 
For instance, many DL-based algorithms have been proposed to improve and accelerate near-optimal detection schemes. Nguyen \emph{et al.} \cite{nguyen2019deep} employed a DL model, namely FS-Net, to initialize the highly reliable solution for the tabu search (TS) detection scheme, and also proposed an early termination scheme to further accelerate the optimization process. Compared with the original TS scheme, the DL-aided TS detector can reduce the computational complexity by approximately 90\% at an SNR of 20 dB with similar performance. DL was also employed to generate the initial radius for the sphere decoding (SD) detector \cite{askri2019dnn}. 

\subsection{State-of-the-art}
\label{SubSec:StateOTA}
Owing to the importance of AI for physical layer signal processing, a number of surveys and magazine articles have been published on this topic over the past few years. DL techniques for solving physical layer signal processing problems such as modulation, channel coding, detection, and end-to-end learning were reviewed  \cite{wang2017deep}. However, this survey mainly focused on reviewing DL techniques and did not include many up-to-date studies as it was published quite a long time ago.
The concept of end-to-end DL was first introduced in 2017 \cite{o2017introduction} to model the entire physical communication as an autoencoder DNN. This discovery constituted a major breakthrough in the design of communication systems and has been widely employed in many research efforts. 
A chapter in a recent book \cite{erpek2020deep} described the benefits and the use of end-to-end learning for channel estimation, signal identification, and wireless security. 
Qin \emph{et al.} \cite{qin2019deep} demonstrated the applications of DL to the optimization of individual signal processing blocks in the physical layer (e.g., signal compression and detection) and also end-to-end design.
He \emph{et al.} \cite{he2019model} discussed the significance of model-driven DL techniques in physical layer design and illustrated use cases for receiver design, signal detection, and channel estimation. 
A brief on model-driven deep unfolding for MIMO signal detection and beamforming was presented \cite{balatsoukas2019deep}.  
Furthermore, Zappone \emph{et al.} \cite{zappone2019wireless} discussed model-based, AI-based, or hybrid methods and presented examples for designs of the wireless physical layer. A brief with demonstration of modulation and classification was carried out \cite{zhou2020deep}. 

Another line of work included various surveys and tutorials on the applications of AI to wireless networking. In particular, the use of AI for Internet of Things (IoT) applications and massive connectivity, privacy, and security was reviewed 
\cite{Sharma2020TowardMTC, Hussain2020ML_IoT}. Gu \emph{et al.} \cite{Gu2020Machine} conducted a survey on AI applications for 
optical communications and networking. AI-based solutions for cybersecurity problems (e.g., misuse detection and anomaly detection)
were discussed \cite{Xin2018MachineLearning, Mishra2019ADetailed}. 
Fadlullah \emph{et al.} \cite{Fadlullah2017StateOfDL} reviewed the adoption of AI for network traffic control systems.
The use of machine learning (ML) techniques for designing traffic classification strategies was studied \cite{Pacheco2019Towards}. 
ML techniques for applications including computational offloading, mobile big data, and mobile crowdsensing at the network edge were reviewed \cite{Pham2020ASurvey_MEC}. 
Xie \emph{et al.} \cite{Xie2019ASurvey_ML} discussed opportunities and challenges arising from the use of ML techniques for software-defined networking (SDN). 
A tutorial on artificial neural networks (ANN) for wireless networking was presented \cite{Chen2019Artificial}.
Mao \emph{et al.} \cite{Mao2018DeepLearning} conducted a survey of mobile networking from the mobile big data perspective for which a top--down approach was used. 
Another survey on DL for 5G mobile and wireless networking was presented \cite{Zhang2019DeepLearning}. The use of deep reinforcement learning (DRL) for wireless communications and networking
was reviewed \cite{Luong2019ApplicationsDRL}. Federated learning (FL) in mobile edge networks was surveyed \cite{Lim2020FederatedLearning}.
Four main types of ML (i.e., supervised learning, unsupervised learning, DL, and reinforcement learning) and their application to wireless networks were considered \cite{Wang2020ThirtyYears}. 
The use of swarm intelligence for next-generation wireless networks was recently reviewed in \cite{pham2020swarm}. 
Table~\ref{Table:Summary_Previous_Surveys} summarizes existing surveys and tutorials on AI for wireless networking and radio signal processing.

\begin{table*}[h]
	\caption{Summary of existing surveys and tutorials on AI techniques for wireless networking and signal processing.}
	\label{Table:Summary_Previous_Surveys}
	\centering
	\begin{tabular}{|c|c|c|c|c|p{13.0cm}|}
		\hline 
		\multirow{2}{*}{\textbf{Paper}}  & \multicolumn{4}{c|}{AI Models} & \multirow{2}{*}{\textbf{Applications}}  \\ 
		
		\cline{2-5}
		{} & ML & DL & DRL & FL & {} \\
		\hline
		\hline
		
		\cite{Musumeci2019AnOverview} & \checkmark &  &  &  &  IoT applications, e.g., smart health, smart city, smart transportation, and smart industry. \\ \hline
		
		\cite{o2017introduction} &  & \checkmark &  &  & Proposal of end-to-end learning and example of modulation classification.  \\ \hline
		
		\cite{wang2017deep} &  & \checkmark &  &  & DL for modulation, channel coding, detection, and end-to-end learning. \\ \hline
		
		\cite{erpek2020deep} &  & \checkmark &  &  & End-to-end learning for wireless networks: channel  estimation, signal identification, and wireless security. \\ \hline
		
		\cite{qin2019deep} &  & \checkmark &  &  & Applications of DL for block optimization and end-to-end design. \\ \hline
		
		\multirow{1}{*}{\cite{he2019model}} &  & \multirow{1}{*}{\checkmark} &  &  & Model-driven DL and demonstrations of receiver design, signal detection, and channel estimation.   \\ \hline
		
		\cite{balatsoukas2019deep} &  & \checkmark &  &  & Applications of deep unfolding for MIMO systems: signal detection and beamforming. \\ \hline
		
		\cite{zappone2019wireless} & \checkmark & \checkmark & \checkmark & \checkmark & Discussions and examples of model-based, AI-based, and hybrid methods for wireless networks. \\ \hline
		
		\cite{Sharma2020TowardMTC} & \checkmark & \checkmark &  &  & ML techniques for solving challenges in massive machine-type communications. \\ \hline
		
		\cite{Hussain2020ML_IoT} &  & \checkmark & \checkmark &  & Security preservation and threats in IoT. \\ \hline
		
		\cite{Xin2018MachineLearning} & \checkmark & \checkmark &  &  & Intrusion detection in wireless networks. \\ \hline	
		
		\cite{Mishra2019ADetailed} & \checkmark & \checkmark &  &  &  Network intrusion detection systems. \\ \hline
		
		\multirow{1}{*}{\cite{Fadlullah2017StateOfDL}}  &  & \multirow{1}{*}{\checkmark} &  &  & Traffic control in network systems, e.g., sensor networks, flow prediction, social networks, and cognitive radio. \\ \hline
		
		\cite{Pacheco2019Towards} & \checkmark &  &  &  & Network traffic classification. \\ \hline 
		
		\cite{Xie2019ASurvey_ML} & \checkmark &  &  &  & Software-defined networking, e.g., traffic classification, routing, security, and quality of service (QoS) prediction. \\	\hline

		\cite{Chen2019Artificial} & \checkmark & \checkmark & \checkmark &  & 5G applications, e.g., UAV communications, wireless virtual reality, self-organized networks, and IoT. \\ \hline
		
		\cite{Mao2018DeepLearning} &  & \checkmark & \checkmark &  & Mobile big data applications, e.g., physical coding, spectrum allocation, and routing protocols.  \\ \hline
		
		\multirow{1}{*}{\cite{Zhang2019DeepLearning}} &  & \multirow{1}{*}{\checkmark} & \multirow{1}{*}{\checkmark} &  & DL for 5G applications, e.g., network security, network control, localization, mobility analysis, and data analytics. \\ \hline
		
		\multirow{1}{*}{\cite{Luong2019ApplicationsDRL}}  &  &  & \multirow{1}{*}{\checkmark} &  & Network access, caching and offloading, security and privacy, resource scheduling, and data collection. \\ \hline
		
		\cite{Lim2020FederatedLearning}  & & & & \checkmark & Edge computing applications, e.g., cyberattack detection, edge caching, and user association. \\ \hline
		
		\cite{Wang2020ThirtyYears}  & \checkmark & \checkmark & \checkmark &  & ML applications for wireless networks. \\ \hline
		
	\end{tabular}
\end{table*}

\subsection{Contributions and Organization of this Paper}
\label{SubSec:Contributions}
Notwithstanding the plethora of surveys on AI applications for research topics, we are still unaware of any comprehensive survey on the use of AI techniques for intelligent radio signal processing. Existing surveys (e.g., \cite{Hussain2020ML_IoT, Gu2020Machine, Fadlullah2017StateOfDL, Xie2019ASurvey_ML}) are limited to the scope of mobile networking and communications. Furthermore, most existing studies focus on certain AI techniques and their applications to wireless research such as channel encoding and decoding \cite{o2017introduction, Erpek2020}, unfolding DL for MIMO systems \cite{balatsoukas2019deep}, DL for wireless networks \cite{wang2017deep, he2019model, qin2019deep}, and tracking and localization \cite{zafari2019survey}. In contrast, our aim was to provide a comprehensive survey of AI applications for various aspects of wireless physical signal processing. 
In this vein, we first provide the fundamentals of AI techniques, including ML, DL, and FL and discuss the need to apply AI approaches to design intelligent methods to process radio signals. 
Then, we review AI applications pertaining to four different key signal processing areas, namely modulation classification, signal detection, channel estimation, and MIMO beamforming optimization. We also highlight a number of challenges and future research directions in the area of intelligent radio signal processing. 
Our contributions can be summarized as follows.
\begin{itemize}
	\item We present an overview of AI techniques with potential application to radio signal processing. Specifically, in Section~\ref{Sec:Fundamentals} we present the fundamentals of AI, ML, DL, DRL, and FL. In addition, we summarize the motivations and advantages of using AI techniques for radio signal processing.  
	
	\item We survey the application of AI techniques to four signal processing themes, namely, modulation classification, signal detection, channel estimation, and MIMO beamforming optimization. We present these themes along with basic information to aid readers to design intelligent signal processing frameworks.
	
	\item We provide a set of research challenges 
	and also highlight a number of potential directions along which future investigations would ideally need to follow to enable performance improvement of intelligent radio signal processing methods.  
\end{itemize}


The remainder of this survey is organized as follows. Section~\ref{Sec:Fundamentals} presents the fundamentals of AI techniques with application to the intelligent processing of radio signals in wireless and communication networks. Section~\ref{Sec:Modulation_Classification} discusses the AI applications for AMC. In Section~\ref{Sec:Signal_Detection}, state-of-the-art AI applications for signal detection are reviewed. Section~\ref{Sec:BF_and_CE} reviews the literature on AI techniques for channel estimation and MIMO beamforming.
In Section~\ref{Sec:Challenges}, we present the challenges associated with and unresolved challenges arising from existing research devoted to AI for radio signal processing and further highlight potential research directions. The survey is concluded in Section~\ref{Sec:Conclusion}. 

\section{Artificial Intelligence: Background Information}
\label{Sec:Fundamentals}	
In this section, we provide an overview of ML, DL, DRL, and FL. 

\subsection{Machine Learning: Preliminaries}
AI is one of the most pioneering sciences and has been in development since 1956 when the name AI was adopted by McCarthy and colleagues. The foundations of AI are based in many long-standing disciplines, e.g., philosophy, mathematics, economics, neuroscience, psychology, computer engineering, control theory, cybernetics, and linguistics \cite{russell2002artificial}. Today, AI is a thriving field which has found many applications in the field of engineering.
ML, which is the principal AI discipline, allows patterns to be mined/learned from raw data to gain knowledge. In general, ML can be classified into three main types: supervised learning, unsupervised learning, and reinforcement learning (RL). 

Supervised learning is concerned with mapping known inputs with known outputs given a training set including both inputs and outputs. Basically, supervised learning can be divided into two types: regression and classification, the respective output values of which are continuous and discrete. Examples of popular supervised learning techniques are support vector machine (SVM), K-nearest neighbor (KNN), Na\"{\i}ve Bayesian model, and decision trees \cite{Wang2020ThirtyYears}. 
On the other hand, in unsupervised learning, the labels of the output are not included in the training data and the goal of unsupervised learning is to learn useful representations and properties from the input data. Increasing efforts have been made to utilize unsupervised learning for designing wireless and communication networks.
In fact, massive amounts of unstructured and unlabeled data are generated by wireless devices and emerging applications. Moreover, annotating the ground truth for a large number of examples has a huge associated cost. 
The performance of unsupervised learning models is typically inferior to that of models that use supervised learning. Lastly, RL learns from interactions with the environment, i.e., the learning agent continuously interacts with the environment and adopts good policies to make decisions so as to maximize the reward. Two main features of RL are trial-and-error learning (i.e., using error information and evaluative feedback to update actions/policies) and delayed reward (i.e., an action does not only affect the immediate reward but also the future reward) \cite{sutton2018reinforcement}. 

\subsection{Deep Learning}
The appropriate presentation of handcrafted features extracted from raw data is a prerequisite for conventional ML algorithms, whereas DL is able to directly learn a complicated model from raw data by using a neural network \cite{goodfellow2016deep}. The input and output are presented at the first layer (i.e., the visible layer) and the last layer (i.e., the output layer), respectively. Hidden layers are used to increase the level of feature abstraction, i.e., they calculate representational features at multiscale resolutions. Specifically, the visible layer receives the input data and then extracts simple features, which are further abstracted by the subsequent hidden layers, and the output layer uses additional functions to transform the features received from the last hidden layer into the output. Owing to developments in computing infrastructure, big data, and data science, DL has been recognized as a crucial technology and found many practical applications, e.g., image and speech recognition, natural language processing, drug discovery, self-driving vehicles, and mobile communications and networking. 

Depending on the structure of the neural network, different ANN architectures have been developed for various learning tasks and data modalities. In the following, we briefly introduce important ANN architectures that have been widely used in the studies we reviewed.

\subsubsection{Feedforward Neural Networks (FNN)} An FNN is composed of a visible layer, an output layer, and one or more hidden layers. The information traverses the neural network and feedback connections from the outputs are not used. Similar to a general neural network, the basic components of an FNN are neurons (i.e., the nodes in the network), weights (i.e., the numerical values representing connections), and activation functions, which are used to determine the output of neurons in the neural network. Currently, most  ANN models use the rectified linear unit (ReLU) $ f(z) = \max\{0,z\} $, the logistic sigmoid $ f(z) = 1/(1 + \exp(-z)) $, and the hyperbolic tangent (tanh) $ f(z) = \text{tanh}(z) $ for nonlinear transformation. Other activation functions that have been investigated to improve ANNs include the adaptive piecewise linear \cite{AgostinelliHSB14} and the Swish \cite{RamachandranZL18}.

\begin{figure} [t]
	\centering
	\includegraphics[width=0.70\linewidth]{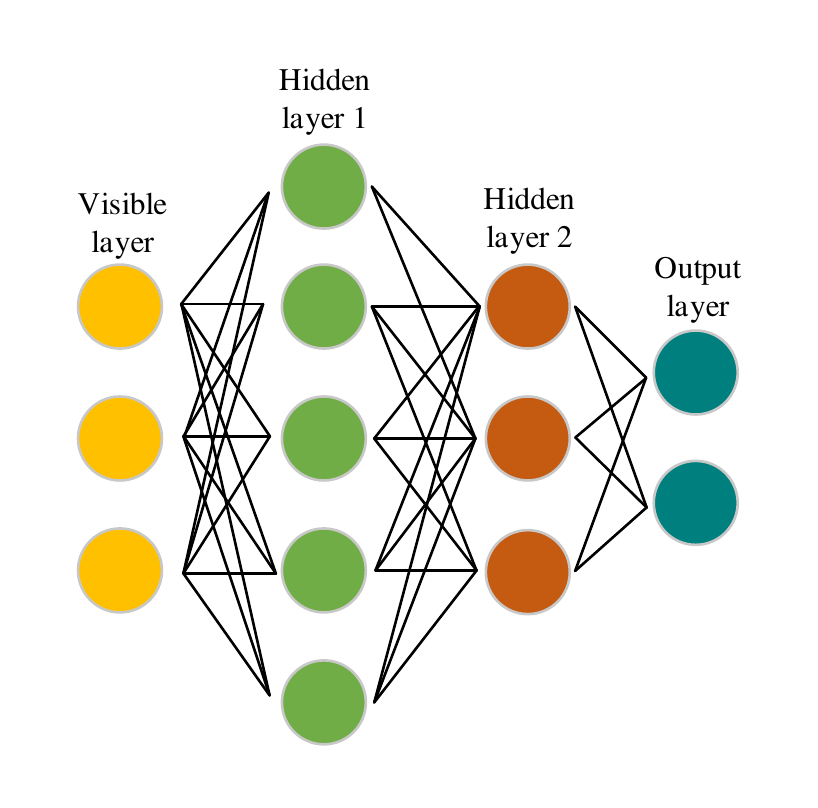}
	\caption{Architecture of an FNN with two hidden layers.}
	\label{Fig:Architecture_FNN}
\end{figure}

Previously \cite{hornik1991approximation}, it was found that any continuous function can be approximated by an FNN composed of one hidden layer and a finite number of neurons. 
To optimize the neural network, several approaches have been employed including first-order gradient-descent algorithms (e.g., backpropagation and its variants such as Quickpro and resilient propagator), second-order minimization algorithms (e.g., conjugate gradient, quasi-Newton, Gauss--Newton), and metaheuristics (e.g., the whale optimization algorithm and Harris Hawks optimization) \cite{Pham2020SumRate}. 
We invite interested readers to refer to \cite{Ojha2017Metaheuristic} for a survey on metaheuristic optimization for FNNs and \cite[Section 6.5]{goodfellow2016deep} for an overview of the backpropagation method and derivative-based algorithms. An example of FNN architecture with two hidden layers is shown in Fig.~\ref{Fig:Architecture_FNN}. In the following, the two terms, FNN and DNN, are usually used interchangeably. 

\subsubsection{Convolutional Neural Networks (CNN)} 
\begin{figure} [t]
	\centering
	\includegraphics[width=0.925\linewidth]{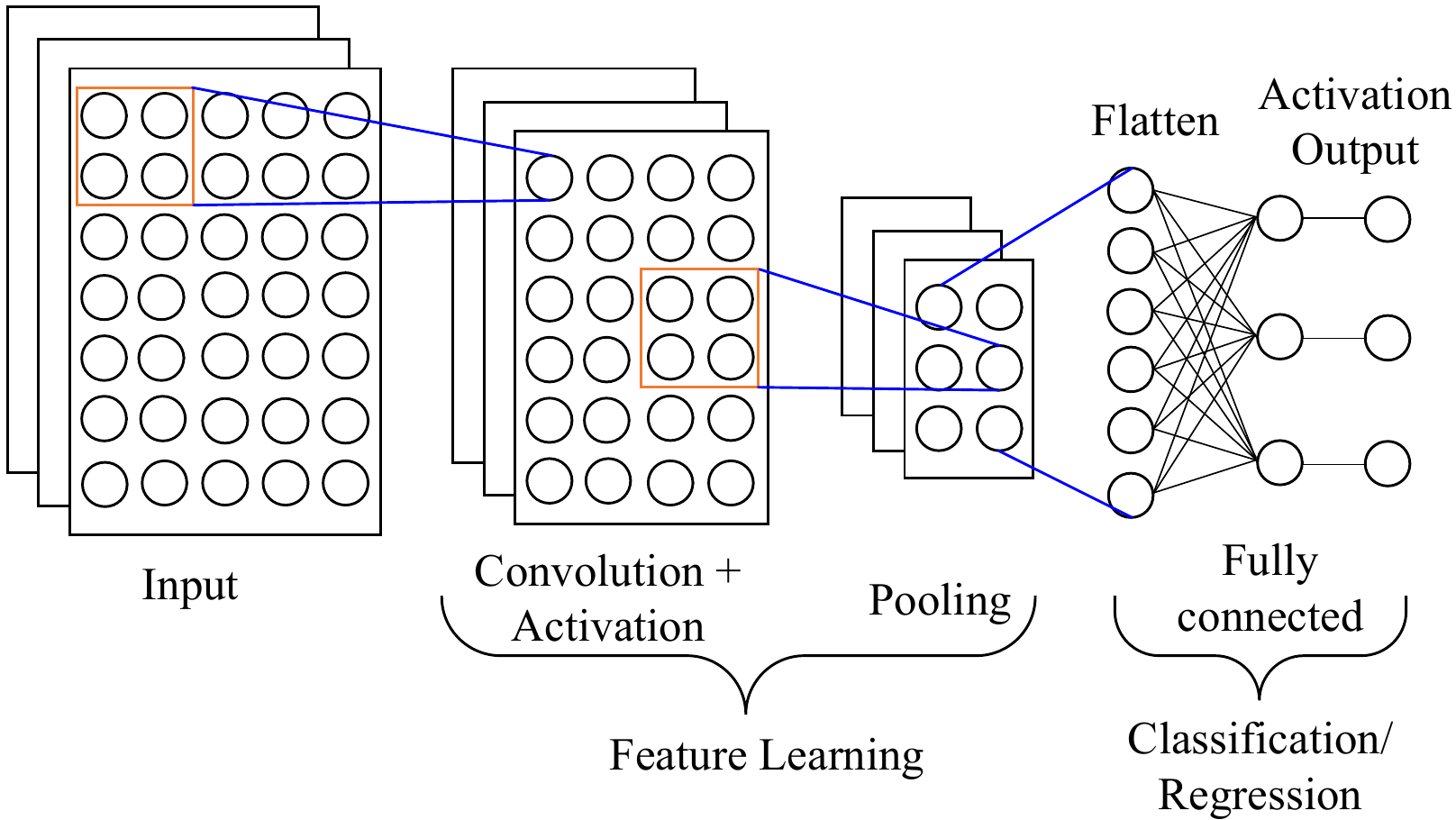}
	\caption{Architecture of a CNN.}
	\label{Fig:Architecture_CNN}
\end{figure}

In general, CNNs are suitable for processing high-dimensional unstructured data such as images, where many backbone CNNs are initially introduced for image classification. The first advantage of a CNN is its \textit{sparse interactions} feature, which is enabled by setting the size of the kernel to be smaller than that of the input, thus improving the storage requirements and statistical efficiency \cite{goodfellow2016deep}. The second advantage originates from the \textit{parameter sharing} concept used in CNNs, i.e., the kernel is applied across the entire input to create the feature map. In addition, the convolution and pooling operations render CNNs \textit{invariant} and \textit{equivariant} to translations of the input. Finally, CNNs are proficient in automatically extracting high-level representational features for mining intrinsic information. As shown in Fig.~\ref{Fig:Architecture_CNN}, the CNN architecture has three basic components: convolution, a nonlinear activation function, and pooling (i.e., down-sampling in the literature). To date, different efficient CNN architectures have been proposed such as GoogleNet, ResNet, Inception-ResNet-v2, SENet, and EfficientNet-B7 \cite{pmlr_v97_tan19a}. 

The convolution operation is the core feature of CNNs and has two parts: input and kernel (i.e., filter). The kernel in a convolutional layer is specified by a predefined kernel spatial size, where its depth size is identical to the number of input channels. The feature map generated by the convolutional operation may have different spatial dimensionality compared with the input. In particular, zero padding and valid padding are used to maintain and change the dimensionality, respectively. Notably, the size of the kernel (width and height) indicates the number of neurons in the input used to infer a neuron in the output feature map. For example, a kernel of size $ 5 \times 4 $ implies that 20 neurons are used to calculate an output neuron via the dot product of kernel weights and input elements. The next component of a CNN is the activation function, which does not change the size of the input it receives and processes. 
The pooling layer, the last principal component of a CNN, usually has the function of reducing the spatial dimension of feature maps.
This family of layers, including max and average pooling, has a similar operating principal to that of the convolutional layer without learnable parameters. 
For instance, the max-pooling layer returns the maximum value of entries of the input with a receptive field, the so-called pool size.
Therefore, a pooling map can be considered as a lower-resolution version of the feature map when it is down-sampled along the vertical and/or horizontal dimensions. As pooling operates over spatial regions, pooling can help the features to become invariant to small translations of the input. Common pooling methods are average pooling, max pooling, and L2-pooling. 

\subsubsection{Recurrent Neural Networks (RNN)} 
\begin{figure} [t]
	\centering
	\includegraphics[width=0.925\linewidth]{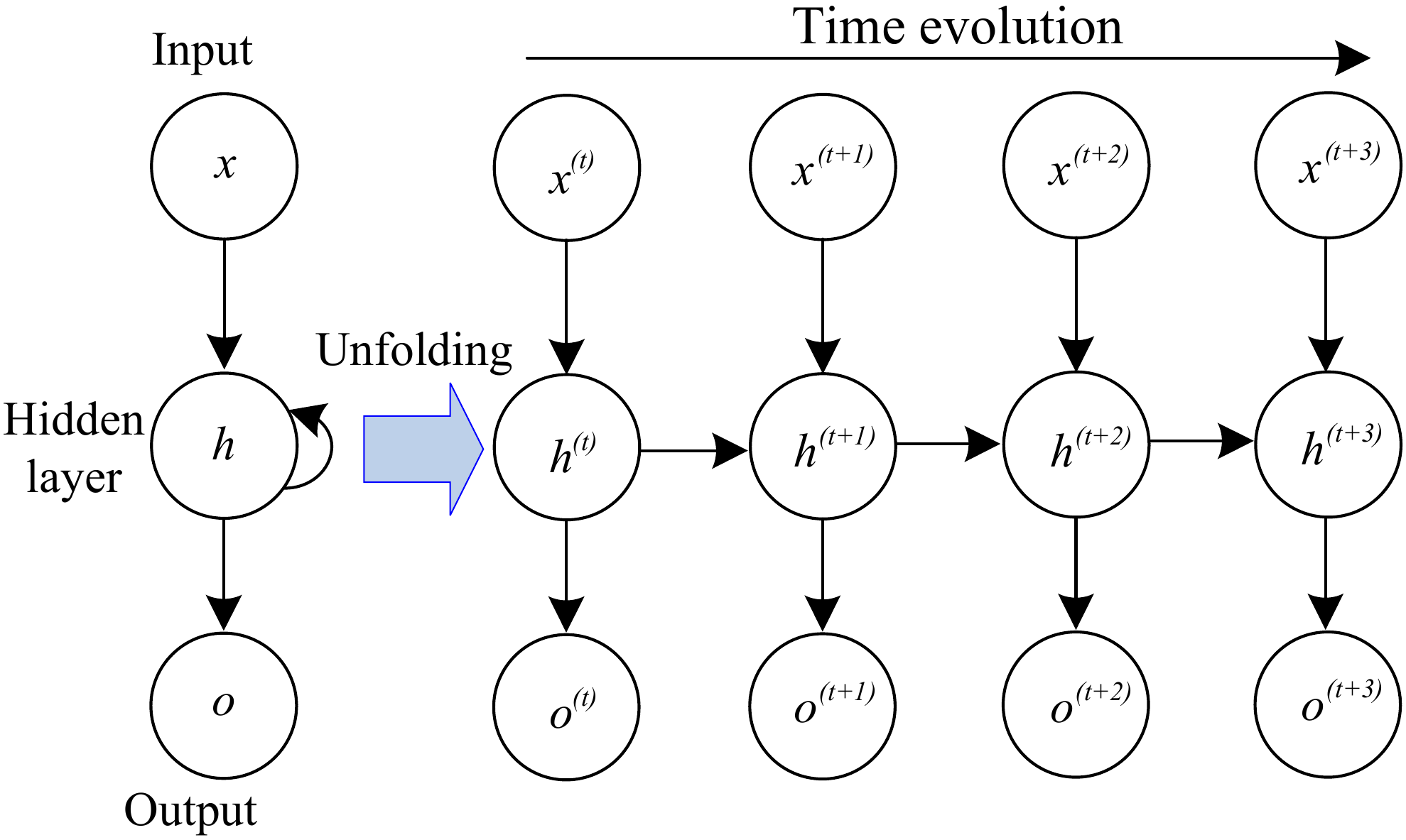}
	\caption{Architecture of an RNN.}
	\label{Fig:Architecture_RNN}
\end{figure}

While CNNs are suitable for processing high-dimensional data, RNNs are usually used to process sequential data, i.e., in situations in which the prediction depends on not only the current sample but also on previous samples. Each neuron in RNNs has the capability to memorize the output, which is fed into the neuron as a subsequent input. An illustration of an RNN with one hidden layer and a length sequence of four inputs is shown in Fig.~\ref{Fig:Architecture_RNN}. In particular, the value $ h^{(t)} $ to be predicted at time $ t $ is a function $ g^{(t)}(\cdot) $ of the input sequence $ \left(x^{(t)}, x^{(t-1)}, \dots, x^{(1)}, x^{(0)}\right) $.
To apply a backpropagation algorithm to RNNs, the unfolding concept is used to transform an RNN into a computational graph, which has a repetitive structure and thus enables the sharing of learning parameters across the neural network. Mathematically, the unfolding recurrence at the time $ t $ can be modeled as $ h^{(t)} = g^{(t)}\left(x^{(t)}, x^{(t-1)}, \dots, x^{(1)}, x^{(0)}\right) $.

Echo state networks, liquid state machines, gated RNNs, and long short-term memory are popular variants of RNNs. RNNs have been used to solve many practical problems, e.g., speech recognition, human activity recognition, and bioinformatics \cite{goodfellow2016deep}. In wireless and communication networks, RNNs have also found many applications, for example, a bidirectional neural model was used \cite{Ale2019Online} to learn the proactive caching policy at the network edge, and a variant of RNNs was employed \cite{Hoang2019Recurrent} for Wi-Fi indoor localization. 

The unfolding concept has been used to improve many iterative algorithms. The key idea is that each iterative cycle of an iterative algorithm is modeled as a layer of the neural network, which is trained to enable the algorithm to converge to the optimum. One such application \cite{samuel2017deep} was based on a DetNet model that was proposed for signal detection by unfolding the projected gradient descent method. The main advantage is that this model obviates the need to determine the network configurations such as the number of hidden layers. Specifically, the number of layers of the neural network is equivalent to the number of iterative cycles of the iterative algorithm, and the number of neurons is specified by the sizes of the input, output, and optimizing variables. Furthermore, deep unfolding incorporated in certain advanced model-based algorithms and transfer learning can improve the model efficiency e.g., faster convergence while requiring a smaller dataset to deliver the same performance \cite{zappone2019wireless}.

Apart from the three types of ANNs described above, numerous other ANNs have been proposed with different design philosophies, e.g., an autoencoder and deep generative models such as the restricted Boltzmann machine and deep belief network. 
An autoencoder, which is a specialized ANN for learning useful properties of the data, is effective with unlabeled examples. A restricted Boltzmann machine is a kind of deep generative model for learning the probability distribution over a set of examples. 
We invite interested readers to refer to a recent book \cite{goodfellow2016deep} and surveys on ANNs and their practical applications including speech recognition, pattern recognition, computer vision, agriculture, arts, and nanotechnology \cite{liu2017survey, abiodun2018state}. 

\begin{figure} [t]
	\centering
	\includegraphics[width=0.925\linewidth]{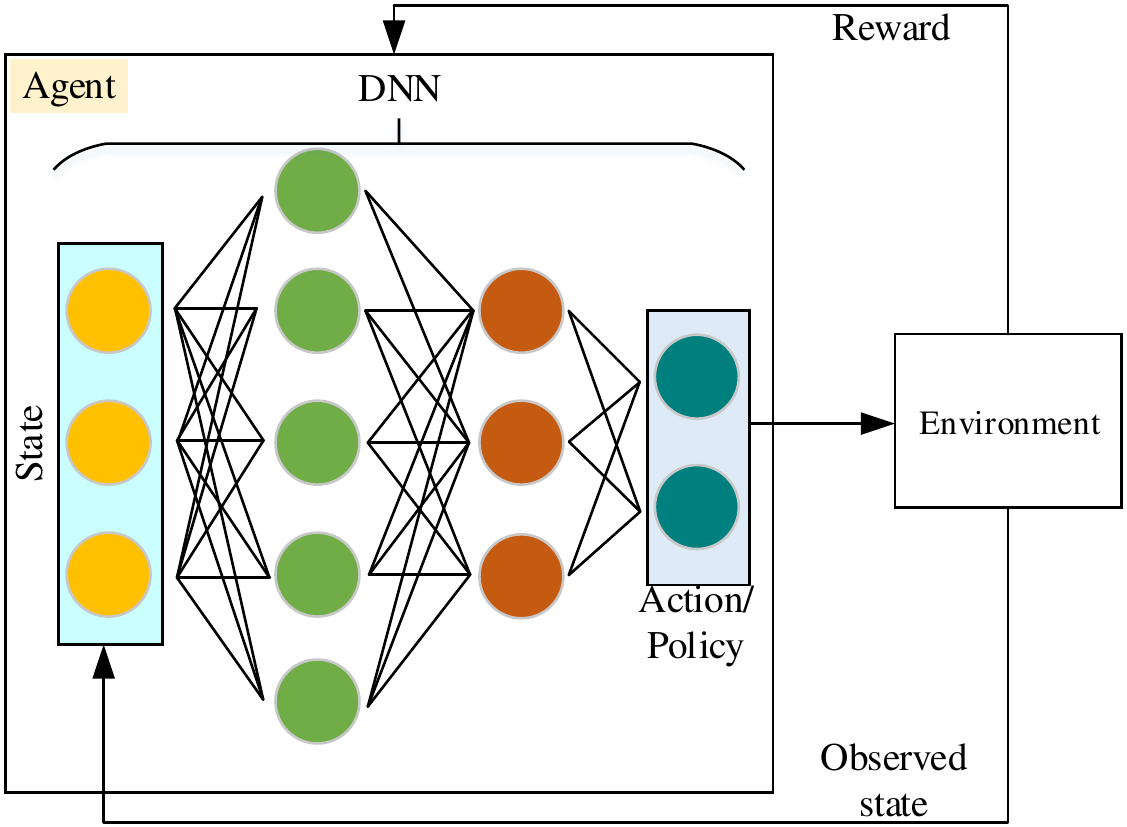}
	\caption{Illustration of the structure of a DRL algorithm.}
	\label{Fig:DRL}
\end{figure}

\subsubsection{Deep Reinforcement Learning}
DRL leverages the strengths of DNNs to improve the performance of RL algorithms \cite{mnih2015human}. Three main approaches exist to solve RL problems: the value function-based approach, policy-based approach, and hybrid actor--critic approach. The value function-based method relies on the estimation of the expected reward of each state, whereas the policy search-based method directly finds the optimal policy. The actor--critic method learns both the policy and value functions, and effectively overcomes the imbalance between variance and bias owing to the policy search and value function methods. For high-dimensional problems, DNNs can be exploited to learn the optimal value function, the optimal policy, or both in case of the actor--critic method \cite{Luong2019ApplicationsDRL}. An illustration of the structure of the DRL algorithm is presented in Fig.~\ref{Fig:DRL}, where DNNs are used to approximate the control policy. Inspired by a proposal \cite{mnih2015human}, DRL has found many successes in various domains and has been widely used in signal processing studies. For further details, we invite interested readers to refer to a recent survey \cite{Luong2019ApplicationsDRL}. 

\subsection{Federated Learning}
\begin{figure} [t]
	\centering
	\includegraphics[width=0.70\linewidth]{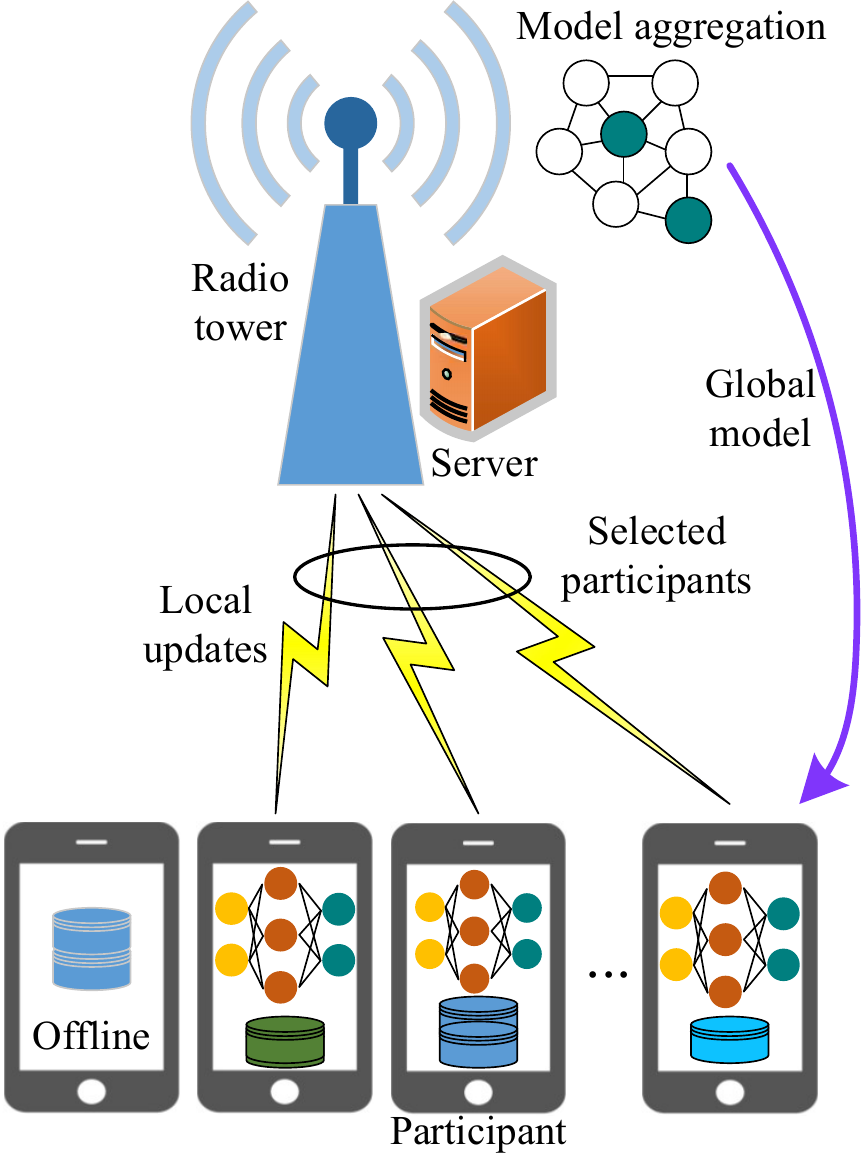}
	\caption{Illustration of an FL system.}
	\label{Fig:Architecture_FL}
\end{figure}

To protect sensitive information as well as preserve individual privacy, Google invented the concept of FL \cite{FederatedLearning}. FL enables an AI model to be trained without requiring all data to be stored and processed at a centralized server, which is typically referred to as the \text{aggregation server} in FL literature. In other words, the data of individual users remain in local storage in their end devices (i.e., the participant) and do not need to be transmitted to the server \cite{Pham2021UAV}. In FL, the server receives locally computed models from a number of devices, which are then aggregated to update the global model. In this way, FL can preserve data privacy and user security, although FL still relies on the trust of the aggregation server. 
FL has found many promising applications in various fields. The notable success of FL in Google's next word prediction application has motivated the adoption of FL for many other applications \cite{FederatedLearning} including smart retail, multiparty database querying, smart healthcare, and vehicular networks \cite{yang2019federated}. The FL system is illustrated in Fig.~\ref{Fig:Architecture_FL}.

Since the first FL paper from Google appeared in 2017, a large number of studies on FL have been conducted over the last few years. For a communication perspective, interested readers may refer to review articles \cite{lim2020federated,wang2020convergence} and references therein. Lim \emph{et al.} \cite{lim2020federated} first discussed the challenges of FL implementations, including communication cost, resource allocation, and privacy and security issues. In this paper, we further review existing FL applications at the network edge such as cyber-attack detection, edge caching and computational offloading, user association, and vehicular networks. The interplay between edge computing and DL has been reviewed in terms of \textit{intelligent edge} and \textit{edge intelligence} \cite{wang2020convergence}. 

\section{Modulation Classification}
\label{Sec:Modulation_Classification}
This section presents a review of applications of AI techniques for automatic modulation classification.

\subsection{Fundamentals of Modulation Classification}

The last decades have seen tremendous advancement and development of innovative communication standards and technologies to satisfy the ever-increasing demand of many wireless applications and services.
Dense network deployment with aggressive spectrum reuse to meet the growing mobile traffic demand has resulted in various undesirable effects such as signal distortion and co-channel interference. 
Signal recognition and modulation classification allow us to more effectively monitor and manage spectrum usage and sharing, which can potentially enhance the network performance~\cite{AMC-Naderpour2020Feature}.
AMC, a fundamental process to analyze the characteristics of radio signals in the physical layer, plays a vital role in intelligent spectrum monitoring and management and is typically deployed in AI-powered wireless communication systems.
This is because it enables the system to blindly identify the modulation format of an incoming radio signal at the receiver~\cite{AMC-Dobre2007Survey}. 
From the ML perspective, modulation classification can be framed as a multiclass decision-making problem, where the intrinsic radio characteristics are obtained using conventional feature engineering algorithms for learning a trainable classification model.
The ability to correctly distinguish advanced modulations (e.g., high-order digital modes) under harmful transmission environments, such as a multipath fading channel with additive noise, remains a challenging research topic~\cite{AMC-Aslam2012KNN} and has received considerable attention from the signal processing and communication communities.

Modern communication systems employ different advanced analog and digital modulation techniques to achieve good tradeoff between spectrum efficiency and transmission reliability.
Fundamentally, an analog modulation technique encodes an analog baseband signal onto a high-frequency periodic waveform (i.e., carrier signal), whereas digital modulation techniques allow a digital low-frequency baseband signal to be transmitted over a high-frequency carrier waveform.
These two modulation families can modify different waveform characteristics of the carrier signal, including the amplitude, frequency, phase, and a combination of amplitude and phase. 
At the receiver, the considered modulated signal must be assigned to the most appropriate modulation class by exploiting certain radio characteristics and a trained classifier.
The complex envelope of received radio signal $y(n)$ can be written as follows:
\begin{equation}
y(n) = x(n,H_k) + g(n),
\label{eqn:01}
\end{equation}
where $g(n)$ is the additive white Gaussian noise (AWGN).
The noiseless signal $x\left ( n,H_k \right )$ under transmission channel effects can be expressed as follows:
\begin{equation}
x\left ( n,H_k \right )=Ae^{2{\pi}{f_o}n\kappa + \zeta_n}\sum_{k=-\infty }^{\infty }{x\left ( k \right )h\left ( n\kappa-k\kappa+\xi_\kappa \kappa \right )},
\label{eqn:02}
\end{equation}
where $A$ is the signal amplitude, ${f_o}$ is the carrier frequency offset, $\kappa$ is the symbol spacing (or interval), $h\left ( \cdot  \right )$ refers to the synthetic effect of the residual baseband channel, $\zeta_n$ is the varying phase offset, $x\left ( k \right )$ refers to the symbol sequence of the original data over a specific modulation scheme, and $\xi_\kappa$ is the timing error (or timing offset between the transmitter and the receiver).
In general, an AMC scheme is developed to accurately predict the modulation format of $x\left ( n,H_k \right )$ that is performed by a trained classifier that effectively learns the informative features of $y\left (n \right )$ by using some specific ML algorithm.
However, this classification task is challenging because it is necessary to process many high-order modulation schemes, considering synthetic channel deterioration.

\subsection{State-of-the-Art AMC Methods}
Numerous methods have been proposed for modulation classification in communications, where AI methods have been widely used to improve the performance in terms of classification accuracy and processing speed.
Based on the progressive development of AI in the last decades, especially the recent explosion of DL, AMC methods that have been reported in the literature can be grouped into two major categories as follows:
\begin{itemize}
	\item \textit{Conventional approaches}: Various methods in this group have employed conventional AI techniques and traditional ML algorithms, which can be further divided into two sub-classes: likelihood-based and feature-based approaches.
	For the likelihood-based approaches, the output of modulation classification is determined with the aim of maximizing the probability of a received signal associated with a certain modulation scheme. The underlying distribution parameters of the scheme are estimated by using expectation/conditional maximization (ECM) algorithms~\cite{AMC-Hameed2009Likelihood}.
    Formulated as a composite problem for hypothesis testing, the maximum-likelihood modulation classification draws the decision as follows:
    \begin{equation}
	\mathcal{\hat{H}} = \argmax_{\mathcal{H}_{ij}}\ln p\left ( y_1,\dots,y_T|\mathcal{H}_i \right ),
	\label{eqn:03}
	\end{equation}
    where $\mathcal{H}_{ij}$ is the hypothesis model associated with the modulation format $\mathcal{M}_i$ ($i=1,\dots,K$), where $K$ is the number of modulation formats, deduced from the observation signal $y_j$ ($j=1,\dots,T$), where $T$ is the number of observations, and $\ln p\left ( y|\mathcal{H}_i \right )$ refers to the log-likelihood function~\cite{AMC-Zhu2014Likelihood}.
    In fact, the likelihood-based approaches can achieve optimal performance with perfect knowledge based on information of signal and channel models~\cite{AMC-WenWei2000Likelihood}, but they are computationally expensive in terms of parameter estimation~\cite{AMC-Soltanmohammadi2013Likelihood}.
    
    Compared with likelihood-based approaches, feature-based methods have been widely deployed in practical systems thanks to their easy implementation, lower complexity, and stronger robustness with various transmission channel scenarios. 
    A typical ML framework requires feature engineering and classification processes, where certain handcrafted feature extractors (i.e., descriptors) are used to mine radio characteristics and traditional classifiers can be employed to learn the modulation patterns in the supervised manner.
    Certain methods in this subclass have achieved a good trade-off between model accuracy and complexity by using advanced feature selection schemes and sophisticated ML algorithms~\cite{AMC-Hameed2009Likelihood}.
    
    \item \textit{Innovative DL-based approaches}: Inspired by great success in the fields of image processing and computer vision~\cite{AMC-Hao2019Bio,AMC-Huynh2020TII,AMC-Huynh2020INS}, the DL technique has been exploited for modulation classification, wherein several deep network architectures, such as RNN, long short-term memory (LSTM), and CNN, have been considered.
    Compared with traditional ML, DL has important advantages because it can automatically learn high-level features for more effective modulation discrimination and it can effectively process wireless big data~\cite{AMC-OShea2016CNN}. 
    With an appropriately built computing platform with graphics processing units (GPUs), the execution speed of both learning and inference (i.e., prediction) processes can be accelerated significantly to satisfy the high reliability and low latency requirements of emerging wireless applications and services.
\end{itemize}
\subsubsection{Conventional AMC approaches}

In the last decades, many conventional AMC methods have been proposed to enable dynamic spectrum access and intelligent spectrum management, where the expectation-maximization (EM) algorithms were employed to build maximum in likelihood-based classifiers~\cite{AMC-Zhang2017ICC,AMC-Zhang2017Likelihood,AMC-Zhang2018Likelihood,AMC-Zheng2018Likelihood,AMC-Tian2018Likelihood,AMC-Chen2019Likelihood,AMC-Abdul2019Likelihood}.
Zhang \textit{et al.}~\cite{AMC-Zhang2017ICC} took advantage of the EM algorithm to estimate the maximum-likelihood of the unknown for modulation classification in a cooperative multiuser scenario.
For each hypothesis, the EM algorithm performs an expectation step (E-step) and a maximization step (M-step) at each iterative step to estimate the unknown channel amplitude $a^{t+1}_{c,km}$ and phase $\phi^{t+1}_{c,km}$ that are associated with the radio signal encoded by the modulation format $c$ and transmitted from the $m$-th user to the $k$-th receiver.
The calculations in these two steps can be expressed as follows:
\begin{align}
\mathrm{E-step:}~J\left ( \bm{\theta} |\bm{\theta} ^{\left ( t \right )}_c \right ) = &\mathbb{E}_{\bm{z}|\bm{x},\bm{\theta} ^{\left ( t \right )}_c}\left [ \ln p\left ( \bm{z}|\bm{\theta}  \right ) \right ], \label{eqn:04_1}\\
\mathrm{M-step:}~\bm{\theta} ^{\left ( t+1 \right )}_c = & \argmax_{\bm{\theta}} J\left ( \bm{\theta} |\bm{\theta} ^{\left ( t \right )}_c \right ), \label{eqn:04}
\end{align}
where $J\left ( \bm{\theta} |\bm{\theta} ^{\left ( t \right )}_c \right )$ refers to the expected value of the log likelihood function of unknown parameters $\bm{\theta}$ with respect to the conditional distribution $\bm{z}$ given the received samples $\bm{x}$ and the current estimates of parameters $\bm{\theta} ^{\left ( t \right )}$.
By decoupling the multivariate maximum-likelihood problem into multiple separated optimization problems, the proposed method can estimate the complete data and unknown parameters more effectively.
The Cramér-Rao lower bounds (CRLBs) for estimating unknown multipath channels were applied to enable the EM algorithm to reach the performance upper bound of modulation classification~\cite{AMC-Zhang2017Likelihood}.
To distinguish continuous phase modulation signals, a maximum-likelihood-based classifier~\cite{AMC-Zhang2018Likelihood} was introduced with the Baum--Welch (BW) algorithm to estimate the unknown fading channel coefficients.
In the E-step of the EM algorithm, the BW method was applied to iteratively maximize the auxiliary function as follows:
\begin{equation}
    J\left ( \bm{\theta} |\bm{\theta} ^{\left ( t \right )}_c \right )=\sum_{v}p_c\left ( \bm{x},\bm{v}|\bm{\theta} ^{\left ( t \right )}_c \right )\log p_c\left ( \bm{x},\bm{v}|\bm{\theta} \right ),
    \label{eqn:05}
\end{equation}
where $\bm{v}$ refers to the hidden variables obtained by a hidden Markov model (HMM).

For identification of the modulation parameters, including the signal constellation and the number of subcarriers, in orthogonal frequency division multiplexing (OFDM) with index modulation, Zheng \textit{et al.}~\cite{AMC-Zheng2018Likelihood} recommended two likelihood-based classifiers of an average likelihood ratio test (ALRT) and a hybrid likelihood ratio test (HLRT). 
For the known channel state information (CSI) scenario, the HLRT classifier first determines the indices of active subcarriers and estimates transmitted symbols subsequently, whereas the ALRT classifier averages out the set of subcarrier indices with higher complexity.
On the contrary, a blind AMC only employs the HLRT classifier to distinguish interesting subcarriers corresponding to each hypothesis of modulation parameter combination via an energy-based detector.
To accelerate the convergence process of the ECM algorithm to blindly estimate the channel parameters in flat fading and nonGaussian noise impairments, Chen \textit{at el.}~\cite{AMC-Chen2019Likelihood} upgraded the squared extrapolation method by adding a parameter-checking scheme.
The enhanced method also derives a convergence point of log-likelihood function more reliably compared with the original one.

The design of AMC strategies for MIMO systems is more challenging than those for single-input single-output systems because the incoming signal at the receiver is a mixture of multiple signals transmitted by different antennas.
Therefore, the effects of channel impairment on the modulation signals are dissimilar.
As a result, it is difficult to properly estimate channel characteristics via EM algorithms.
Another approach involved grouping the received signals with the same observation interval outline and mining the hidden relationship between uncorrelated modulation classes. This enabled the modulation classification to be studied as a multiple-clustering problem, where the final modulation decision making is accomplished by evaluating the maximum-likelihood of multiple clusters corresponding to modulations in a given dataset~\cite{AMC-Tian2018Likelihood}.
The learning efficiency can be increased by recovering the centroids of all clusters by a constellation-structure-based reconstruction algorithm for parameter reduction with good convergence performance. 
Adaptive CSI estimation for modulation classification in MIMO systems was introduced by Abdul Salam \textit{et al.}~\cite{AMC-Abdul2019Likelihood} to jointly exploit the Kalman filter (KF) and an adaptive interacting multiple model (IMM).
The IMM--KF output was subsequently analyzed by a quasi-likelihood ratio test (QLRT) algorithm for modulation identification.
It is worth noting that EM is derived for recursively computing estimates in IMM--KF and making decisions by the QLRT-based classifier.

\begin{figure*}[!t]
    \captionsetup[subfigure]{justification=centering}
	\centering
	\subfloat[ \label{amc-coacm}]{\includegraphics[width=0.06\linewidth]{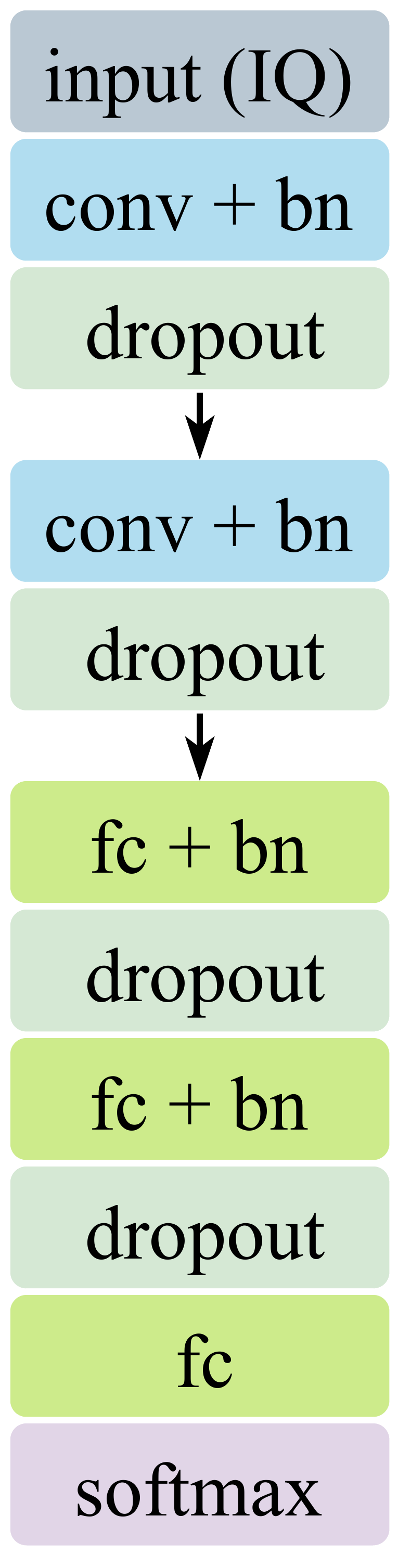}} \hspace{4pt}
	\subfloat[ \label{amc-vtcnn}]{\includegraphics[width=0.06\linewidth]{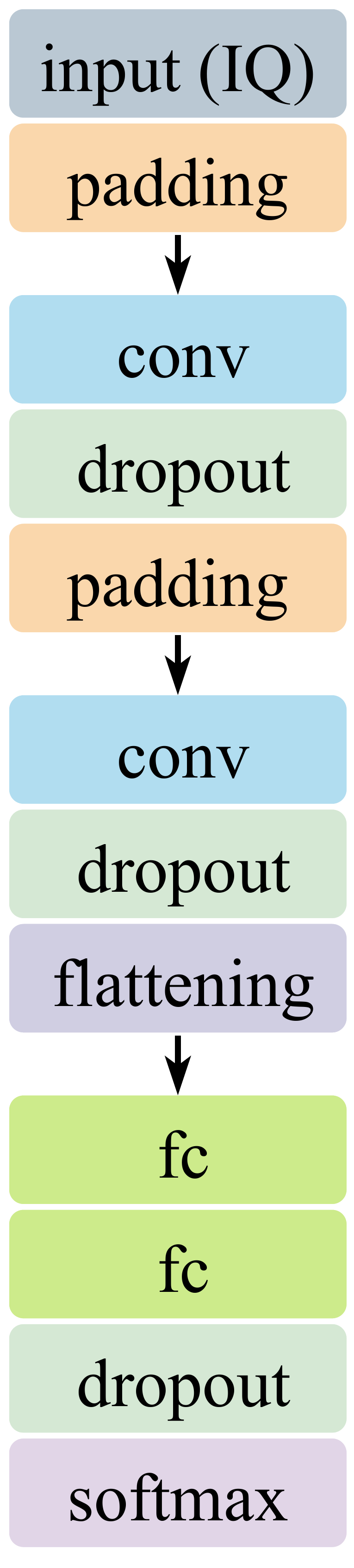}} \hspace{4pt}
	\subfloat[ \label{amc-2branch}]{\includegraphics[width=0.11\linewidth]{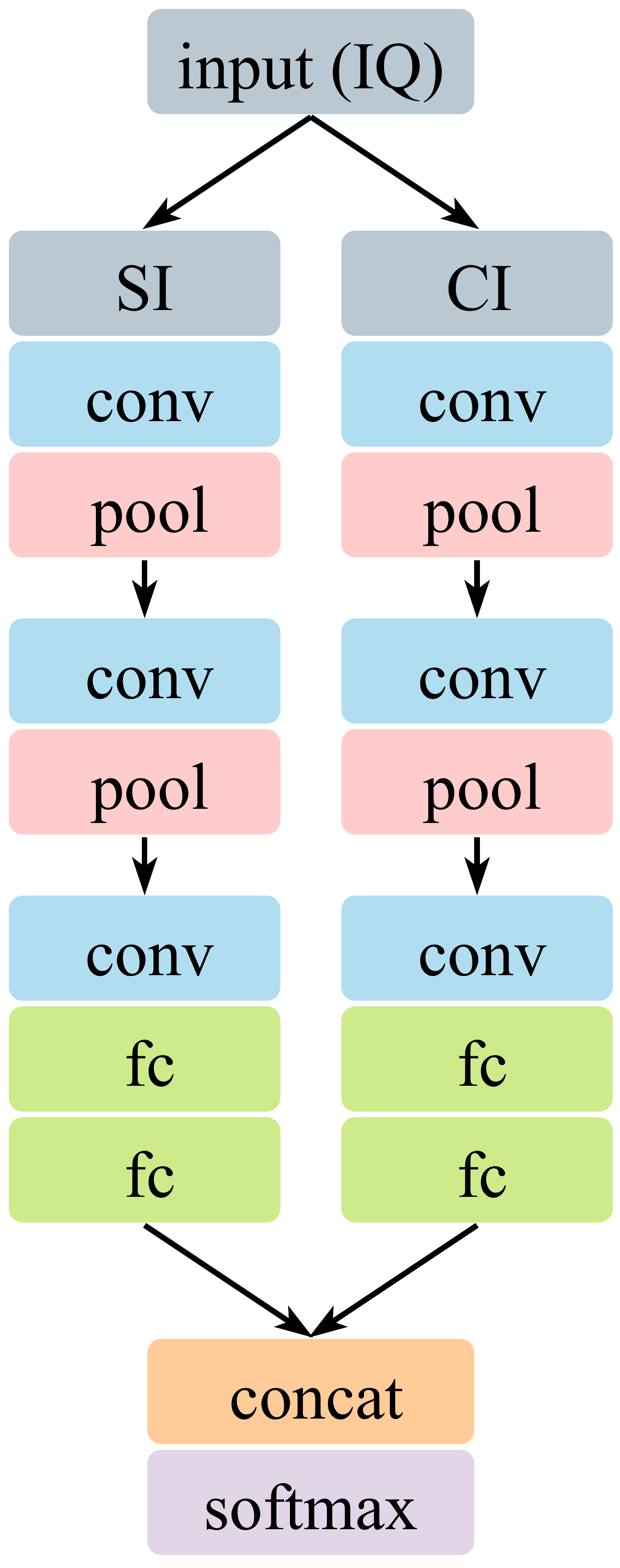}} \hspace{4pt}
	\subfloat[ \label{amc-drcnns}]{\includegraphics[width=0.14\linewidth]{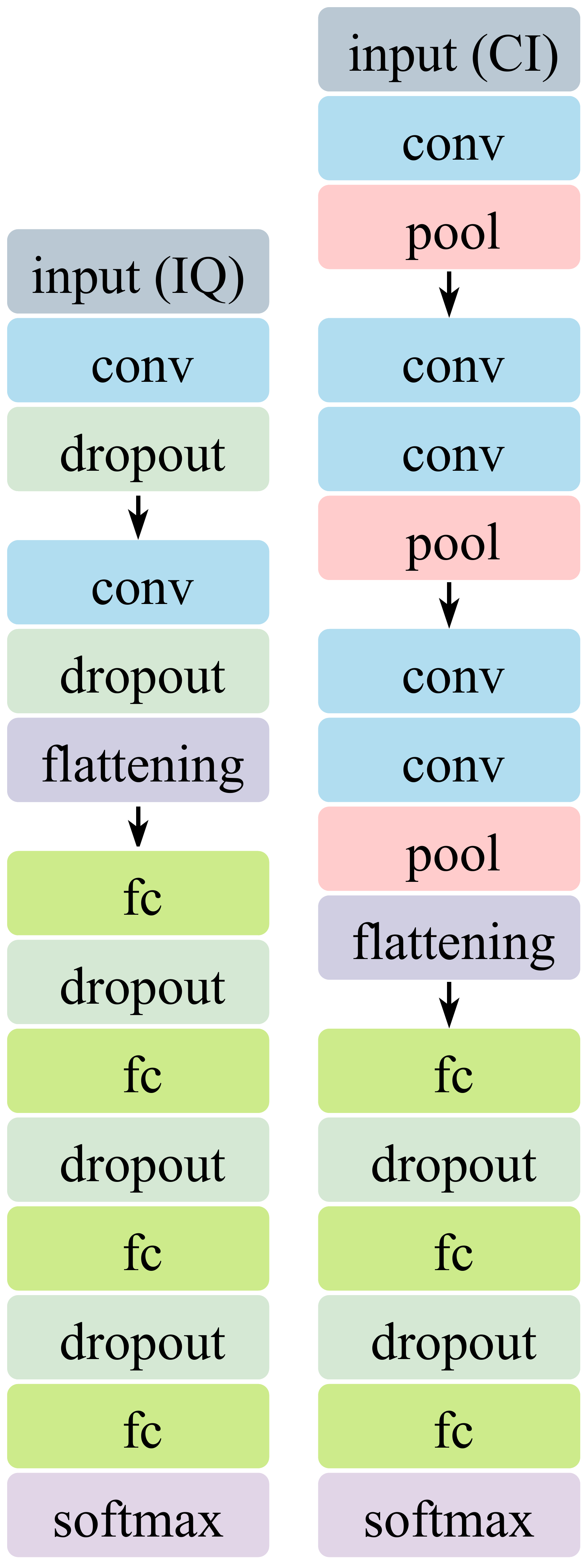}} \hspace{4pt}
	\subfloat[ \label{amc-vgg}]{\includegraphics[width=0.06\linewidth]{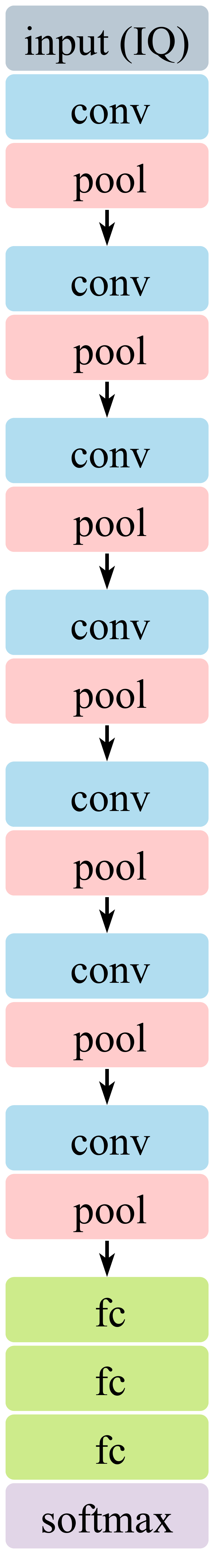}} \hspace{4pt}
	\subfloat[ \label{amc-resnet}]{\includegraphics[width=0.06\linewidth]{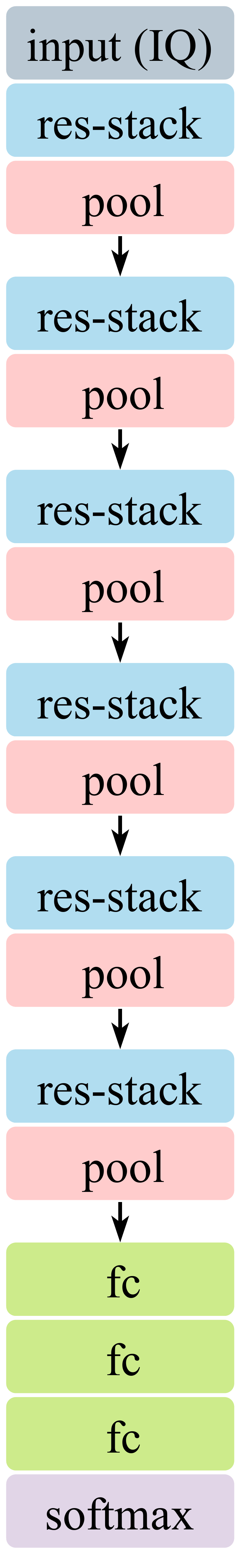}} \hspace{4pt}
	\subfloat[ \label{amc-cnnamc}]{\includegraphics[width=0.13\linewidth]{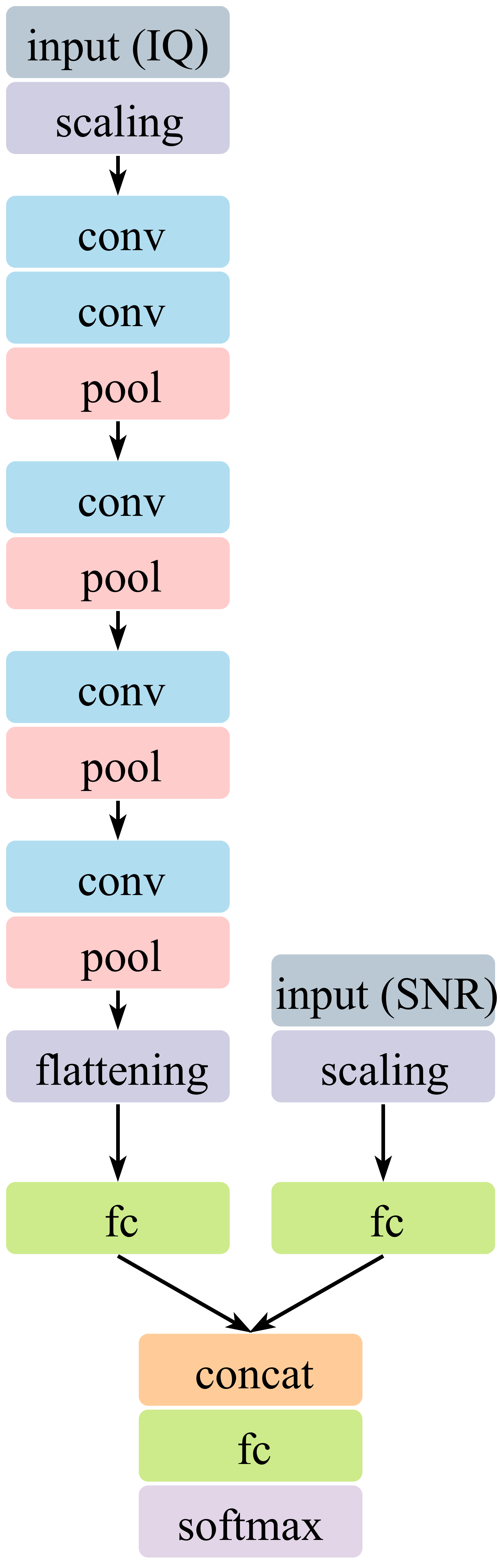}} \hspace{4pt}
	\subfloat[ \label{amc-mcnet}]{\includegraphics[width=0.12\linewidth]{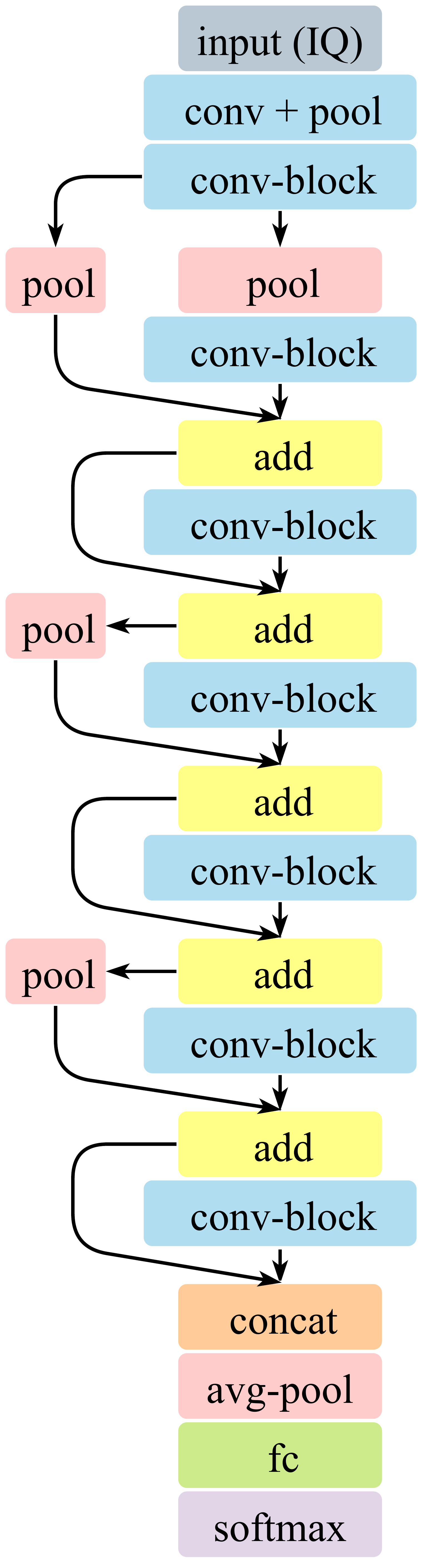}} \hspace{4pt}
	\subfloat[ \label{amc-block}]{\includegraphics[width=0.12\linewidth]{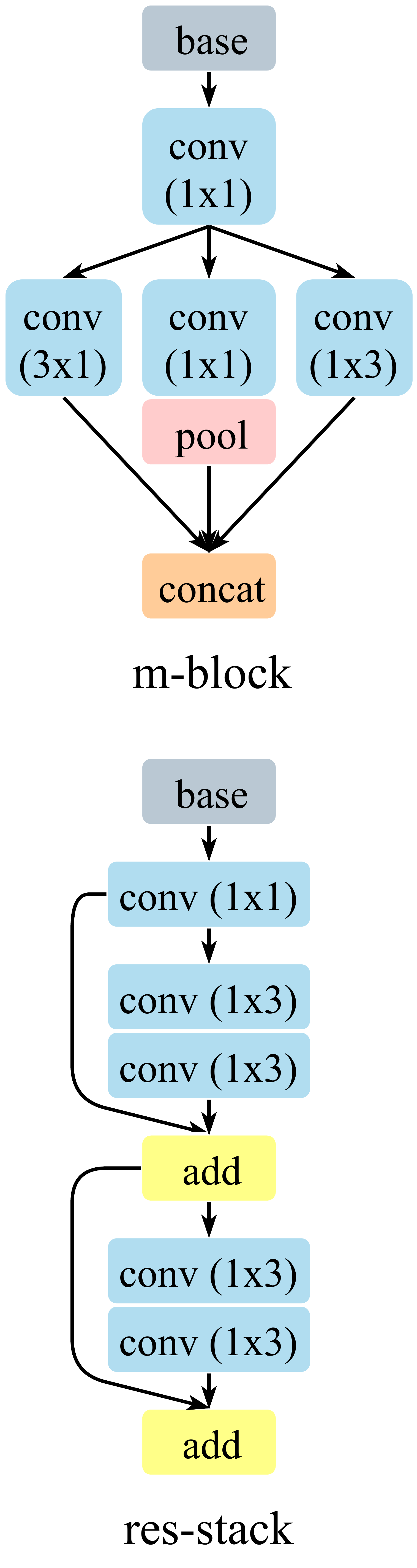}}
	\caption{CNN architectures of several state-of-the-art modulation classification approaches: (a) Co-AMC~\cite{AMC-Wang2020DeepMIMO}, (b) VTCNN2~\cite{AMC-Lin2020NNPruning}, (c) Two-branch CNN~\cite{AMC-Wu2019ConstellationFusion}, (d) CNN1 (for IQ data) and CNN2 (for CI data) in a hierarchical classifier~\cite{AMC-Wang2019IQConstellation}, (e) VGG~\cite{AMC-OShea2018DeepSig}, (f) ResNet~\cite{AMC-OShea2018DeepSig}, (g) CNN-AMC~\cite{AMC-Meng2018DL} with a supplementary input of SNR values, MCNet~\cite{AMC-Huynh2020MCNet}, and (i) the structures of m-block and res-stack in MCNet and ResNet. The input of these networks can be the signal sequence of IQ samples, an image of the constellation diagram (CI), and an image of the spectrogram (SI). Notations presented in figures: conv (convolutional layer), bn (batch normalization layer), pool (max-pooling layer), avg-pool (average pooling layer), concat (depthwise concatenation layer), fc (fully connected layer or dense layer), add (elementwise addition layer). Customized layers are defined as follows: padding layer for adding zero values to input borders, flattening layer for collapsing the spatial dimensions of input to the channel dimension (or the depth dimension of feature maps), scaling layer for normalizing data values into a specific range. Each conv is followed by an activation layer such as ReLU, leaky ReLU, and exponential linear unit (ELU).}
	\label{Fig:AMC_CNN_Architecture}
\end{figure*}

Apart from likelihood-based approaches, numerous AMC methods that follow a typical ML framework with two principal steps, feature extraction and model learning, have been introduced.
These feature-based methods mostly rely on sophisticated handcrafted feature engineering techniques for an improved description of radio characteristics, and conventional classifiers are employed for learning modulation patterns from extracted features.
High-order cumulants (HOCs) of the amplitude, phase, real, and imaginary components of received signals were calculated for radio characteristic representation~\cite{AMC-Han2017SVM}.
To flexibly accommodate different channel scenarios, linear SVM (LSVM) and the approximate maximum-likelihood (AML) algorithms were developed assuming that the channel condition is known, whereas a backpropagation neural network (BPNN) was considered for unknown channel conditions.
Remarkably, phase and frequency offsets were estimated from high-order moments (HOMs) to enhance the performance of modulation classification in the unknown channel scenario. 
In another study, Huang \textit{et al.}~\cite{AMC-Huang2017Cumulant} performed independent component analysis (ICA) to select highly effective HOCs of arbitrary orders and lags.
Furthermore, a maximum-likelihood-based multicumulant classification (MLMC) algorithm was proposed to identify the most appropriate modulation format by maximizing the posterior probability of the multicumulant vector.
The elementary cumulant and cyclic cumulant calculated at the second and fourth orders were fused in a hierarchical hypothesis-based classification framework~\cite{AMC-Majhi2017Feature} to improve the system performance in terms of the classification rate for a flat fading channel.
Despite being faster than conventional EM-based methods, this approach still requires huge amounts of computing resources for macro- and micro-classifiers in hierarchical association.

Hybrid approaches that combine the likelihood-based and feature-based classifiers have been proposed to achieve a good trade-off between accuracy and processing speed.
For example, Abu-Ramoh \textit{et al.}~\cite{AMC-Abu2018Momemt} derived a maximum-likelihood classifier capable of handling HOMs as features.
Compared with conventional algorithms that manipulate the received sequence of modulation symbols, this approach can exploit few statistical moments to evaluate the likelihood function.
As a result, the proposed AMC method induces lower complexity if a large number of modulations are considered for classification.
This hybrid strategy for modulation classification was also extended ~\cite{AMC-Abdelbar2018Cumulant} by leveraging a mixture of HOCs and HOMs at the second and fourth orders to design the maximum-likelihood classifier.
Notably, a hierarchical classification framework is considered to improve the accuracy; however, the overall system complexity significantly increases in situations in which three binary classifiers are required to process four effortless modulations. 
Zhang \textit{et al.}~\cite{AMC-Zhang2018Dictionary} built a dictionary set of high-order statistics for learning modulation patterns using block coordinate descent dictionary learning, where the modulation format of a signal is given by referring the sparse representation of statistical features to the dictionary.

Sophisticated feature extraction algorithms to more accurately describe the radio signal characteristics have recently been proposed.
To overcome the limitations of global features, including HOCs and HOMs, Xiong \textit{et al.}~\cite{AMC-Xiong2019Infocom} recommended two novel signal signatures: the first is modulation-specific transition features in the time domain, and the second is sequential features from the Fisher kernel.
With Gaussian mixture model dictionary learning, exploitation of these local features results in an improvement of up to 30$\%$ in terms of classification accuracy compared to conventional approaches.
Another approach involved capturing the normalized HOCs on the frequency domain by using discrete Fourier transform (DFT) for modulation classification in an OFDM system.
Simulations were used to show that modulation classification strategies based on HOCs of DFT are more effective than those calculated in the time domain; however, the computational complexity significantly increases as the number of cumulant values increases~\cite{AMC-Gupta2019Cumulant}.

\subsubsection{Innovative DL-based AMC approaches}
Over the last few years, DL~\cite{AMC-LeCun2015} has emerged as a vital ML tool for many applications ranging from natural language processing to vision recognition and bioinformatics.
With many advantages such as automatic learning of high-level features and the effective exploitation of big data, DL has achieved remarkable success in many applications where it has become a core technique of model learning and pattern analysis.
For AI-powered communications, DL is being exploited to address many challenging design tasks including network traffic control~\cite{Fadlullah2017StateOfDL} and intelligent resource allocation~\cite{AMC-Lee2020}.
For AMC, DNN~\cite{AMC-Shah2019RBF,AMC-Ali2017Autoencoder,AMC-Shah2020FeaSelection} has been recommended to replace traditional classifiers for learning statistical features.
For example, two sparse autoencoder-based DNNs were developed~\cite{AMC-Shah2019RBF,AMC-Ali2017Autoencoder} to improve the accuracy of high-order and intraclass digital modulations. 
Although their performance is slightly higher than that of LSVM and approximately maximum-likelihood classifiers, they are computationally more complex because of the requirement to compute a large number of neurons in hidden layers.
Selection of the most relevant HOC features for learning a sparse autoencoder DNN~\cite{AMC-Shah2020FeaSelection} makes it possible to substantially reduce the overall complexity of the classifier without performance loss.
LSTM, an advanced architecture of RNN that exploits the long-term dependencies between temporal attributes in sequential data, was further studied for modulation classification~\cite{AMC-Hu2020DNN}.
Three stacked LSTM layers configured in the underlying architecture allows the network to capture the temporal relation of in-phase and quadrature (IQ) samples while remaining flexible by accepting variable length input.

\begin{table*}[!ht]
	\caption{Summary of state-of-the-art AMC methods for communication systems.}
	\label{Table:Summary_AMC_Methods}
\resizebox{\textwidth}{!}{	
	\begin{tabular}{|c|c|c|p{7cm}|p{6.5cm}|}
		\hline 
		Category & Year  & Paper & Highlights & Limitations  \\ 
		\hline
		\hline
		
		\multirow{15}{*}{\makecell{Likelihood- \\ based}} & \multirow{4}{*}{2017}  & \multirow{2}{*}{\cite{AMC-Zhang2017ICC}} & Decoupling the interactive multivariate maximum-likelihood problem into multiple separated optimization problems. & Expensive computational complexity of multiple optimizations. \\
		
		\cline{3-4} 
		
		&   & \cite{AMC-Zhang2017Likelihood} & Improving \cite{AMC-Zhang2017ICC} with CRLBs of the joint unknown estimates. & Conventional accuracy on few effortless modulations. \\ 
		
		\cline{2-5} 
		
		&  \multirow{8}{*}{2018} & \multirow{2}{*}{\cite{AMC-Zhang2018Likelihood}} & Formulating continuous phase modulation signals as HMM variables. & Forward--backward algorithm in HMM is more expensive. \\ 
		& & & Applying BW for unknown fading estimation. &\\

		\cline{3-5} 
		
		&   & \multirow{3}{*}{\cite{AMC-Zheng2018Likelihood}} & Two classifiers based on ALRT and HLRT for known/unknown CSI scenarios. & Low accuracy of blind AMC (without CSI information). Poor trade-off between classification rate and computational cost. \\ 

		\cline{3-5} 
		
		&   & \multirow{2}{*}{\cite{AMC-Tian2018Likelihood}} & Clustering received modulation signals of same observation characteristics. & System complexity progressively increases along the number of modulations. \\ 
		& & & Constellation-structure-based centroid reconstruction. & \\
		
		\cline{2-5} 
		
		&  \multirow{3}{*}{2019} & \multirow{2}{*}{\cite{AMC-Chen2019Likelihood}} & Squared extrapolation method with a parameter checking scheme & \multirow{2}{*}{Insubstantial classification at low SNRs.}\\ 
		
		\cline{3-5} 
		
		&   & \multirow{2}{*}{\cite{AMC-Abdul2019Likelihood}} & Estimating CSI robustly via an IMM--KF model.  &  Performance is sensitive to the parameter initialization 
		\\ 
		& & & QLRT-based classifier. & \\
		\hline	
		
		\multirow{15}{*}{\makecell{Feature- \\ based}} & \multirow{7}{*}{2017}  & \multirow{2}{*}{\cite{AMC-Han2017SVM}} & LSVM and AML algorithms for a known channel condition. BPNN for an unknown channel condition.& Conventional performance of blind modulation classification (without channel information). \\
		& & & HOMs for phase and frequency offsets estimation. & \\
		
		\cline{3-5} 
		
		&   & \multirow{2}{*}{\cite{AMC-Huang2017Cumulant}} & MLMC classifies modulation based on posterior probability. ICA for HOC feature selection. & Feature efficiency strongly depends upon the quality of channel estimation. \\ 
		
		\cline{3-5} 
		
		&   & \multirow{2}{*}{\cite{AMC-Majhi2017Feature}} & Hierarchical hypothesis-based classification framework. & High computational complexity. \\ 
		& & & Fusing elementary and cyclic cumulants. & \\
		
		\cline{2-5} 
		
		&  \multirow{6}{*}{2018} & \multirow{1}{*}{\cite{AMC-Abu2018Momemt}} & HOMs-based maximum-likelihood classification algorithm. & Expensive 
		cost if considering few modulations. \\ 
		
		\cline{3-5} 
		
		&   & \multirow{2}{*}{\cite{AMC-Abdelbar2018Cumulant}} & Mixture of HOMs and HOCs for likelihood-based hierarchical classification framework. & Complicated classification framework with three binary classifiers. \\ 
		
		\cline{3-5}
		
		&   & \multirow{2}{*}{\cite{AMC-Zhang2018Dictionary}} & Organizing a dictionary of high-order statistics in sparse representation. & \multirow{2}{*}{High memory consumption for dictionary ensemble.} \\ 
		
		\cline{2-5} 
		
		&  \multirow{2}{*}{2019} & \multirow{2}{*}{\cite{AMC-Xiong2019Infocom}} & Local transition features and Fisher-based sequential features. Gaussian mixture model dictionary learning.& Accuracy is sensitive to the dictionary size specified in dictionary initialization.\\ 
		
		\cline{3-5} 
		
		&   & \multirow{1}{*}{\cite{AMC-Gupta2019Cumulant}} & Normalizing HOCs in the frequency domain via DFT. & Computing the DFT is computationally expensive. \\ 
		
		\hline
		
		\multirow{30}{*}{\makecell{DL- \\ based}} & \multirow{2}{*}{2017}  & \multirow{2}{*}{\cite{AMC-Ali2017Autoencoder}} & $k$-Sparse autoencoder for low-complexity input reconstruction with DNN.& Requires a large number of symbols to train the classification model. \\
		
		\cline{2-5} 
		
		&  \multirow{4}{*}{2018} & \multirow{3}{*}{\cite{AMC-OShea2018DeepSig}} & Introducing the RadioML 2018.01A dataset of modulation classification. &  Following network backbones for image classification without architecture fine tuning.\\ 
		& & & Analyzing the performance of VGG and ResNet for modulation classification. &   \\
		
		\cline{3-5} 
		
		&   & \multirow{1}{*}{\cite{AMC-Meng2018DL}} & CNN-AMC for processing long symbol-rate signals. & Requires SNR information. \\
		
		\cline{2-5}
		
		&  \multirow{12}{*}{2019} & \multirow{2}{*}{\cite{AMC-Shah2019RBF}} & Cooperative classifier with radial basic function network (RBFN) and sparse autoencoder DNN.& Initializing many neurons in hidden layers inducing a heavy load.\\ 
		\cline{3-5} 
		
		&   & \multirow{2}{*}{\cite{AMC-Zheng2019FusionCNN}} & Introducing a CNN-based multilevel fusion architecture for effectively learning intrinsic information from coarse to fine. & High computation and memory consumption for processing multiple CNN streams. \\  \cline{3-5} 
		
		&   & \multirow{2}{*}{\cite{AMC-Huang2019CCNN}} & Plotting IQ samples into a scattered diagram for constellation image with contrast enhancement.  &   \\ 
		\cline{3-4}
		&   & \cite{AMC-Huang2019CFCN} & Formulation of a synthetic loss function with contrastive loss. & \\ 
		
		\cline{3-4}
		&   & \multirow{1}{*}{\cite{AMC-Zeng2019Spectrum}} & Transformation of IQ data to spectrogram image by STFT. &  Additional computing resource for data transformation.\\  
		
		\cline{3-4} 
		&   & \multirow{2}{*}{\cite{AMC-Wang2019IQConstellation}} & Hierarchical classification with two CNNs for handling IQ data and constellation image. & Performance is overly sensitive to the output size (a.k.a. resolution) of constellation and spectrogram images.\\ 
		 
		\cline{3-4} 
		&   & \multirow{2}{*}{\cite{AMC-Wu2019ConstellationFusion}} & A compact-sized two-stream CNN to simultaneously learn the constellation diagram and cyclic spectra. &  \\  
		
		\cline{2-5}		
		&  \multirow{8}{*}{2020} & \multirow{1}{*}{\cite{AMC-Shah2020FeaSelection}} & HOC feature selection for learning sparse autoencoder DNN. & Taking into account a limited number of modulations.\\ 
		\cline{3-5} 
		&  & \multirow{2}{*}{\cite{AMC-Hu2020DNN}} & Deploying LSTM for learning long-term dependencies of IQ samples. & Poor accuracy when the channel condition is unknown. Extremely high training complexity.\\ 
		\cline{3-5} 
		&  & \multirow{1}{*}{\cite{AMC-Lin2020NNPruning}} & Cost-efficiency CNN-based modulation classifiers. & \multirow{2}{*}{Only applicable to convolutional layers.}\\ 
		\cline{3-3} 
		&  & \multirow{1}{*}{\cite{AMC-Wang2020LightAMC}} & Accelerating the processing speed with pruning technique. & \\ 
		\cline{3-5} 
		&   & \multirow{2}{*}{\cite{AMC-Huynh2020MCNet}} & Introducing MCNet with multiple associated blocks. & \multirow{2}{*}{High-order modulation classification is less robust.}  \\ 
		&  &  & Each block is specified by asymmetric kernels. &  \\
		\cline{3-5} 
		&  & \multirow{2}{*}{\cite{AMC-Wang2020DeepMIMO}} & Cooperative CNN-based approach for MIMO system with a weighted averaging decision rule. & Conventional accuracy of four-modulation classification under non-channel impairment condition. \\
				
		\hline
	\end{tabular}
}
\end{table*}

Among several DL architectures, CNN~\cite{AMC-OShea2018DeepSig,AMC-Meng2018DL,AMC-Lin2020NNPruning,AMC-Wang2020LightAMC} is more useful than DNN and RNN thanks to its ability to learn multiscale representational features from high-dimensional and unstructured data.
In addition to releasing RadioML 2018.01A, a rich modulation classification dataset containing more than 2.5 million radio signals covering up to 24 analog and digital modulation formats in a wide range SNR $\left [ -20:2:30 \right ]$ dB, O’Shea \textit{et al.}~\cite{AMC-OShea2018DeepSig} investigated the classification performance of two CNNs inspired by VGG~\cite{AMC-VGG} and ResNet~\cite{AMC-ResNet} originally proposed for image classification.
Compared with the baseline approach, which calculates high-order statistics for learning an ensemble model of gradient boosted trees (XGBoost), the accuracy of these CNNs is significantly superior at different SNR levels under synthetic channel impairments, such as carrier frequency offset, symbol rate offset, delay spread, and additive noise.
Notably, by exploiting skip connection in residual stacks, ResNet classifies modulations more precisely than VGG at high SNRs.
In addition, the accuracy of these two CNNs is investigated under different parameter configurations, in particular, the number of convolutional layers (in VGG), the number of residual stacks (in ResNet), and the signal length (i.e., the number of IQ samples in a partitioned signal), to analyze the performance sensitivity.
Meng~\textit{et al.}~\cite{AMC-Meng2018DL} introduced an end-to-end CNN, namely, CNN-AMC, for identifying the modulation of a long symbol-rate signal sequence, in which supplementary information in the form of the SNR is incorporated in fully connected layers via a concatenation operation to improve the accuracy.
Even though CNN-AMC has the potential to obtain remarkable accuracy, the training and prediction processes have very high computational complexity because of the huge number of connections between a flatten layer and a fully connected layer.
A compact-sized CNN, namely, VTCNN2~\cite{AMC-Lin2020NNPruning}, was developed for cost-efficient modulation classification in edge devices, in which a pruning technique is applied to optimize the processing speed.
This method allows the network to ignore the low-impact parameters (i.e., weight and bias) of the convolutional layers.
Despite achieving a good tradeoff between accuracy and computing cost (measured by the number of floating point operations), the network size is still heavy because of the large number of trainable parameters distributed across the two fully connected layers.
The pruning technique was further studied and led to the proposal of LightAMC~\cite{AMC-Wang2020LightAMC}, a CNN-based AMC method, to significantly accelerate the processing speed with negligible accuracy loss in IoT applications and unmanned aerial vehicle (UAV) systems.

Advanced CNN-based modulation classification methods have been recommended for performance enhancement by using a specialized novel structure of convolutional layers~\cite{AMC-Huynh2020WCNC,AMC-Huynh2020MCNet} and using fusion mechanisms~\cite{AMC-Wang2020DeepMIMO,AMC-Zheng2019FusionCNN}.
An efficient CNN, namely, MCNet~\cite{AMC-Huynh2020MCNet}, was introduced for robust automatic modulation recognition under various channel impairments, in which the network architecture is specialized by several processing blocks associated via skip connection to prevent MCNet from experiencing a vanishing gradient and preserve the information identity by using many nonlinear operations.
Moreover, to gain rich features and reduce the number of trainable parameters, each block in MCNet is configured by different one-dimensional asymmetric kernels (i.e., filter).
Skip connection was also studied to design several specific blocks for feature learning in MBNet~\cite{AMC-Tunze2020ICT}, Chain-Net~\cite{AMC-Huynh2020GLOBECOM}, SCGNet~\cite{AMC-Tunze2020TVT}, and RefNet~\cite{AMC-Huynh2021ICCE}.
Feature-level fusion and decision-level fusion models, that is, early fusion and late fusion, were cleverly exploited in recent CNN-based modulation classification methods to counter channel deterioration.
For example, Wang~\textit{et al.} introduced a decision-level fusion model for processing different incoming signals received by multiple antennas in a MIMO system, where a five-layer CNN performs the function of feature extraction in the proposed cooperative modulation classification method, namely Co-AMC~\cite{AMC-Wang2020DeepMIMO}.
The CNN induces the classification scores of MIMO signals that are cooperated via a weighted averaging decision rule to infer the final class of modulation.
Moreover, a multilevel fusion architecture~\cite{AMC-Zheng2019FusionCNN} was introduced with three fusion mechanisms (including feature-based, confidence-based, and majority voting-based fusion) to take advantage of meaningful information ranging from coarse to fine. 
These fusion models concurrently handle multiple CNN streams, in which each stream takes into consideration a fixed-length signal partitioned from a long sequence of IQ samples.
Although the overall classification performance is improved, the computation is highly complex, including the computational cost and memory utilization. This prevents the potential application of this method for low-latency communication services.

Apart from processing IQ samples directly, several modulation classification methods have used deep models for the graphical presentations of signals, such as a constellation diagram~\cite{AMC-Huang2019CCNN,AMC-Huang2019CFCN,AMC-Wang2019IQConstellation,AMC-Huynh2020ICT,AMC-Doan2020GLOBECOM} and spectrogram~\cite{AMC-Zeng2019Spectrum,AMC-Wu2019ConstellationFusion}. 
Huang~\textit{et al.}~\cite{AMC-Huang2019CCNN,AMC-Huang2019CFCN} designed a compressive CNN to learn the visual features of different modulation patterns from constellation diagrams.
To improve the classification accuracy, both a regular constellation (RC) image (i.e., plotting the real and imaginary parts of the modulation signal as scattered points on a two-dimensional diagram) and contrast enhanced grid (i.e., RC image with probability distribution of scattered points) were jointly exploited via a fusion module specified in a single CNN.
Furthermore, a synthetic loss function was formulated from cross-entropy loss, L2 regularization, and contrastive loss to maximize the difference between interclass features.
Zeng~\textit{et al.}~\cite{AMC-Zeng2019Spectrum} used short-time discrete Fourier transform (STFT), a fundamental time-frequency analysis algorithm, for visualizing the spectrum of frequencies of the modulation signal.
The set of transformed spectrogram images is then processed by a conventional CNN by adopting architecture with four convolutional layers for learning high-level features.
Hybrid approaches~\cite{AMC-Wang2019IQConstellation,AMC-Wu2019ConstellationFusion} were recommended for simultaneously using IQ data and image data.
A hierarchical framework with two classifiers was proposed by Wang~\textit{et al.}~\cite{AMC-Wang2019IQConstellation} who used a dataset consisting of IQ samples for the first CNN-based classifier for inter-group modulation discrimination. A dataset consisting of constellation diagrams was leveraged for the second CNN-based classifier for intergroup modulation identification.
In another study~\cite{AMC-Wu2019ConstellationFusion}, a compact-sized two-branch CNN with two processing streams organized in parallel was proposed for identifying the modulation format of a signal.
The meaningful features that were independently extracted from the image representations of the constellation diagram and cyclic spectra were intensively fused at the end of the network for classification.
Even though these methods based on constellation images or spectrogram images are more accurate compared with IQ-based approaches, they require more computational resources for data transformation, visualization, and storage. 
The CNNs of DL-based modulation classification approaches are described in Fig.~\ref{Fig:AMC_CNN_Architecture}, wherein most of them, typically designed to accept IQ data and image data as their input, are based on a simple network architecture (a straightforward connected structure of convolutional layers, activation layers, and pooling layers).

\begin{table*}[t]
	\caption{Summary of simulation configuration of DL-based modulation classification approaches.}
	\label{Table:Summary_AMC_Datasets}
	\begin{tabular}{|c|c|c|c|c|c|c|p{8cm}|}
		\hline 
		\multirow{2}{*}{Paper} & \multirow{2}{*}{Model} & \multicolumn{3}{c|}{Channel impairments}  & No. & \multirow{2}{*}{SNR (dB)} & \multirow{2}{*}{Dataset remarks}  \\ 
		\cline{3-5}
		& & Flat & Multipath & AWGN & modes & & \\
		\hline
		\hline
\cite{AMC-Ali2017Autoencoder}	&	DNN	&	$\checkmark$	&		&	$\checkmark$	&	4	&	$\left [ 0:1:15 \right ]$	&	2000 1000-symbol signals of $\{$4PSK, 16PSK, 16QAM, 128QAM$\}$.	\\	\hline
\multirow{5}{*}{\cite{AMC-OShea2018DeepSig}}	&	\multirow{5}{*}{CNN}	&		&	\multirow{5}{*}{$\checkmark$}	&	\multirow{5}{*}{$\checkmark$}	&	\multirow{5}{*}{24}	&	\multirow{5}{*}{$\left [-20:2:30  \right ]$}	&	RadioML 2018: 2,555,904 IQ signals of $\{$BPSK, QPSK, 8PSK, 16PSK, 32PSK, 16APSK, 32APSK, 64APSK, 128APSK, 4ASK, 8ASK, 16QAM, 32QAM, 64QAM, 128QAM, 256QAM, OOK, GMSP, OQPSK, FM, AM-SSB-WC, AM-SSB-SC, AM-DSB-WC, AM-DSB-SC$\}$.	\\	\hline
\multirow{2}{*}{\cite{AMC-Meng2018DL}}	&	\multirow{2}{*}{CNN}	&	\multirow{2}{*}{$\checkmark$}	&		&	\multirow{2}{*}{$\checkmark$}	&	\multirow{2}{*}{7}	&	\multirow{2}{*}{$\left [-6:2:10  \right ]$}	&	341,000 IQ signals of $\{$BPSK, 4PSK, 8PSK, 16QAM, 16APSK, 32APSK, 64QAM$\}$.	\\	\hline
\cite{AMC-Shah2019RBF}	&	DNN	&		&	$\checkmark$	&	$\checkmark$	&	4	&	$\left [0:1:15  \right ]$	&	96,000 IQ signals of $\{$GMSK, GFSK, CPFSK,OQPSK$\}$.	\\	\hline
\multirow{2}{*}{\cite{AMC-Zheng2019FusionCNN}}	&	\multirow{2}{*}{CNN}	&		&	\multirow{2}{*}{$\checkmark$}	&	\multirow{2}{*}{$\checkmark$}	&	\multirow{2}{*}{12}	&	\multirow{2}{*}{$\left [-20:2:30  \right ]$}	&	624,000 IQ signals of  $\{$BPSK, QPSK, 8PSK, OQPSK, 2FSK, 4FSK, 8FSK, 16QAM, 32QAM, 64QAM, 4PAM, 8PAM$\}$.	\\	\hline
\cite{AMC-Huang2019CCNN}	&	CNN	&		&		&	$\checkmark$	&	5	&	$\left [-5:2:15  \right ]$	&	2,750,000 CIs of $\{$BPSK, QPSK, 8PSK, 16QAM, 64QAM$\}$.	\\	\hline
\cite{AMC-Huang2019CFCN}	&	CNN	&		&		&	$\checkmark$	&	5	&	$\left [-5:2:15  \right ]$	&	1,9250,000 CIs of $\{$BPSK, QPSK, 8PSK, 16QAM, 64QAM$\}$.	\\	\hline
\multirow{2}{*}{\cite{AMC-Zeng2019Spectrum}}	&	\multirow{2}{*}{CNN}	&		&	\multirow{2}{*}{$\checkmark$}	&	\multirow{2}{*}{$\checkmark$}	&	\multirow{2}{*}{11}	&	\multirow{2}{*}{$\left [-20:2:18  \right ]$}	&	RadioML 2016: 220,000 SIs of $\{$BPSK, QPSK, 8PSK, 16QAM, BFSK, CPFSK, PAM4,WB-FM,AM-SSB,AM-DSB$\}$.	\\	\hline
\multirow{2}{*}{\cite{AMC-Wang2019IQConstellation}}	&	\multirow{2}{*}{CNN}	&		&	\multirow{2}{*}{$\checkmark$}	&	\multirow{2}{*}{$\checkmark$}	&	\multirow{2}{*}{8}	&	\multirow{2}{*}{$\left [-8:2:18  \right ]$}	&	84,000 IQ samples and 84,000 CIs of $\{$BPSK,  QPSK, 8PSK, GFSK, CPFSK, PAM4, 16QAM, 64QAM$\}$.	\\	\hline
\cite{AMC-Wu2019ConstellationFusion}	&	CNN	&		&	$\checkmark$	&	$\checkmark$	&	11	&	$\left [-20:2:18  \right ]$	&	RadioML 2016	\\	\hline
\cite{AMC-Hu2020DNN}	&	LSTM	&		&	$\checkmark$	&	$\checkmark$	&	4	&	$\left [0:2:20  \right ]$	&	440,000 IQ signals of $\{$BPSK, QPSK, 8PSK, 16QAM$\}$.	\\	\hline
\cite{AMC-Lin2020NNPruning}	&	CNN	&		&	$\checkmark$	&	$\checkmark$	&	11	&	$\left [-20:2:18  \right ]$	&	RadioML 2016	\\	\hline
\cite{AMC-Wang2020LightAMC}	&	CNN	&	$\checkmark$	&		&	$\checkmark$	&	4	&	$\left [-10:1:10  \right ]$	&	505,000 IQ signals of $\{$BPSK, QPSK, 8PSK, 16QAM$\}$.	\\	\hline
\cite{AMC-Huynh2020MCNet}	&	CNN	&		&	$\checkmark$	&	$\checkmark$	&	24	&	$\left [-20:2:18  \right ]$	&	RadioML 2018	\\	\hline
\cite{AMC-Wang2020DeepMIMO}	&	CNN	&	$\checkmark$	&		&	$\checkmark$	&	4	&	$\left [-10:2:10  \right ]$	&	1,320,000 IQ signals of $\{$BPSK, QPSK, 8PSK, 16QAM$\}$.	\\	\hline
	\end{tabular}
\begin{tabular}{p{8cm} p{8cm}}
	\\
	\multicolumn{2}{l}{\textbf{Modulation abbreviation}} \\
    PSK: Phase-shift keying &  ASK: Amplitude-shift keying  \\ 
    BPSK: Binary phase-shift keying & FSK: Frequency-shift keying \\
    QPSK: Quadrature phase-shift keying &  OOK: On–off keying\\
    OQPSK: Offset quadrature phase-shift keying & GFSK: Gaussian frequency-shift keying \\
    CPFSK: Continuous phase frequency-shift keying &  GMSK: Gaussian minimum-shift keying \\ 
    SSB-WC: Single-sideband modulation with carrier & AM: Amplitude modulation  \\
    DSB-WC: Double-sideband modulation with carrier &  QAM: Quadrature amplitude modulation \\
    SSB-SC: Single-sideband suppressed-carrier modulation & PAM: Pulse-amplitude modulation\\
    DSB-DC: Double-sideband suppressed-carrier modulation & WB-FM: Wide band frequency modulation\\
    
	\end{tabular}
\end{table*}

\subsection{Summary and Takeaway Points}
In this section, we reviewed state-of-the-art AMC methods, which are categorized as being either conventional (including likelihood-based and feature-based approaches) or innovative (including DL-based approaches), as summarized in Table~\ref{Table:Summary_AMC_Methods}.
Most of the likelihood-based approaches are very computationally costly in terms of parameter estimation under unknown channel conditions, whereas the feature-based methods achieve moderate performance in terms of their classification rate because of the sensitivity of handcrafted statistical features and limited learning capacity of traditional classifiers.
To overcome these limitations in conventional methods, researchers have exploited the excellent advantages of DL, such as automatic learning of high-level features and effective handling of big communication data to enhance the performance of modulation classification.
Additionally, CNN-based approaches are applicable to numerous digital and analog modulations under different channel impairments, such as a flat fading channel, multipath fading channel with attenuation, and additive noise, as summarized in Table~\ref{Table:Summary_AMC_Datasets}.
Interestingly, a DNN not only accepts a sequence of IQ samples (partitioned with a fixed length) as its input, but also accepts other transformed data (for instance, a constellation diagram in the form of a scattered plot and a spectrogram image via time-frequency analysis).
Other than a few notable CNN models, such as ResNet~\cite{AMC-OShea2018DeepSig} and MCNet~\cite{AMC-Huynh2020MCNet}, which were introduced for discriminating 24 challenging modulations, several other neural networks do not effectively leverage the powerful learning capability of CNN for effortless classification, for example, Co-AMC~\cite{AMC-Wang2020DeepMIMO} processes four low-order digital modulations based on a huge dataset containing more than 1.3 million samples.
Despite the superiority of deep learning over conventional approaches, certain aspects deserve further investigation when developing a DL model for modulation classification:
\begin{itemize}
	\item Deploying many fully connected layers without global average pooling~\cite{AMC-Meng2018DL} can rapidly increase the network size (usually measured by the number of trained parameters), leading to extremely high computational complexity.
	\item Configuring kernels of various sizes~\cite{AMC-Huynh2020MCNet}, such as unit $1\times1$, symmetric $n \times n$, and asymmetric $1 \times n$, potentially enriches the representational feature maps.
	\item Sophisticated techniques such as skip connection and dropping out can be leveraged to prevent the network from experiencing vanishing gradient descent and overfitting.
	\item Deep fusion frameworks with multiple processing streams to process different types of input data~\cite{AMC-Wu2019ConstellationFusion} are recommended to more effectively learn intrinsic radio characteristics.
	\item Balancing the accuracy and computational cost should be an important design objective to meet the requirements of modern communication services.
\end{itemize}

\section{Signal Detection}
\label{Sec:Signal_Detection}
In this section, we discuss applications of AI techniques for intelligent signal detection.

\subsection{Fundamentals of Signal Detection}

In MIMO communication systems with $N$ transmit and $M$ receive antennas, the received baseband signal can be expressed as 
\begin{align*}
	\vy = \mH \vs + \vn, \nb \label{system_model}
\end{align*}
where $\vs = [s_1, s_2, \ldots, s_N]^T$ and $\vy = [y_1, y_2, \ldots, y_M]^T$ represent the transmitted and received signal vectors, respectively, with $(\cdot)^T$ denoting the transpose of a vector. Further, $\mH$ of size $M \times N$ represents the channel matrix between the transmitter and receiver, and $\vn$ is a Gaussian noise vector. The goal of signal detection is to determine $\vs$ from $\vy$. This can be achieved via classical detection schemes such as the optimal maximum likelihood, near-optimal sphere decoding (SD) \cite{hassibi2005sphere}, tabu search (TS) \cite{nguyen2019qr, nguyen2019groupwise}, suboptimal linear zero-forcing (ZF), minimum mean square error (MMSE), and successive interference cancellation (SIC) receivers. Furthermore, interest in the development of ML-based detectors (MLDs) has recently been growing. In the following subsections, we review selected fundamental classical detection schemes and discuss ML-based approaches for signal detection.

\subsubsection{Optimal maximum-likelihood detector}
The optimal maximum-likelihood solution is obtained by an exhaustive search as follows:
\eqn {
	\sopt = \underset{\vs \in \setA^N}{\arg\min} \norm {\vy - \mH \vs}^2, \label{ML_solution}
}
where $\setA$ is an alphabet containing all possible transmitted signals $s_n, n=1,\ldots,N$. The computational complexity of the maximum-likelihood detector increases exponentially with $N$, which is prohibitive even for a small value of $N$. To overcome this challenge, near-optimal reduced-complexity detection schemes have been proposed, such as SD \cite{hassibi2005sphere} and TS \cite{nguyen2019qr, nguyen2019groupwise}.

\subsubsection{Linear detectors}

In linear detectors, the discrete alphabet $\setA$ is relaxed to a continuous space, allowing closed-form solutions of \eqref{ML_solution} to be found by solving the nonconstrained convex optimization problem $\min \norm {\vy - \mH \vs}^2$. The obtained solution is then quantized to the nearest vector in $\mathcal{A}^N$ \cite{albreem2019massive}. The ZF and MMSE receivers are two typical linear detectors, the solutions of which are given by
\begin{align*}
	\sZF &= \quan{(\mH^H \mH)^{-1} \mH^H \vy}, \nb \label{ZF_solution}\\
	\sMMSE &= \quan{(\mH^H \mH + \sigma^2 \mI)^{-1} \mH^H \vy}, \nb \label{MMSE_solution}
\end{align*}
respectively, where $\sigma^2$ is the variance of Gaussian noise, and $\quan{\cdot}$ is the element-wise quantization operator that quantizes elements in $(\cdot)$ to the nearest elements in $\setA$. The ZF detector performs poorly in the case of ill-conditioned channels due to noise enhancement. By contrast, the MMSE detector reduces noise enhancement and attains improved performance with respect to ZF. Both the ZF and MMSE receivers have low computational complexity. However, their performance is far from optimal, especially in square systems, i.e., when $N \approx M$.

\subsubsection{MLD}
The key idea of MLD is to model and train an ML algorithm such that its output $\sML$ can approximate the transmitted signal vector $\vs$ with high accuracy. In general, an ML-based solution for signal detection can be formulated as 
\begin{align*}
	\sML = \quan{ \Pi (\vx, \mathcal{P}) }, \nb \label{DL_solution}
\end{align*}
which represents a nonlinear transformation with the input vector $\vx$ and the trainable parameter set $\mathcal{P}$, followed by quantization. It is observed from \eqref{DL_solution} that the performance of an MLD depends on the input signal vector $\vx$, nonlinear function $\Pi$, and the learnable parameter set $\mathcal{P}$. In particular, $\vx$ contains information about the received signals and CSI if available. Furthermore, the nonlinear transformation $\Pi$ and trainable set $\mathcal{P}$ are determined by the underlying ML model and training process, which are the deciding factor for the learning ability and the accuracy of the ML model. These configurations, which result in various MLDs with different performance and computational complexity, are reviewed in the next subsection. 

\subsection{State-of-the-Art MLDs}

Various studies have considered the application of ML to signal detection, leading to numerous MLDs. Using different ML tools, a training process, and CSI models, existing MLDs can be classified as follows:
\begin{itemize}
	\item \textit{ML tools for signal detection}: Various ML techniques have been considered for signal detection. Among them, DNN is the most widely used owing to its powerful learning capability \cite{he2018model, he2020model, khani2019adaptive, xue2020modular, baek2019implementation, liu2018deep, samuel2019learning, corlay2018multilevel, gao2018sparsely, shlezinger2020deep, farsad2017detection, tan2018low, xue2018unsupervised, ye2017power, gao2018comnet, yao2019deep}. However, its computational cost and energy consumption are generally high \cite{xue2020modular} owing to its large number of neurons and layers, especially in the case of those developed for large-scale systems. Other ML tools, such as CNN \cite{xia2020mimo, baek2019implementation, farsad2017detection, he2020generic, fan2019cnn}, recurrent NN (RNN) \cite{baek2019implementation, farsad2017detection}, extreme learning machine (ELM) \cite{yan2017signal, liu2019online}, auto encoder (AE) \cite{yan2017signal, balevi2019one}, and ensemble learning \cite{ha2018signal}, were also leveraged for signal detection.

	\item \textit{CSI requirement}: In classical signal detection schemes, CSI is crucial for obtaining the estimate of the transmitted signal, as seen in \eqref{ML_solution}--\eqref{MMSE_solution}. However, it becomes optional in MLDs. Specifically, while both the received signal $\vy$ and CSI, i.e., $\mH$, are taken as the input of ML algorithms in \cite{he2018model, he2020model, khani2019adaptive, xue2020modular, liu2018deep, samuel2019learning, corlay2018multilevel, gao2018sparsely, ye2017power, gao2018comnet, huang2018cascade, fan2019cnn}, only $\vy$ is required for the MLDs in \cite{baek2019implementation, shlezinger2020deep, farsad2017detection, he2020generic, xue2018unsupervised, yan2017signal, balevi2019one, liu2019online, yao2019deep}. The omission of CSI can simplify the communication system, in which the channel estimation block is removed, or it can reduce the size of the input signal vector. As a result, a considerable reduction in overall computational complexity as well as power consumption can be attained. However, this could result in potential performance degradation in some scenarios, especially in block-fading channels \cite{baek2019implementation} and in multiple-antenna systems, where the CSI is crucial for removing the intersymbol interference.
	
	\item \textit{Training approaches}: Unlike the classical methods in which a hand-engineered detection scheme is applied in an online method, an MLD needs to train an ML model before it is used for signal detection. For example, in a DNN-based detector, the weights and biases of the DNN are trained to minimize the distance between the ML-based solution and the labels, i.e., $\sML$ and $\vs$ \cite{samuel2017deep, samuel2019learning, gao2018sparsely, nguyen2019deep}. In particular, the training can be carried out either offline \cite{xia2020mimo, xue2020modular, baek2019implementation, liu2018deep, samuel2019learning, corlay2018multilevel, gao2018sparsely, farsad2017detection, tan2018low, he2020generic, xue2018unsupervised, ye2017power, gao2018comnet, balevi2019one, yao2019deep, huang2018cascade, fan2019cnn} or online  \cite{khani2019adaptive, shlezinger2020deep, liu2019online}. In offline training, the computational complexity of the training process can generally be ignored, and the parameters of an ML model can be readily optimized by using a sufficiently large amount of training data. However, this training method is only suitable for certain channel models such as an independent and identically distributed (i.i.d.) Rayleigh fading channel. In contrast, the method may become less impractical in real-world communication systems, where the channel characteristics change rapidly or the channel statistics are unavailable, e.g., in molecular communication \cite{farsad2017detection}. In these scenarios, a good solution would be to apply an online training method~(e.g., \cite{khani2019adaptive, shlezinger2020deep, liu2019online}, at the cost of increased latency and computational complexity.
\end{itemize}

The aspects listed above distinguish existing MLDs based on their configurations. However, we found it to be more beneficial to classify MLDs based on the way ML techniques are leveraged for signal detection. This is useful not only for analyzing and synthesizing existing MLDs, but also for providing methodologies for further development in this area. Our literature review shows that ML techniques can be leveraged for signal detection in three ways, namely, black-box, unfolding, and classical detector-based MLDs. First, an ML tool can be modeled as a \textit{black-box MLD}, i.e., it outputs the estimate of the transmitted signal vector $\vs$ as an independent detector. Second, in \textit{unfolding MLDs}, each layer of the DNN is constructed based on the operations in each iteration of the classical projected gradient descent (PGD) algorithms. Finally, existing iterative or near-optimal detection algorithms can be further optimized by using ML, resulting in so-called \textit{classical detector-based MLDs}. These three groups of MLDs are discussed in the following.

\begin{figure}[t]
	\centering
	\includegraphics[width=1.00\linewidth]{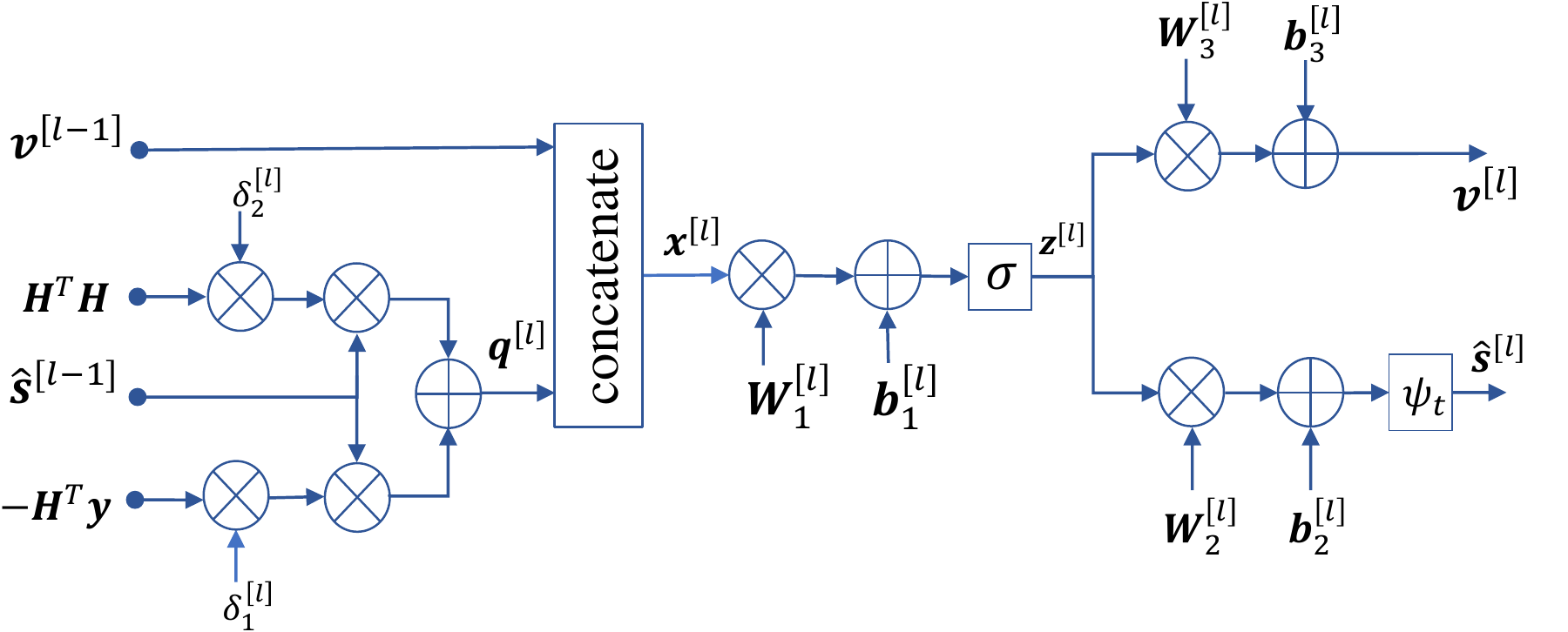}
	\caption{Illustration of layer $l$ of the DetNet \cite{samuel2017deep, samuel2019learning, nguyen2019deep}.}
	\label{fig_DetNet}
\end{figure}

\subsubsection{Black-box MLDs}

Inspired by the learning capability of ML techniques, certain detector designs attempted to replace the classical detectors with black-box MLDs \cite{xue2020modular, baek2019implementation, liu2018deep, farsad2017detection, xue2018unsupervised, ye2017power, balevi2019one, liu2019online, yao2019deep}. We note that in this work, the term \emph{black-box} reflects the fact that an ML model can learn and output the desired solution, which is $\sML$ for signal detection, without any expert knowledge. The ingenuity of this application lies in choosing an appropriate ML tool and optimizing the model configurations, e.g., the number of layers, number of nodes in each layer, and activation functions for the underlying DNN.

DNNs were used for black-box MLDs in OFDM systems \cite{ye2017power, yao2019deep}. Unlike the classical OFDM receivers that first require conducting CSI estimation, the DNN-based detectors proposed in \cite{ye2017power} and \cite{yao2019deep} perform the signal detection directly. In other words, the DNN is a black box that takes $\vy$ as input and outputs $\sML$, without requiring the estimated CSI. Notably, these black-box MLDs are shown to outperform the conventional least-square (LS) and MMSE receivers, which compute the estimates of transmitted signals based on CSI, as shown in \eqref{MMSE_solution}. An end-to-end OFDM communication system was modeled as a single AE \cite{balevi2019one}, in which the DNN-based detector was trained offline irrespective of the channel. The performance of the black-box MLD was also investigated with the use of ELM \cite{liu2019online}, RNNs, and CNNs \cite{farsad2017detection}. In particular, the ELM-based black-box MLD in \cite{liu2019online} outperformed the DNN-based MLDs proposed in \cite{ye2017power}, with less complexity and excellent robustness under different multipath fading channels. Furthermore, the CNN- and RNN-based black-box MLDs in \cite{farsad2017detection} were evaluated using experimental data collected by a chemical communication platform, for which the channel model was unknown and which was difficult to model analytically.

A common observation from the studies discussed above is that the developed black-box MLDs can perform well without the CSI. This is advantageous compared with the classical detectors when the CSI is either not available or the wireless channel is fast fading. Furthermore, the black-box MLDs achieve not only improved performance, as shown in \cite{yao2019deep, farsad2017detection, liu2019online}, but also reduced complexity with respect to the simple linear receivers \cite{xue2020modular}. The reason for this reduction in complexity is that these black-box MLDs only perform matrix multiplications and additions, whereas computationally expensive matrix inversions or factorizations (such as singular-value decomposition or QR decomposition) are required for most conventional MIMO receivers including linear ZF, MMSE, and SIC. However, it is worth noting that most of these black-box MLDs are proposed for general OFDM systems. By contrast, the signal detection in multiple-antenna systems requires more sophisticated MLDs, which are discussed in the next subsections.

\subsubsection{Unfolding MLDs}
 
While various ML tools are leveraged for black-box MLDs, unfolding MLDs employ DNNs because they can learn complicated nonlinear functions. However, the ingenuity of detectors in this group is that, instead of directly using the well-known fully connected DNN (FC-DNN), they employ unfolding layer-by-layer architectures. In these architectures, the layers have the same structure and are constructed following the classical PGD algorithm with different weights because of their different input signals.

Intuitively, \eqref{ML_solution} can be solved by the iterative PGD optimization method. Motivated by this, a series of unfolding MLDs, including the \textit{detection network} (DetNet) \cite{samuel2017deep, samuel2019learning}, \textit{sparsely-connected DNN} (ScNet) \cite{gao2018sparsely}, \textit{fast-convergence sparsely connected DNN} (FS-Net) \cite{nguyen2019deep}, \textit{multilayer DNN} (Twin-DNN) \cite{corlay2018multilevel}, and \textit{Cascade DNN} (Cascade-Net) \cite{huang2018cascade} were proposed for signal detection. In these schemes, $\sML$ is updated over $L$ layers of the DNN as follows:
\begin{align*}
	\hat{\vs}^{[l]} 
	&= f \left( \vs - \delta^{[l]} \frac{\partial \norm {\vy - \mH \vs}^2}{\partial \vs} \right)_{\vs = \hat{\vs}^{[l-1]}} \\
	&= f \left( \hat{\vs}^{[l]} - \delta^{[l]} \mH^T \vy + \delta^{[l]} \mH^T \mH \hat{\vs}^{[l-1]} \right), \numberthis \label{unfolding}
\end{align*}
where $f(\cdot)$ denotes a nonlinear projection operator, $\delta^{[l]}$ is the step size, and $l=1,\ldots,L$. Inspired by \eqref{unfolding}, DetNet was introduced \cite{samuel2019learning, samuel2017deep}. The $l$th layer of DetNet is illustrated in Fig. \ref{fig_DetNet}, and its operations are summarized as follows:
\begin{align*}
	\vq^{[l]} &= \hat{\vs}^{[l-1]} - \delta_1^{[l]} \mH^T \vy + \delta_2^{[l]} \mH^T \mH \hat{\vs}^{[l-1]}, \numberthis \label{d_q}\\
	\vx^{[l]} &= \left[\vv^{[l-1]}, \vq^{[l]} \right]^T, \numberthis \label{d_x}\\
	\vz^{[l]} &= \sigma \left(\mW_1^{[l]} \vx^{[l]} + \vb_1^{[l]} \right), \numberthis \label{d_z}\\
	\hat{\vs}^{[l]} &= \psi_t \left(\mW_2^{[l]} \vz^{[l]} + \vb_2^{[l]}\right), \numberthis \label{d_s}\\
	\vv^{[l]} &= \mW_3^{[l]} \vz^{[l]} + \vb_3^{[l]}, \numberthis \label{d_v}
\end{align*}
where $\hat{\vs}^{[0]} = \vv^{[0]} = \mathbf{0}$, with $\mathbf{0}$ being an all-zero vector of an appropriate size, and $\{ \mW_1^{[l]}, \mW_2^{[l]}, \mW_3^{[l]}, \vb_1^{[l]}, \vb_2^{[l]}, \vb_3^{[l]}, \delta_1^{[l]}, \delta_2^{[l]} \}$ are the training parameters, including the weights, biases, and step size, in the $l$th layer of DetNet. Furthermore, $\sigma(\cdot)$ represents the rectified linear unit (ReLU) activation function, and $\psi_t(\cdot)$ guarantees that the amplitudes of the elements of $\hat{\vs}^{[l]}$ are in an appropriate range of desired signals. The final solution of DetNet is $\sML = \mathcal{Q}\left(\hat{\vs}^{[L]}\right)$. Although DetNet achieves promising performance, it has several drawbacks. Specifically, the significance of the intermediate signal vector $\vv^{[l]}$ is not clear and considerably enlarges the size of the input vector $\vx^{[l]}$, thereby complicating the network architecture. Consequently, the computational complexity of DetNet is extremely high. Furthermore, although the performance of DetNet is shown to be good for the case $N \ll M$, subsequent work \cite{gao2018sparsely, nguyen2019deep} showed the network performance to be far from optimal for square systems, i.e., $N \approx M$. 

\begin{table*}[t]
	\renewcommand{\arraystretch}{1.00}
	\caption{Summary of existing MLDs}
	\label{tab_summary_MLD}
	\begin{tabular}{|c|c|c|c|p{9.5cm}|}
		\hline
		\textbf{MLD group} & \textbf{Detector} & \textbf{Paper} & \textbf{ML models} & \textbf{Interesting observations} \\
		\hline
		\hline
		
		\multirow{5}{*}{\makecell{Black-box\\MLDs}} & \multirow{5}{*}{ML model}
		&\multirow{5}{*}{{\makecell{\cite{xue2020modular}, \cite{baek2019implementation},\\ \cite{liu2018deep}, \cite{farsad2017detection},\\ \cite{xue2018unsupervised}, \cite{ye2017power},\\ \cite{balevi2019one}, \cite{liu2019online},\\ \cite{yao2019deep}}}}
		
		& \multirow{4}{*}{\makecell{DNN, \\CNN, RNN, \\ELM, AE }}
		& $\bullet$ No CSI requirement\\
		&&&&$\bullet$ Suitable for fast varying channels or unavailable CSI\\
		&&&&$\bullet$ Outperforms linear receivers\\
		&&&&$\bullet$ Matrix inversion and factorization are unnecessary\\
		&&&&$\bullet$ Generally proposed for OFDM rather than for MIMO systems \\
		\hline
		
		\multirow{12}{*}{\makecell{Unfolding \\ MLDs}} 
		& \multirow{12}{*}{\makecell{Projected\\GD-based}} 
		& \multirow{5}{*}{\cite{samuel2017deep, samuel2019learning}}
		& \multirow{5}{*}{DetNet}
		&  $\bullet$ Performs well for $N \ll M$, BPSK, QPSK, and i.i.d. Rayleigh fading channels\\
		&&&&$\bullet$ Performance degrades for $N \approx M$ and higher-order modulation \\ 
		&&&&$\bullet$ High computational complexity caused by complicated network architecture and use of an intermediate input vector \\
		&&&&$\bullet$ Difficult to train because of many learning variables\\
		\cline{3-5} 
		
		&& \multirow{2}{*}{\cite{gao2018sparsely}}	& \multirow{2}{*}{ScNet}
		&  $\bullet$ Simplifies the network connections of DetNet \\
		&&&&$\bullet$ Improved performance, reduced complexity, and fewer learning variables with respect to DetNet \\
		\cline{3-5} 
		&& \multirow{4}{*}{\cite{nguyen2019deep}}	& \multirow{4}{*}{FS-Net} &  $\bullet$ Simplifies the input vector and network connection, and optimizes the loss function of ScNet and DetNet\\
		&&&&$\bullet$ Improved performance, reduced complexity, and fewer learning variables with respect to DetNet and ScNet \\
		\cline{3-5} 
		&& \multirow{2}{*}{\cite{corlay2018multilevel}}	& \multirow{2}{*}{Twin-DNN} & $\bullet$ Improved performance with respect to DetNet \\
		&&&&$\bullet$ Much higher complexity than DetNet owing to the requirement for two parallel DetNets and a ZF solution\\
		\cline{3-5} 
		&& \cite{huang2018cascade}	& cascade-Net &  $\bullet$ Outperforms both DetNet and the classical ZF detectors\\
		\hline

		\multirow{18}{*}{\makecell{Classical\\ detector-based \\ MLDs}} 
		& \multirow{7}{*}{\makecell{Iterative\\algorithm}} 
		& \multirow{3}{*}{\cite{he2018model}}	& \multirow{3}{*}{OAMP-Net} &  $\bullet$ Easier and faster to train compared with DetNet\\
		&&&&$\bullet$ Improved performance with respect to the classical OAMP scheme\\
		&&&&$\bullet$ Performance degradation in correlated channels\\
		\cline{3-5} 
		&& \cite{he2020model}	& OAMP-Net2 &  $\bullet$ Outperforms DetNet, OAMP-Net, and classical MMSE-based SIC\\
		\cline{3-5} 
		&& \multirow{2}{*}{\cite{khani2019adaptive}}	& \multirow{2}{*}{MMNet} &  $\bullet$ Generalizes well to correlated channels\\
		&&&&$\bullet$ Outperforms OAMP-Nets with lower complexity\\
		\cline{2-5} 
		
		& \multirow{4}{*}{TS} &\multirow{4}{*}{\cite{nguyen2019deep}} & \multirow{4}{*}{FS-Net} & $\bullet$ Leverages the FS-Net solution to generate an initial solution, modify the search procedure, and terminate the TS algorithm early \\
		&&&&$\bullet$ $90 \%$ complexity reduction with almost no performance loss with respect to the classical TS schemes \\
		\cline{2-5} 
		
		& \multirow{7}{*}{SD} 
		& \multirow{4}{*}{\cite{askri2019dnn, mohammadkarimi2018deep, weon2020learning}}	& \multirow{4}{*}{FC-DNN} &  $\bullet$ Use FC-DNNs to predict initial radii for SD\\
		&&&&$\bullet$ Complexity reduction of approximately $60\%$ at moderate SNRs with marginal performance loss \\
		&&&&$\bullet$ Computationally expensive and time-consuming training phase for performing the classical SD scheme to collect training data\\
		\cline{3-5} 
		&& \multirow{4}{*}{\cite{nguyen2020application}}	& \multirow{4}{*}{FS-Net} &  $\bullet$ Uses FS-Net to generate initial candidate, facilitating candidate ordering, layer ordering, and early pruning in SD/KSD schemes\\
		&&&& $\bullet$ Complexity reduction of $90\%$ at moderate SNRs with no performance loss\\
		&&&& $\bullet$ Unnecessary to perform the classical SD to collect training data\\
		\hline
	\end{tabular}
\end{table*}

ScNet \cite{gao2018sparsely} and FS-Net \cite{nguyen2019deep} were proposed to overcome the drawbacks of DetNet. ScNet focuses on simplifying the network architecture of DetNet by removing the intermediate vector $\vv^{[l]}$ and redundant connections in the network, whereas FS-Net reduces the number of training parameters and optimizes the loss function to accelerate the convergence of DetNet and ScNet. Although the network architecture of DetNet was significantly simplified, ScNet and FSNet improved the performance remarkably. Specifically, an improvement of approximately $2$-dB and $3$-dB in SNR was achieved by ScNet and FS-Net, respectively, with respect to DetNet, with the computational complexity being approximately $40\%$ of that of DetNet \cite{nguyen2019deep}. Unlike ScNet and FS-Net, Twin-DNN \cite{corlay2018multilevel} uses two parallel DetNets with different input vectors $\hat{\vs}^{[0]}$: the first is $\sZF$ and the other is randomly generated. The solution of Twin-DNN is the more accurate of the two output vectors of the two DetNets, motivated by ensemble learning. As a result, Twin-DNN improves the performance at the cost of approximately double the complexity of DetNet. A detailed comparison of DetNet, ScNet, FSNet, and Twin-DNN in terms of their network architecture, performance, and computational complexity was reported \cite{nguyen2019deep}. Another variant of DetNet is Cascade-Net, which was proposed for single-antenna systems \cite{huang2018cascade}. In Cascade-Net, a DNN is cascaded with a ZF preprocessor to prevent the network from converging to a saddle point or local minimum point. Simulation results in \cite{huang2018cascade} show that Cascade-Net performs much better than DetNet and the classical ZF detector.

\subsubsection{Classical detector-based MLDs}

Although the black-box and unfolding MLDs discussed above outperform simple detectors, their performance is still far from optimal, especially in challenging scenarios such as square systems with high-order QAM signaling. To overcome this limitation, we can incorporate ML algorithms with classical hand-engineered detectors such as iterative or near-optimal detection schemes. This incorporation is useful to further optimize the classical detection schemes in terms of their computational complexity and/or performance.

\paragraph{Iterative algorithm-based MLDs}
 
He \textit{et al.} \cite{he2018model} proposed an iterative orthogonal approximate message passing (OAMP) algorithm-based network (OAMP-Net). In the $l$th layer of OAMP-Net, the output $\hat{\vs}^{[l]}$ is updated as follows:
 \begin{align*}
 	\vz^{[l]} &= \hat{\vs}^{[l-1]} + \gamma^{[l]} \mA^{[l]} (\vy - \mH \vz^{[l-1]}), \nb \label{OAMP_z} \\
 	\vs^{[l]} &= \mean{\vs | \vz^{[l]}, \tau^{[l]}}, \nb \label{OAMP_s}
 \end{align*}
 where \eqref{OAMP_s} represents the MMSE denoiser, and
 \begin{align*}
 	\mA^{[l]} &= \frac{N v^{[l]^2} \mH^H (v^{[l]^2} \mH \mH^H + \sigma^2 \mI)^{-1}}{\text{trace}(v^{[l]^2} \mH^H (v^{[l]^2} \mH \mH^H + \sigma^2 \mI)^{-1} \mH)}, \nb \label{OAMP_A}\\
 	v^{[l]^2} &= \frac{\norm{\vy - \mH \hat{\vs}^{[l]}}^2 - M\sigma^2}{\text{trace}(\mH^H \mH)}, \\
 	\tau^{[l]^2} &= \frac{1}{2N} \text{trace} ((\mI-\mA^{[l]})(\mI-\mA^{[l]})^H) v^{[l]^2} \\ &\hspace{1cm} + \frac{\theta^{[l]}  \sigma^2}{4N} \text{trace} (\mA^{[l]} \mA^{{[l]}^H}), \nb \label{OAMP_tau}
 \end{align*}
 with $\vs^{[0]} = \mathbf{0}$ and $\tau^{[0]} = 1$. We note that in the classical OAMP scheme, $\gamma^{[l]}$ and $\theta^{[l]}$ in \eqref{OAMP_z} and \eqref{OAMP_tau}, respectively, are both set to one. By contrast, they emerged as learnable variables and were optimized by using training to provide appropriate step sizes for updating the mean and variance of the MMSE denoiser. Compared to DetNet, OAMP-Net is easier and faster to train because only a few adjustable parameters need to be optimized. Furthermore, OAMP-Net achieves an SNR improvement of approximately $2$-dB with respect to the classical OAMP scheme.
 
More recently, OAMP-Net2 was proposed \cite{he2020model} as an improved version of OAMP-Net. Specifically, two additional learnable parameters were added to the denoiser \eqref{OAMP_s} in OAMP-Net2 to construct the nonlinear estimator of $\vs^{[l]}$ to satisfy the divergence-free requirement. This update improves the SNR of OAMP-Net2 with approximately $3$-dB with respect to OAMP-Net \cite{he2020model}. Furthermore, OAMP-Net2 was demonstrated to outperform prior detection schemes such as DetNet and MMSE-based SIC. Although OAMP-Nets perform impressively for i.i.d. Gaussian channels, it is shown in \cite{he2018model} and \cite{khani2019adaptive} that they may not perform very well for realistic channels with spatial correlations \cite{khani2019adaptive}. Moreover, OAMP-Nets require performing matrix inversion in each iteration, making them even more computationally expensive than DetNet \cite{he2020model}. To overcome these limitations of OAMP-Nets, Khani \emph{et al.} proposed the MMNet \cite{khani2019adaptive}. Similar to OAMP-Nets, MMNet follows an iterative approach. However, instead of being computed as in \eqref{OAMP_A}, $\mA^{[l]}$ is considered as an $N \times M$ trainable weight matrix. This is more advantageous than OAMP-Nets in the following respects. First, it allows $\vz^{[l]}$ to be obtained with flexible trainable parameters optimized for each channel realization, thereby facilitating online training to adapt to channel variation. Second, $\mA^{[l]}$ can be obtained without performing computationally expensive matrix inversion as in \eqref{OAMP_A} for OAMP-Nets. The simulation results in \cite{khani2019adaptive} showed that, in practical 3GPP channels, MMNet achieves an improvement of $3$-dB in SNR compared to OAMP-Nets with less computational complexity by a factor of $10-15$.
 
\paragraph{Near-optimal detector-based MLDs}

The two well-known near-optimal detection schemes are TS and SD. The DL-aided TS algorithm was introduced in \cite{nguyen2019deep}. Specifically, Nguyen \emph{et al.} proposed employing FS-Net to generate the highly reliable initial solution of the TS scheme. Furthermore, in this algorithm, an adaptive early termination and a modified searching process are performed based on the predicted approximation error, which is determined from the FS-Net-based initial solution, to enable the final solution to be reached earlier. The simulation results in \cite{nguyen2019deep} demonstrated that the proposed DL-aided TS algorithm reduces the complexity by approximately $90\%$ with respect to the existing classical TS algorithms, while maintaining almost the same performance.

DL-aided SD schemes were recently proposed \cite{askri2019dnn, mohammadkarimi2018deep, weon2020learning, nguyen2020application}. Specifically, a DNN was used to learn the initial radius for SD \cite{askri2019dnn, mohammadkarimi2018deep, weon2020learning}. While a single radius is used in \cite{askri2019dnn}, multiple radii are employed in \cite{mohammadkarimi2018deep}. Furthermore, as an improvement of existing schemes \cite{askri2019dnn, mohammadkarimi2018deep}, Weon \textit{et al}. \cite{weon2020learning} proposed a learning-aided deep path prediction scheme for sphere decoding in large multiple-antenna systems. In particular, the minimum radius for each sub-tree is learned by a DNN, resulting in a more significant complexity reduction with respect to the prior DL-aided SD schemes in \cite{askri2019dnn} and \cite{mohammadkarimi2018deep}. In all three DL-aided SD schemes mentioned above, the common idea is to predict radii for the sequential SD. This approach has certain limitations in the offline learning phase, as well as during online application to SD. First, in the DNN training phase \cite{askri2019dnn, mohammadkarimi2018deep,weon2020learning}, conventional SD needs to be performed first to generate training labels, i.e., the radii. Consequently, these DNNs are time-consuming and computationally complex to train, especially in the case of large MIMO systems. Second, although the radius plays an important role in the search efficiency of conventional Fincke--Pohst SD, it becomes less significant in the Schnorr--Euchner SD, especially for high SNRs, for which a relatively reliable radius can be computed using the conventional formula \cite{hassibi2005sphere}. To overcome these limitations, the fast DL-aided SD (FDL-SD) and fast DL-aided $K$-best SD (KSD) (FDL-KSD) algorithms were proposed \cite{nguyen2020application}. The idea of FDL-SD and FDL-KSD is to use FS-Net \cite{nguyen2019deep} to generate a highly reliable initial candidate for the search in SD/KSD, which facilitates a candidate/layer-ordering scheme and an early rejection scheme to significantly reduce the complexity of the conventional SD schemes. The simulation results in\cite{nguyen2020application} showed that, for moderate SNRs, FDL-SD reduces the complexity by more than $90\%$ with respect to the classical Fincke--Pohst SD scheme, compared with those of the DL-aided SD schemes \cite{askri2019dnn, mohammadkarimi2018deep, weon2020learning}, which achieve a reduction of approximately $60\%$.

\subsection{Summary and Takeaway Points}
In this section, we reviewed the typical existing MLDs in the literature. Based on the application of ML models to the detection process, they are divided into three groups: \emph{black-box MLDs}, \emph{unfolding MLDs}, and \emph{classical detector-based MLDs}, as summarized in Table \ref{tab_summary_MLD}. In the first group, ML tools are used to design independent detectors, and they are trained to learn the transmitted signal vector $\vs$ without requiring any expert knowledge. Various ML models, including DNN, CNN, RNN, ELM, AE, and ensemble learning are used for black-box MLDs. DNNs are the most widely used for the other two groups of MLDs. In unfolding MLDs, the FC-DNN is unfolded to construct an unfolding layer architecture following the PGD algorithm. By contrast, the last group of MLDs leverages the learning capability of DNNs to further optimize well-known classical detectors such as OAMP, TS, and SD. While black-box and unfolding MLDs are shown to outperform linear receivers with lower complexity, the OAMP-, TS-, and SD-based MLDs guarantee near-optimal performance with reduced complexity owing to the aid of DL.

Depending on the system configurations such as the size and modulation scheme, an appropriate MLD can be chosen for signal detection. For example, black-box MLDs and unfolding MLDs perform relatively well for a MIMO system with $N \ll M$ and low-order modulation such as BPSK and QPSK. By contrast, in the case of $N \approx M$ and higher-order QAM signaling, the black-box and unfolding MLDs experience substantial performance loss. In this case, the incorporation of ML with classical iterative or near-optimal detection schemes can be a more appropriate choice to guarantee good performance for those challenging systems. Furthermore, we note that different groups of MLDs have different computational complexities. Therefore, for an optimal performance-complexity tradeoff, the performance and complexity of both the ML model and hand-engineered algorithm would have to be considered for the design of nonblack-box MLDs.

\section{Beamforming and Channel Estimation}
\label{Sec:BF_and_CE}
In this section, we review applications of AI techniques for MIMO beamforming and channel estimation.  

\subsection{Fundamentals of Beamforming}

Beamforming is an effective mean of improving the quality of the received signals in wireless communication systems. It can be realized via precoding at the transmitter and/or combining at the receiver. 
In this subsection, without loss of generality, we provide the fundamentals of precoding design. Specifically, we consider the problem of beamforming design for a single-cell downlink system, where the base station (BS) is equipped with $N$ antennas and serves $K$ single-antenna users. The received signal at user $k$ can be given by
\begin{align*}
	y_k = \vh_k^H \sum_{k=1}^{K} \vf_k s_k + n_k, \nb \label{system_model_BF}
\end{align*}
where $s_k$ is the transmitted signal with $\mean{\abs{s_k}^2}=1$, $\vh_k \in \setC^{N \times 1}$ is the channel vector, and $n_k \sim \mathcal{CN}(0, \sigma^2)$ is an AWGN noise sample at user $k$. Furthermore, $\vf_k \in \setC^{N \times 1}$ represents the beamforming vector for user $k$. The received SINR at user $k$ is given as
\begin{align*}
\sinr = \frac{\abs{\vh_k^H \vf_k}^2}{\sum_{i=1, i\neq k}^{K} \abs{\vh_i^H \vf_i}^2 + \sigma^2}. \nb \label{sinr}
\end{align*}

Beamforming can be implemented by analog, digital, or hybrid analog/digital beamforming (HBF) architectures, leading to different optimization problems for the beamforming design. In particular, sum-rate maximization (SRM) is the most widely considered problem. Therefore, in the following, SRM is used for stating the beamforming design problem, without loss of generality. The optimization of beamformers based on other objective metrics such as SINR and BER performance are also discussed in the subsection on state-of-the-art beamforming designs. 

\paragraph{Digital beamforming}
For digital beamforming (DBF), the SRM problem can be expressed as:
\begin{align*}
\pdbf: \max_{\mF} \sum_{k=1}^{K} \log_2 \left(1 + \sinr \right), \text{s.t. }  \norm{\mF}_F^2 \leq P \nb \label{Pdbf} ,
\end{align*}
where $P$ is the total power budget at the transmitter, and $\mF = \left[ \vf_1, \ldots, \vf_K \right]$ is the beamforming matrix. 

\paragraph{Analog beamforming}
In analog beamforming (ABF), analog circuitry with phase shifters and/or switches is used for signal processing. Therefore, the entries of the ABF matrix are required to have a constant modulus of $\frac{1}{\sqrt{N}}$, and their phases are adjustable in a given space depending on the resolution of the phase shifters. Therefore, ABF vectors belong to a feasible codebook $\mathcal{F}$, leading to the following SRM problem:
\begin{align*}
\pabf: \max_{\mF} \sum_{k=1}^{K} \log_2 \left(1 + \sinr \right), \text{s.t. } \mF \in \mathcal{F} \nb \label{Pabf}.
\end{align*}

\paragraph{Hybrid analog/digital beamforming}
For HBF, we have $\vf_k = \mF^{RF} \vf_k^{BB}$, where $\mF^{RF}  \in \setC^{N \times N_{RF}}$ denotes the analog precoding matrix with $N_{RF}$ being the number of RF chains, and $\vf_k^{BB} \in \setC^{N_{RF} \times 1}$ denotes the digital precoding vector for the $k$th user. Therefore, the SRM problem can be rewritten as
\begin{subequations}
	\begin{align*}
		\phbf: \quad \max_{\mF^{BB}, \mF^{RF}} \quad & \sum_{k=1}^{K} \log_2 \left(1 + \sinr \right) \nb \label{obj} \\
		\text{s.t. } \quad & \norm{\mF^{RF} \mF^{BB}}_F^2 \leq P, \mF^{RF} \in \mathcal{F}. \nb \label{constraint}
	\end{align*}
\end{subequations}

The SRM problems $\pabf, \pdbf,$ and $\phbf$ for beamforming design are challenging because of their nonconvexity. Existing classical hand-engineered algorithms for the SRM problems only achieve suboptimal solutions but remain computationally complex because of their complex matrix operations, such as matrix inversions and factorization and iterations. To overcome this challenge, various ML-aided beamformers (MLBs) that were introduced recently are discussed in the following subsection.

\subsection{State-of-the-art MLBs}

\begin{table*}[t]
	\renewcommand{\arraystretch}{1.00}
	\caption{Summary of existing MLBs}
	\label{tab_summary_MLB}
	\begin{tabular}{|c|c|c|c|p{9cm}|}
		\hline
		\textbf{MLB group} & \textbf{Sub-group} & \textbf{Paper} & \textbf{ML model} & \textbf{Interesting observation} \\
		\hline
		\hline
		
		\multirow{4}{*}{\makecell{Black-box\\MLBs}} & \multirow{2}{*}{Unsupervised}
		&\multirow{2}{*}{{\makecell{\cite{lin2019beamforming},  \cite{huang2018unsupervised}}}}
		
		& \multirow{2}{*}{\makecell{DNN}}
		& $\bullet$ Directly generate the beamforming matrix\\
		&&&&$\bullet$ ML models are trained using unsupervised learning.\\
		
		\cline{2-5} 
		& \multirow{2}{*}{Supervised} &\multirow{2}{*}{{\makecell{\cite{alkhateeb2018deep,anton2019learning,ramon2005beamforming,long2018data}}}}
		
		& \multirow{2}{*}{\makecell{DNN, CNN, \\ SVM, KNN, SVC }}
		& $\bullet$ Output an indicator of the beamformer\\
		&&&&$\bullet$ ML models are trained using supervised learning. \\
		\hline

		\multirow{10}{*}{\makecell{Feature-based\\ MLBs}} 
		& \multirow{2}{*}{\makecell{DFT-based \\ factorization}} 
		& \multirow{2}{*}{\cite{li2019joint}}	& \multirow{2}{*}{DNN} &  $\bullet$ Beamforming vector is modeled as $\vf_k = (\mathcal{F}^h)_n \otimes (\mathcal{F}^v)_m$.\\
		&&&&$\bullet$ Used to predict the codeword indices $n$ and $m$\\
		\cline{2-5} 
		
		& \multirow{3}{*}{\makecell{BM-IC\\algorithm-based}} 
		& \multirow{3}{*}{\cite{zhou2018deep}}	& \multirow{3}{*}{DNN} &  $\bullet$ Used to approximate the conventional BM-IC algorithm\\
		&&&&$\bullet$ Performance comparable to that of the BM-IC algorithm, but requires much less computational time\\
		\cline{2-5} 
		
		& \multirow{5}{*}{\makecell{Optimization\\-based}} 
		& \multirow{5}{*}{\makecell{\cite{xia2019deep, huang2019fast}}}	& \multirow{5}{*}{CNN, DNN} &  $\bullet$ Use a CNN to predict scalars, which are solutions of optimization problems, i.e., SINR balancing, power minimization, and SRM problems.\\
		&&&&$\bullet$ Use a beamforming recovery module to construct the beamforming matrix\\
		&&&&$\bullet$ Performance comparable to or higher than that of the WMMSE scheme with much lower complexity\\
		\hline
		
		& \multirow{5}{*}{\makecell{Passive \\beamfomring\\for RIS}} 
		& \multirow{5}{*}{\makecell{\cite{taha2019deep, Gao2020Unsupervised}}}	& \multirow{5}{*}{DNN} &  $\bullet$ Use a DNN with supervised or unsupervised learning to predict the achievable rates associated with RIS interaction vectors or to output the phase shifts of RIS\\
		&&&&$\bullet$ Need a beamforming recovery module to construct the diagonal passive beamforming matrix from the interaction vectors or the set of phase shifts\\
		\hline
		
		\multirow{2}{*}{\makecell{RL-based\\ MLBs}} 
		& \multirow{2}{*}{\makecell{N/A}} 
		& \multirow{2}{*}{\makecell{\cite{maksymyuk2018deep}, \\ \cite{mismar2019deep, wang2020precodernet, sun2018deep, moon2019online, Feng2020Deep, Taha2020Deep}}}	& \multirow{2}{*}{RL, NN} 
		&  $\bullet$ Formulate the beamforming design problem using the concepts of RL\\
		&&&&$\bullet$ Can be applied to various networks such as MIMO, D2D, F-RAN, RIS-aided wireless networks\\
		\hline

		\multirow{3}{*}{\makecell{FL-based\\ MLBs}} 
		& \multirow{4}{*}{\makecell{N/A}} 
		& \multirow{4}{*}{\cite{yang2020federated, elbir2020federated, Ma2020Distributed}}	& \multirow{4}{*}{FL, CNN} 
		&  $\bullet$ Useful to reduce the overhead for CSI\\
		&&&&$\bullet$ More tolerant to the imperfections and corruptions of CSI\\
		&&&&$\bullet$ Enhanced passive beamforming performance and user privacy in RIS-aided wireless systems\\
		&&&&$\bullet$ Lower computational complexity\\
		\hline
	\end{tabular}
\end{table*}

The objective of MLBs is to leverage the learning capability of ML models, such as SVM, CNN, DNN, DRL, and FDL, to directly generate the beamformer \cite{lin2019beamforming, ramon2005beamforming, alkhateeb2018deep, huang2019deep, anton2019learning, huang2018unsupervised} or to suggest its design by learning its important features \cite{li2019joint, zhou2018deep, xia2019deep, huang2019fast, long2018data, maksymyuk2018deep}. In the former application, an ML tool is modeled as a black box that outputs either the precoding/combining vector, their index in a codebook, or the precoded/combined signals. We refer to MLBs with this application of ML as \emph{black-box MLBs}. The term \textit{black box} refers to the learning capability of ML to output the desired solution, which is the beamforming vector/matrix or its indicator, without expert knowledge. By contrast, the latter application leverages both expert knowledge and the learning capability of ML models. Details of these MLB groups are discussed in the following.

\subsubsection{Black-box MLB}

Black-box MLB has found widespread use \cite{lin2019beamforming, ramon2005beamforming, alkhateeb2018deep, huang2019deep, anton2019learning, huang2018unsupervised, long2018data}, and has been used in combination with various ML models, such as DNNs \cite{lin2019beamforming, alkhateeb2018deep, huang2019deep, huang2018unsupervised, anton2019learning}, SVM \cite{ramon2005beamforming}, KNNs \cite{anton2019learning}, and the support vector classifier (SVC) \cite{anton2019learning}. In this group, the task of the ML model is to generate the precoder directly \cite{lin2019beamforming, ramon2005beamforming, huang2018unsupervised} or to output an indicator of the precoder \cite{alkhateeb2018deep, anton2019learning, long2018data, ramon2005beamforming}. An important observation is that, for black-box MLBs that output the beamforming vector/matrix directly, unsupervised learning is generally used. In contrast, supervised learning is more widely used for black-box MLBs that generate beamformer indicators. In the following, we discuss these two subgroups, which reveal interesting observations on the use of learning methods.

\paragraph{Black-box MLBs based on unsupervised learning} 

The black-box MLBs in this subgroup employ DNNs that are trained using unsupervised learning \cite{lin2019beamforming, huang2018unsupervised} owing to the unavailability of training labels. Specifically, unlike MLDs, where the label is set exactly to the transmitted signal, it is difficult to find an appropriate label for MLB design. It is possible to use classical beamforming algorithms to generate the label. However, in this case, the performance of the resulting MLB is limited by that of the classical beamforming scheme. Therefore, a DNN was trained to enable it to learn to optimize the analog beamformer thus maximizing the sum rate, or equivalently, to solve $\pabf$ \cite{lin2019beamforming}. A similar method was proposed \cite{huang2018unsupervised} to generate the digital precoder, i.e., to solve $\pdbf$. The proposed black-box MLB was demonstrated to improve the computational efficiency significantly with performance close to that of the classical weighted MMSE (WMMSE) algorithm \cite{huang2018unsupervised}.

\paragraph{Black-box MLBs based on supervised learning}
Typical black-box MLBs in this subgroup were reported by several studies \cite{alkhateeb2018deep, anton2019learning, ramon2005beamforming, long2018data}. Specifically, by modeling the beamforming selection as a classification problem, the optimal beamformer in the codebook $\mathcal{F}$ can be predicted by DNN, KNN, and SVC models \cite{anton2019learning}, and an SVM model \cite{ramon2005beamforming, long2018data}. In particular, simulation results confirmed that DNN-assisted black-box MLB performs almost optimally and outperforms both KNN- and SVC-aided methods \cite{anton2019learning}. Furthermore, as long as sufficient training data are used, the derived classification model can select the optimal analog precoder with low complexity \cite{long2018data}. Unlike \cite{anton2019learning, ramon2005beamforming, long2018data}, in which the classification problem is considered, the DNN in \cite{alkhateeb2018deep} is trained to predict the achievable rates corresponding to multiple precoders in the codebook $\mathcal{F}$, allowing the one with the highest predicted rate to be chosen for application. Furthermore, Huang \emph{et al.} \cite{huang2019deep} developed a DNN-based black-box MLB to realize end-to-end hybrid precoding in mmWave massive MIMO systems. In other words, the DNN is trained to output the precoded signals, which are ready for transmission. This method is capable of minimizing the BER and enhancing the spectrum efficiency of the mmWave massive MIMO, which achieves superior performance in hybrid precoding compared with conventional schemes while substantially reducing the computational complexity. 

\begin{figure*}[t]
	\centering
	\includegraphics[width=0.75\linewidth]{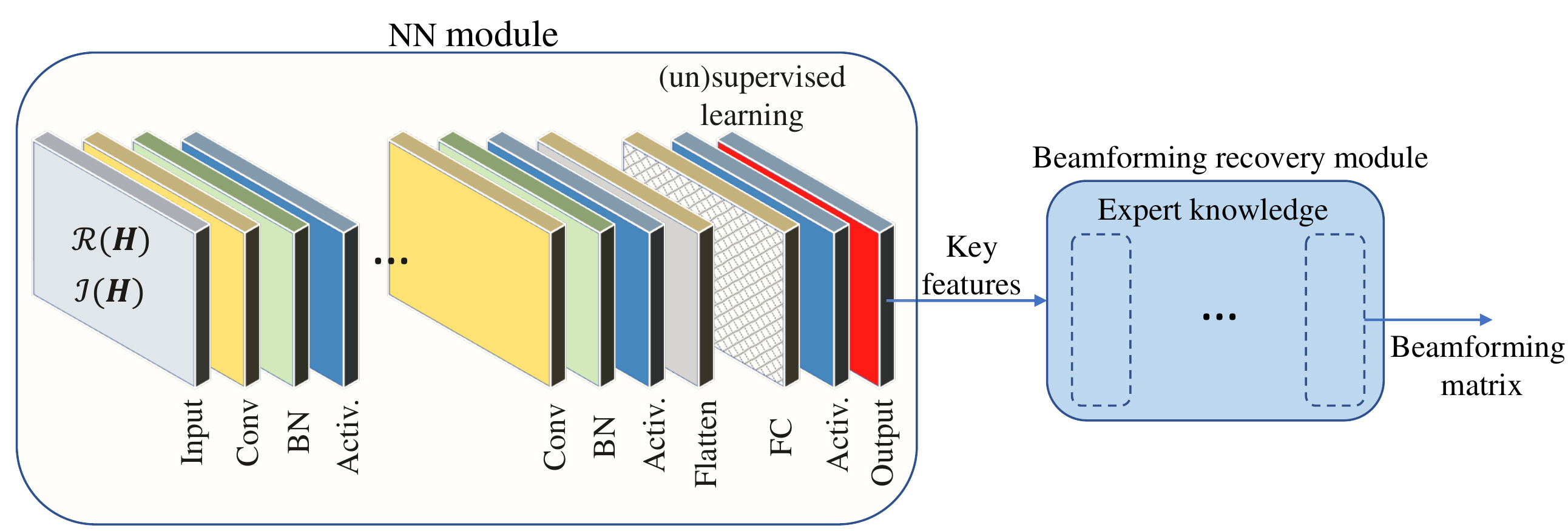}
	\caption{Feature-based DLB framework \cite{xia2019deep} with two main modules: NN and beamforming recovery. The NN module consists of input, convolutional (conv), batch normalization (BN), activation (activ.), flatten, fully connected (FC), and output layers. The NN module outputs the key features, which allow the beamforming matrix to be constructed by leveraging the expert knowledge in the beamforming recovery module.}
	\label{fig_DLBF}
\end{figure*}
\subsubsection{Feature-based MLB}

The common objective of MLB schemes in this group is to leverage expert knowledge to transform the task of predicting the beamformer to predicting their key features, which are sufficient to construct the beamformer. This method is beneficial for both the training and application phases. Specifically, instead of learning a beamforming vector/matrix  \cite{lin2019beamforming, ramon2005beamforming, huang2018unsupervised}, the ML model only needs to learn certain key features, which are usually scalars. The considerably reduced number of learning variables enables the ML model to be trained more effectively to increase the accuracy. Furthermore, the smaller-sized output and reduced number of learning variables also simplify the network architecture, leading to a reduction in the overall computational complexity in the online application phase. 

Typical work in this group entailed equipping the transmitter with a uniform planar antenna array with $N_R$ rows and $N_C$ columns of antenna elements \cite{li2019joint}. Based on this structure, the beamforming vector can be factorized as $\vf_k = (\mathcal{F}^h)_r \otimes (\mathcal{F}^v)_c$. Here, $\mathcal{F}^h$ and $\mathcal{F}^v$ are the discrete Fourier transform (DFT)-based codebooks for the horizontal and vertical dimensions, respectively, and the subscript $(\cdot)_i$ indicates the codeword index in the codebook. Because the DFT codebooks are available, only the codeword indexes $r$ and $c$ are required to obtain $\vf_k$. To this end, a DNN can be trained to predict $r$ and $c$ \cite{li2019joint}. Another application of DNN to beamforming was reported by \cite{zhou2018deep}, which solved the SRM problem by proposing the use of a DNN to approximate the conventional beam management and interference coordination (BM-IC) algorithm. The performance of this proposed DNN-based BM-IC scheme was shown to be comparable to the conventional BM-IC algorithm, but requiring less computational time.

CNN has strong feature extraction as well as approximation abilities. In addition, compared to DNN, CNN has reduced the number of learning parameters by sharing weights and biases. Therefore, it is employed in \cite{xia2019deep, huang2019fast} to obtain key features of the beamformer. This approach is illustrated in Fig.~\ref{fig_DLBF}. Three CNN-based MLB schemes were proposed for three typical optimization problems, i.e., SINR balancing, power minimization, and SRM problems \cite{xia2019deep}. The common procedure in \cite{xia2019deep} to solve these problems was to relax them to virtual equivalent problems such that the solution, i.e., the beamforming matrix, can be expressed as a function of unknown scalars, which are easier to predict. Then, CNNs were employed to predict these scalars, followed by a beamforming recovery module to construct the required beamforming matrix. A similar method to solve the SRM problem was used by proposing a CNN-based beamforming prediction network (BPNet) \cite{huang2019fast}. Specifically, the SRM problem is first separated into power allocation and virtual uplink beamforming design modules, which can be jointly optimized by the BPNet. The simulation results \cite{xia2019deep, huang2019fast} show that the performance of the MLB scheme proposed by Xia \emph{et al.} \cite{xia2019deep} closely approximates that of the WMMSE, whereas the performance of BPNet \cite{huang2019fast} improves with respect to that of the WMMSE, both with much lower computational complexity.

Recently, reconfigurable intelligent surface (RIS) has emerged as a spectral- and cost-efficient approach for wireless communications systems. RISs provide passive beamforming gains to enhance the connection between the BS and MS, especially when the direct link between them are blocked or considerably attenuated by obstacles such as trees and buildings. The passive beamforming design of RISs is usually formulated in the rate/capacity maximization problem, which are non-convex and intractable. Existing hand-engineered schemes for passive beamforming design, such as the alternating optimization or projected gradient approaches, require a high computational complexity. To overcome this problem, various ML-based passive beamforming schemes have been proposed \cite{taha2019deep,Feng2020Deep,Huang2020Reconfigurable,Gao2020Unsupervised,Ma2020Distributed}. In particular, in \cite{taha2019deep, Gao2020Unsupervised}, DNNs are leveraged for the passive beamforming design. Specifically, while a supervised learning strategy is employed for the DNN to predict the achievable rates associated with RIS interaction vectors in \cite{taha2019deep}, unsupervised learning is used to generate the phase shifts of the RIS in \cite{Gao2020Unsupervised}.

\subsubsection{RL-based beamforming}

Several studies applied RL to beamforming design \cite{maksymyuk2018deep, mismar2019deep, wang2020precodernet, sun2018deep, moon2019online}. Unlike ML-based beamforming schemes, the common motivation for using RL for beamforming is to formulate the beamforming design problem based on the RL concepts, i.e., the agent, actions, and rewards, and find the solution from interaction with the defined environment. For example, Wang \emph{et al.} \cite{wang2020precodernet} proposed a novel DRL-based HBF design method named PrecoderNet to design the hybrid precoding matrix and improve the spectral efficiency and BER performance of mmWave point-to-point massive MIMO systems. This proposal is based on the finding that the HBF problem can be modeled as a Markov decision process (MDP) that can be effectively solved based on DRL by defining the system sum rate, i.e., the objective function, as reward, and the beamformer, i.e., the desired solution, as state. PrecoderNet was numerically shown to significantly outperform classical HBF schemes, such as orthogonal matching pursuit (OMP) and MMSE. Mismar \emph{et al.} \cite{mismar2019deep} questioned the existence of a method that could jointly solve the nonconvex optimization problem of beamforming, power control, and interference coordination design to achieve the SINR upper bound without performing an exhaustive search over the entire solution space. Therefore, they utilized the ability of DRL to explore the solution space by learning from interaction. In particular, the proposed DRL algorithm does not require CSI, and hence, channel estimation becomes unnecessary. Furthermore, DRL is also exploited for passive beamforming problem for RIS-assisted wireless communications in \cite{Taha2020Deep, Feng2020Deep, Huang2020Reconfigurable}.

Furthermore, DRL was also leveraged for beamforming in Fog radio access networks (RAN) (F-RAN). Specifically, DRL-based mode selection and resource management for F-RAN was proposed \cite{sun2018deep}. Using DRL, the controller can quickly control the communication modes of user equipment (UE), i.e., cloud RAN (C-RAN) and device-to-device (D2D) modes, and the on--off states of processors in the cloud. After the controller makes the decision, precoding vectors for UE in the C-RAN mode are subsequently optimized in terms of their QoS, power, and computing capability constraints. A similar application of DRL to mode selection and precoding design in F-RAN was investigated in \cite{moon2019online}. Specifically, DRL-based adaptive selection of backhaul and fronthaul transfer modes was proposed with the aim of optimizing the performance of content delivery. Numerical results showed that the proposed DRL-based schemes in \cite{sun2018deep} and \cite{moon2019online} outperformed baseline schemes.

\subsubsection{FL-based beamforming}

Because of the special characteristics of FL, its application to beamforming is discussed separately from the above groups. Particularly, FL is more related to distributed learning across the users to preserve data privacy and save network bandwidth \cite{yang2020federated} rather than to beamforming design. With respect to the beamforming problem, FL is useful for reducing the overhead for the transmission of CSI from the users to the BS, and it can also be combined with an ML technique to achieve the benefits of both ML and FL. For example, Elbir \emph{et al.} \cite{elbir2020federated} designed a CNN to generate analog beamformers at the output. Especially, in this work, instead of using global training as is usually the case in ML schemes, an FL-based framework, in which the model is trained at the BS by collecting only the gradients from users, was employed. The simulation results showed that, compared with ML, FL is more tolerant to the imperfections and corruptions of the CSI, and at the same time, it also has lower computational complexity. By contrast, beamforming was used to improve the performance and the convergence rate for FL via over-the-air computation in \cite{yang2020federated} or to enhance the passive beamforming performance and user privacy in RIS-aided wireless communications \cite{Ma2020Distributed}. This approach resulted in a significant reduction in training loss and an improvement in the training accuracy.

\subsection{Channel Estimation}

Channel estimation is an important task in signal processing and significantly affects the performance of the signal detection and beamforming schemes. In this subsection, for completeness, we review ML-based channel estimation schemes that were recently reported in the literature.


Consider a block-fading MIMO system with the input--output relationship given in \eqref{system_model}. To estimate the channel matrix $\mH$, $P$ pilot signal vectors $\{\vs_1, \ldots, \vs_P\}$, which are known at both the transmitter and receiver, are transmitted. Let $\vy_p$ and $\vn_p$ be the received signal and let the AWGN noise vectors correspond to $\vs_p$. Furthermore, by denoting $\mS = \left[ \vs_1, \ldots, \vs_P \right]$, $\mY = \left[ \vy_1, \ldots, \vy_P \right]$, and $\mN = \left[ \vn_1, \ldots, \vn_P \right]$,  we can write
\begin{align*}
	\mY = \mH \mS + \mN.
\end{align*}
The task of channel estimation is to recover the channel matrix $\mH$ based on knowledge of $\mS$ and $\mY$. The LS and MMSE schemes are two typical classical channel estimators. In the LS scheme, the estimated channel is given as \cite{biguesh2006training}
\begin{align*}
	\hat{\mH}_{\text{LS}} = \mY \mS^H \left(\mS \mS^H\right)^{-1}. \nb \label{LS_est}
\end{align*}
This indicates that the LS estimator only requires the observations $\mY$ without requiring the channel statistics. 
{However, its mean-square error (MSE) is inversely proportional to the SNR, implying that it may be subject to noise enhancement.} By contrast, the MMSE estimator improves the performance with respect to the LS scheme, but requires knowledge of the channel covariance matrix $\mR = \mean{\mH^H \mH}$. In particular, in the MMSE method, the estimated channel is given as \cite{biguesh2006training}
\begin{align*}
	\hat{\mH}_{\text{MMSE}} = \mY \left( \mS^H \mR \mS + \sigma^2 M \mI \right)^{-1} \mS^H \mR. \nb \label{MMSE_est}
\end{align*}
The assumption of knowledge of $\mR$ can be unrealistic in practical scenarios, especially in fast fading channels. Furthermore, the matrix multiplications, additions, and inversion performed by the MMSE estimator in \eqref{MMSE_est} are computationally intensive. In addition, the common limitation of the aforementioned classical estimators is the significant performance degradation when the pilot length is smaller than the number of transmit antennas, i.e., $P < N$. We note that the assumption of $P \geq N$ can be impractical for large-sized systems such as the downlink of massive MIMO systems, where $N$ is very large. Furthermore, the use of a long pilot sequence generates substantial training overhead, thereby reducing the overall spectral efficiency of the system as well as increasing the computational load.

\begin{table*}[t]
	\renewcommand{\arraystretch}{1.00}
	\caption{Summary of MLCEs and selected properties}
	\label{tab_summary_MLCE}
	\begin{tabular}{|c|c|c|c|p{9.4cm}|}
		\hline
		\textbf{MLCE group} & \textbf{Sub-group} & \textbf{Paper} & \textbf{ML model} & \textbf{Interesting observation} \\
		\hline
		\hline
		
		\multirow{5}{*}{\makecell{Black-box\\ MLCEs}} 
		& \multirow{2}{*}{\makecell{Single-stage}} 
		& \multirow{2}{*}{\cite{gao2019deep, sun2018limited}}	& \multirow{2}{*}{DNN} &  $\bullet$ Uses a single ML model to learn and output the channel matrix based on the received signal or its compression, without expert knowledge\\
		\cline{2-5} 
		
		& \multirow{6}{*}{\makecell{Two-stage}} 
		& \multirow{6}{*}{\cite{chun2019deep, liao2019chanestnet, dong2019deep, soltani2019deep, kang2018deep}}	& \multirow{6}{*}{\makecell{CNN, DNN, \\ RNN, AE}} &  $\bullet$ In the first stage, a coarse estimate of the channel is obtained based on classical or ML schemes, which are then processed by another ML model to output the high-accuracy channel.\\
		&&&&$\bullet$ Outperform classical estimators such as LS and MMSE\\
		&&&&$\bullet$ 
		{Generally have higher computational complexity than the single-stage black-box MLCEs and simple classical estimators}\\
		\hline

		\multirow{9}{*}{\makecell{Feature-based\\ MLCEs}} 
		& \multirow{3}{*}{\makecell{Parameter\\ estimation}} 
		& \multirow{3}{*}{\cite{huang2018deep, ma2020sparse}}	& \multirow{3}{*}{DNN} &  $\bullet$ Key parameters such as the gains and angles of the channel are estimated, which are then used to reconstruct the channel matrix based on expert knowledge.\\
		&&&&$\bullet$ Outperform classical channel estimation schemes such as CS and OMP\\
		&&&&$\bullet$ 
		{Requires expert knowledge on the channel model}\\
		\cline{2-5} 
		
		& \multirow{3}{*}{\makecell{Image\\processing-based}} 
		& \multirow{3}{*}{\cite{jin2019channel, he2018deep, wen2018deep, balevi2020massive}}	& \multirow{3}{*}{CNN, DNN} &  $\bullet$ Apply DL-aided image processing techniques for channel estimation\\
		&&&&$\bullet$ Leverage denoising techniques\\
		&&&&$\bullet$ Outperform classical channel estimation schemes such as LS and CS\\
		&&&&$\bullet$ 
		{Require similarity between the channel and the natural image, more suitable to sparse channels}\\
		\cline{2-5} 
		
		& \multirow{3}{*}{\makecell{Classical\\scheme-based}} 
		& \multirow{3}{*}{\cite{neumann2018learning, yang2019deep}}	& \multirow{3}{*}{DNN, CNN} 
		& $\bullet$ Use a classical estimator to generate the input for ML models or approximate a classical scheme using ML models for performance improvement and complexity reduction\\

		\hline
	\end{tabular}
\end{table*}

The application of ML to channel estimation to overcome the limitation of classical hand-engineered channel estimation schemes, especially for massive MIMO systems, has recently attracted research interest. Numerous ML-based channel estimators (MLCEs), which exploit various ML models, learning methods, and algorithms, have been reported in the literature. Generally, because of the heavy learning task associated with channel estimation, DNN and CNN are commonly used. Furthermore, for channel estimation, supervised learning is more widely used than unsupervised learning. In this subsection, in which we review state-of-the-art MLCEs, we focus on their design methodology, i.e., the motivation and how ML is leveraged for channel estimation. Our literature review revealed that ML can be leveraged in two different ways. First, an ML tool can be modeled as an independent channel estimator to directly output the channel matrix/vector. We refer to this group of MLCEs as \emph{black-box MLCEs}. Second, ML can be used to either estimate key features, which are then used to reconstruct the channel matrix, or to support classical well-known channel estimation algorithms. Therefore, we refer to the MLCEs in this group as \emph{feature-based MLCEs}. These typical groups of MLCEs are discussed in the following.

\subsubsection{Black-box MLCEs}

The task of ML in black-box MLCEs is to learn and output the channel matrix $\mH$ without expert knowledge. Because this is a computationally intensive task, complicated NN architectures such as DNN \cite{gao2019deep, chun2019deep, sun2018limited}, CNN \cite{liao2019chanestnet, soltani2019deep, dong2019deep}, and RNN \cite{liao2019chanestnet} are typically used. However, to focus on the design of MLCEs rather than on the ML tools and learning techniques, we classify the black-box MLCEs based on their design architectures.

\paragraph{Single-stage MLCEs} A very straightforward black-box MLCE, which fully exploits the learning ability of a DNN, is the direct-input DNN (DI-DNN) proposed by Gao \emph{et al.} \cite{gao2019deep}. In the DI-DNN, the pilot signals are all set to $1$, and the DNN can learn the channel matrix $\mH$ based only on the received signal $\mY$. In particular, to reflect practical systems in which the resolution patterns of analog-to-digital converters (ADCs) differ, Gao \emph{et al.} considered mixed-resolution ADCs, i.e., ADCs with both high and low resolution components. Therefore, the input of the DNN can be the exact received signal $\mY$ or its quantized version $\mathcal{Q}(\mY)$. 
{The simulation results showed that the proposed scheme is less affected by the error floor. As a result, it outperforms the classical linear MMSE channel estimation method across the entire SNR regimes.} Other than the aforementioned approach \cite{gao2019deep}, an MLCE that uses a transformed version of the received signals as the input of the DNN was proposed \cite{sun2018limited}. Specifically, to minimize user operations and reduce the feedback overhead, the signal received by the user is compressed to a scalar value and provided as feedback to the BS for channel estimation. This scheme requires only limited feedback and is proposed for channel estimation in frequency division duplex massive MIMO systems.

\paragraph{Two-stage MLCEs} A two-stage channel estimation process was proposed in \cite{chun2019deep}. In the first stage, the pilot and received signals are input into a DNN, which learns and outputs the first estimate of the channel matrix. Then, the estimated channel is further improved by another DNN, which, at this time, exploits the data and received signals in the transmission phase. The advantage of this two-stage MLCE compared with the conventional single-stage channel estimator is that the quality of the estimator can be improved, not only in the training phase but also in the transmission phase. Therefore, a larger portion of the coherent time can be allocated for the transmission phase to achieve higher spectral efficiency, as shown by the simulation results in \cite{chun2019deep}. Liao \emph{et al.} \cite{liao2019chanestnet} proposed architecture named ChanEstNet, which uses two networks: a CNN to extract channel response feature vectors, which are then processed by an RNN to obtain the final estimate of the channel. 

The two-stage MLCE \cite{chun2019deep, liao2019chanestnet} fully exploits the ML models in both stages. In contrast, other researchers proposed to perform the first-stage estimation based on conventional schemes \cite{dong2019deep, soltani2019deep}. Specifically, in the first stage, a coarse estimate of the channel matrix was obtained by removing the effects of the beamformers \cite{dong2019deep} or by using an LS filter \cite{soltani2019deep}. The coarse channel estimate is then processed by a CNN to generate an estimate of the channel matrix with higher performance. Both of the proposed MLCEs \cite{dong2019deep, soltani2019deep} outperform the classical MMSE estimator and its variants. The application of two-stage MLCE for wireless energy transfer was also considered \cite{kang2018deep}. Specifically, the use of a shallow FNN to obtain the energy feedback information harvested at the output was proposed \cite{kang2018deep}. Then, a DNN is used to learn and output the channel from the energy feedback information.

\subsubsection{Feature-based MLCE}

Instead of using a single or multiple NNs to generate the channel matrix, as in black-box MLCEs, feature-based MLCEs leverage expert knowledge such as the known channel characteristics and structure. In this way, the ML model only needs to estimate an intermediate parameter, which is then used to reconstruct an estimate of the channel matrix. For example, given the channel model, channel parameters such as the gains or angles can be learned, which can be readily exploited to build the channel matrix. Another approach is to leverage ML-aided image processing schemes for channel estimation based on the similarity between sparse channels and the natural image. These approaches are reviewed in the following paragraphs.

\paragraph{Parameter estimation} The parameters of mmWave channels model were estimated \cite{huang2018deep, ma2020sparse}. Specifically, in a typical uplink channel of a massive MIMO system, the channel between the BS, equipped with $N$ antennas, and the $k$th user can be expressed as
\begin{align*}
	\vh_k = \sum_{i=1}^{N_p} g_{k,i} \va(\theta_{k,i}) = \mA_k \vg_k, \nb \label{sparse_channel}
\end{align*}
where $N_p$ is the number of paths, $g_{k,i}$ is the complex gain of the $i$th path between user $k$ and the BS, and $\mA_k = \left[ \va(\theta_{k,1}), \va(\theta_{k,2}), \ldots, \va(\theta_{k,N_p}) \right]$. Here, $\theta_{k,i}$ denotes the physical angle of arrival (AoA) of the $i$th path, and $\va(\theta_{k,i})$ is a steering vector given as
\begin{align*}
	\va(\theta_{k,i}) = \frac{1}{\sqrt{N}} \left[ 1, e^{j\pi\theta_{k,i}}, \ldots, e^{j(N-1)\pi\theta_{k,i}} \right]^T.
\end{align*}
From \eqref{sparse_channel}, it is observed that the information of the complex gains and AoAs, i.e., $\{g_{k,i}, \theta_{k,i}, i=1,\ldots,P\}$, are key characteristics that model the channel vector $\vh_k$. Motivated by this, Huang \emph{et al.} \cite{huang2018deep} proposed using DNNs to estimate the complex gains and AoDs. This scheme outperforms many existing classical channel estimation schemes such as compressed sensing (CS). Another method of leveraging the channel structure in \eqref{sparse_channel} was proposed \cite{ma2020sparse}. In particular, instead of estimating the parameters of $\vh_k$, a DNN was employed to estimate the amplitudes of the elements of the beamspace channel vector $\tilde{\vh}_{k} = \mA_k^H \vh_k$ \cite{ma2020sparse}. Based on the estimated amplitudes of $\tilde{\vh}_{k}$, the indices of dominant entries in $\vh_k$ are determined. This allows $\vh_k$ to be approximately reconstructed based on the sparsity of $\vh_k$ and the fact that $\mA_k^H \mA_k = \mI$. The simulation results \cite{ma2020sparse} verify the performance improvement of the proposed scheme compared to the classical OMP and distributed grid matching pursuit (DGMP) schemes. 

\begin{figure}[t]
	\centering
	\includegraphics[width=0.765\linewidth]{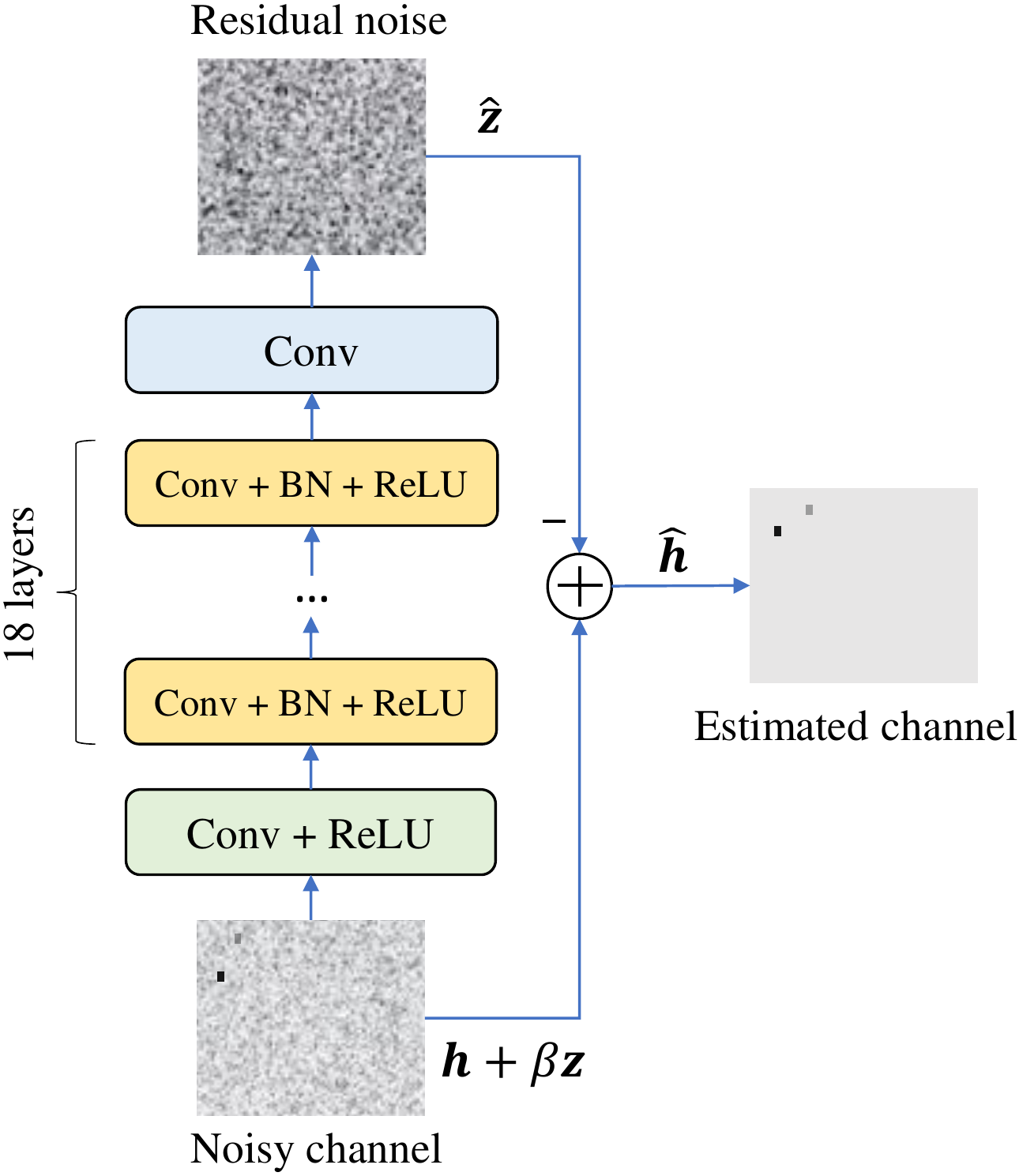}
	\caption{Network architecture of the DnCNN denoiser \cite{he2018deep}, consisting of 20 convolutional layers. The first layer employs 64 filters sized $3\times 3 \times 1$, followed by a ReLU. Each of the next 18 successive layers contains 64 filters sized $3\times 3 \times 64$, each followed by a BN and ReLU. The last layer uses a $3\times 3 \times 64$ filter to reconstruct the signal.}
	\label{fig_DnCNN}
\end{figure}

\paragraph{Image processing-based MLCEs} Unlike the above mentioned studies \cite{ma2020sparse, huang2018deep}, in which the channel parameters were estimated by ML models, the sparse channel matrix was considered as a natural image, and this motivated the application of DL-aided image processing techniques for channel estimation \cite{jin2019channel, he2018deep, wen2018deep, balevi2020massive}. For example, He \emph{et al.} \cite{he2018deep} found that the correlation among the elements of the channel matrix is very similar to that of a 2D natural image: the channel is sparse and the changes between adjacent elements are subtle. Therefore, the authors proposed to leverage the learned denoising-based approximate message passing (LDAMP) network, originated from image recovery, for channel estimation. In the LDAMP network, the CNN-based denoiser (DnCNN), illustrated in Fig. \ref{fig_DnCNN}, can handle Gaussian denoising problems. In this scheme, a noisy channel $\vh + \beta \vz$ is processed by the DnCNN to estimate the residual noise $\hat{\vz}$, which is then subtracted from the input to obtain the estimate of the channel $\hat{\vh}$. Interestingly, the LDAMP network outperforms many CS-based channel estimation algorithms. However, a drawback of the LDAMP scheme is that the DnCNN denoiser is tailored to a specific noise level, and it only performs well for this trained noise level \cite{jin2019channel}. 
{This motivated the proposal of the fast and flexible denoising CNN (FFDNet) \cite{jin2019channel}. In contrast to LDAMP, FFDNet, with a flexible noise level map at the input, is suitable for a wide range of SNRs. As a result, it outperforms the LDAMP scheme across a large range of noise levels.}

Another MLCE \cite{wen2018deep} is also based on the image recovery problem. However, it focuses on the optimization of CSI feedback. In particular, to avoid excessive feedback overhead of the CSI, the authors proposed the CsiNet, which encodes the original channel matrix to compress codewords at the receiver before feeding back to the transmitter for beamforming. At the transmitter, the compressed CSI is decoded by the CsiNet to the original channel. Here, the CsiNet is constructed based on CNNs. The aforementioned schemes \cite{jin2019channel, he2018deep, wen2018deep} are based on supervised learning, which requires a large number of parameters to be trained before online application becomes possible. To overcome this limitation, Balevi \emph{et al.} \cite{balevi2020massive} proposed an untrained DNN based on the deep image prior network for channel estimation. The idea of this scheme is that, instead of denoising the channel matrix, the received signal is first denoised by a DNN, followed by the classical LS channel estimation scheme.

\paragraph{Classical LS/MMSE-based MLCEs} The incorporation of ML in a classical channel estimator was proposed \cite{neumann2018learning, yang2019deep}. Specifically, CNNs were used to approximate the MMSE channel estimator \cite{neumann2018learning}. Alternatively, the LS estimate of the channel was exploited as an input of the DNN-based estimator \cite{yang2019deep}. 
{These combinations of a classical estimator and ML models result in significant performance improvement with respect to conventional hand-engineered estimators. In particular, the MLCE can compensate for the performance loss resulting from insufficient pilot signals \cite{yang2019deep}; hence, it outperforms the linear MMSE estimator. On the other hand, a hierarchical learning algorithm was proposed to avoid convergence to local optima during the training of the NN \cite{neumann2018learning}. As a result, the trained NN can be optimized for a general channel model using stochastic gradient methods. The simulation results \cite{neumann2018learning} indicated that, for a simple channel model, the performance of the proposed learning-based methods is less promising but is significantly improved for a more realistic channel model such as the 3GPP model with multiple paths and different path gains.}

\subsection{Summary and Takeaway Points}

In this section, typical MLBs and MLCEs were discussed on the basis of their designs, which are summarized in Tables \ref{tab_summary_MLB} and \ref{tab_summary_MLCE}, respectively. Whereas black-box MLBs/MLCEs focuses on the learning ability of ML models to learn and output beamforming/channel matrices, feature-based MLBs/MLCEs utilizes both expert knowledge and machine intelligence to construct the beamformer and channel estimator. For the beamforming problem, the exploited expert knowledge can either be the known optimal solution or well-developed hand-engineered iterative algorithms to solve SRM problems. For channel estimation, channel characteristics such as the path gains, angles, or sparsity are leveraged. In particular, DL-aided image processing techniques can be utilized for channel estimation owing to the similarity between a 2D natural image and a sparse channel matrix. The proposed MLBs and MLCEs are shown to have advantages in terms of performance improvement and/or complexity reduction with respect to classical schemes.

Various ML models were deployed for MLBs and MLCEs, ranging from simple SVM, SVC, and KNN models to more sophisticated DNN, CNN, FL, and RL models. Especially, DNNs and CNNs are widely used owing to their powerful feature extraction capabilities as well as their wide application to image processing. Most of the MLBs and MLCEs outperform classical hand-engineered schemes such as WMMSE for beamforming and LS, linear MMSE, OMP, and CS-based methods for channel estimation. MLBs and MLCEs have developed along two major well-exploited directions: designs based on black-box and feature-based beamformers/estimators. The former are generally more computationally complex than the latter because ML models need to learn and output a solution, which is a vector/matrix. However, it is independent from the solution structure. By contrast, feature-based MLBs/MLCEs generally have reduced complexity because of the simpler learning task. However, they depend on the structure and characteristics of the solution. For example, the beamformers in analog, digital, and hybrid beamforming have different constraints on the phases and amplitudes of the entries, or, for channel estimation, the channel models and sparsity vary depending on the communication scenario. These practical aspects challenge the design of MLBs/MLCEs and create room for further studies in these fields.

\section{Research Challenges and Future Directions}
\label{Sec:Challenges}
This section discusses the challenges presented by intelligent radio signal processing and highlights several promising directions. 


\subsection{End-to-End Learning}
Most studies on intelligent radio signal processing focused on modular design; that is, the entire system is composed of blocks such as transmitters, amplifiers, channels, and receivers, each of which has its individual process. However, the number of services and applications that require an end-to-end performance guarantee is expected to increase \cite{kato2020ten, letaief2019roadmap}. Wireless channels are becoming increasingly complex and are affected by many factors, adversaries, channel nonlinearities, and hardware impairments. Additionally, the performance metrics (e.g., rate 0.1-1 Tbps, and end-to-end delay 1 ms), frequency bands (sub-6 GHz, mmWave, THz bands), waveforms, and modulation techniques (e.g., index and spatial modulation) are expected to increase in next-generation networks \cite{saad2020vision}. 
Generally, AI has great potential to provide effective end-to-end solutions.  
Motivated by recent seminal work \cite{o2017introduction}, many studies have been dedicated to improving wireless systems, e.g., modulation identification and channel estimation for spectrum situation awareness and wireless security \cite{Erpek2020}. However, many challenges remain for end-to-end learning when various features and new requirements are taken into consideration. 

\subsection{Distributed, Centralized, or Hybrid Learning} 
Most AI-enabled techniques for signal processing focus on achieving the best performance (e.g., accuracy and spectral/energy efficiency), but the implementation complexity is often neglected. Indeed, when the training server has sufficient computing and storage capabilities, centralized learning is preferred over distributed learning. This is a reasonable choice because large amounts and various types of data can be collected and trained centrally. Meanwhile, problems arising from synchronization, data distribution, and communication dynamics can be safely ignored. The preference for centralized and distributed learning is similar to, for example, centralized optimization vs. distributed game approaches, cloud vs. edge computing, and RAN vs. C-RAN, which has been developed for wireless communications for decades. Centralized learning has three main drawbacks: high computational complexity, security and data privacy (owing to the high concentration of data), and poor scalability. This approach also fails to exploit the large number of computing resources distributed over the network and is not applicable to scenarios in which a centralized entity is not available, e.g., ad hoc and wireless sensor networks. An example of distributed learning for online medium access control in a spectrum sharing network has been discussed \cite{zafaruddin2019distributed}. On the other hand, hybrid learning is a promising solution to these drawbacks as it potentially maintains a balance between complexity and performance. In particular, a learning model can be divided into smaller tasks and then trained hierarchically, e.g., cloud, edge, and end devices. However, selecting the learning mode largely depends on problems and design objectives, and it should be carefully considered for practical implementation. Transfer learning is also a good AI technique to be considered for radio signal processing since learning model can be improved by utilizing retrained AI models using a similar task.

\subsection{Model Compression and Acceleration}
AI-enabled approaches have been shown to have many advantages over conventional approaches, but they are usually computationally and memory intensive. Compared with datasets in computer vision and natural language processing, the datasets used to train signal processing models are typically smaller. However, IoT/edge devices are limited in terms of computing capability and memory and storage capacity. Additionally, because many AI-based services will be available at the network edge, being able to deploy intelligent signal processing algorithms in resource-limited devices (e.g., sensors, industrial IoT and wearable devices, and drones) with comparable performance plays a vital role. For instance, training the MCNet architecture \cite{AMC-Huynh2020MCNet} to classify 24 modulations takes approximately 24 h on a computer with 16 GB memory and a NVIDIA GeForce GTX 1080Ti GPU, and the number of trainable parameters can reach 220,000. The training time would increase markedly and the classification accuracy decreases if the model is trained on a personal computer with less processing power. A promising solution to overcome these challenges is model compression and acceleration. Compressing and accelerating DL techniques can be classified into four main groups \cite{cheng2018model}: 1) parameter pruning and quantization, 2) low-rank factorization, 3) transferred/compact convolutional filters, and 4) knowledge distillation. Over the last few years, DL has employed a large number of techniques and interested readers are invited to refer to a recent survey \cite{deng2020model} for more details. Our observation is that most existing solutions focus on achieving high performance while ignoring the suitability of the proposed DL models for resource-constrained devices.
Another promising direction is to offload the training and inference processes to the edge (i.e., edge intelligence in the literature), i.e., AI techniques are employed at the network edge to intelligently process radio signals. For example \cite{yang2020energy}, the joint task allocation and downlink beamforming problem was optimized to minimize the total energy consumption. Applications of a 6G technology, namely reconfigurable intelligent surface, for edge inference, was reported \cite{hua2019reconfigurable}. 

\subsection{Dataset Generation and Unification} 
The quality of training data greatly affects the performance and prediction accuracy of AI-enabled techniques. Compared with other fields such as computer vision and healthcare, the communication and signal processing communities were not ready to standardize data generation methods and unify the datasets for performance evaluation. The reason for this is that previous network generations can operate effectively with conventional approaches, such as queuing models, convex optimization, and game theory. However, some initiatives and competitions have been launched to resolve this issue, for example, IEEE ICC 2020 for vision-aided wireless systems \cite{ViWi} and IEEE CTW 2020 for user localization \cite{IEEECTW2020}. The Machine Learning for Communications ETI led by the IEEE is one of the best initiatives as it maintains a collection of datasets, papers, reproducible codes, etc., about the use of ML for wireless communications.  
However, the number of datasets is quite small and completely inadequate for a substantial number of network scenarios and problems in wireless systems and radio signal processing. Moreover, the size and quality of these datasets remain questionable as only a few are ready for practical implementation. Additionally, datasets would need to be continuously updated as the network will become highly dynamic, dense, and heterogeneous.

\subsection{Universality and Practicality} 
Despite a few years of development and usage, the number of published papers and preprints are increasing daily, and many studies are devoted to solving the same problem. Moreover, some researchers prefer to use simulated and private datasets, whereas others use public ones. This makes it difficult to compare the proposed algorithms when they are implemented in the same network setting and design objective. For example, the RadioML dataset \cite{AMC-OShea2018DeepSig} was created for the classification of 11 modulation types (8 digital and 3 analog), which have since become 24 modulations\footnote{The dataset is available at \url{https://www.deepsig.ai/datasets}}. Utilizing the same RadioML dataset, Huynh~\emph{et~al.} \cite{AMC-Huynh2020MCNet} applied their DL models to all the modulation types, whereas other researchers only considered a set of selected modulation formats, e.g., 5 types (BPSK, QPSK, 8PSK, 16QAM, 64QAM) \cite{AMC-Huang2019CFCN}, and 11 types (the RadioML2016.10a dataset) \cite{bu2020adversarial}. Additionally, other researchers tested their classification approaches using simulated datasets, wherein emulating realistic channel impairments is difficult \cite{AMC-Abdul2019Likelihood, AMC-Meng2018DL}. The fact is that, the higher the modulation order, the lower the classification accuracy \cite{AMC-Huynh2020MCNet}. Therefore, certain studies only experimented with low-order modulation formats and the applicability to high-order modulations is, therefore, questionable.
All the aforementioned considerations encourage research communities to follow a standard methodology to enhance the practicality. Additionally, existing approaches need to be examined to make sure that they and/or their modifications can support new services and scenarios in the future. 

\subsection{Deep Semisupervised and Active Learning} 
In our summary of recent studies on intelligent signal processing, we observed that they mostly focus on (deep) supervised learning. The main motivations for the use of supervised learning in wireless communications are its high performance (e.g., classification accuracy) and exploitation of available labeled datasets. The use of supervised learning in wireless communications would also be highly desirable; however, labeled datasets for a learning task are not always available and would even be time consuming and costly to construct in many cases. Most recent AI research is based on simulated datasets that were created using communication tools, and wherein channel characteristics and design features are emulated. In many cases, optimization problems are formulated and solved to obtain the locally optimal solutions, which are then considered as labeled instances of the supervised learning model. Therefore, the performance of an unsupervised learning method could possibly be higher than that of its supervised counterpart. An example is the optimization of the weighted sum rate problem, which was achieved by optimizing the transmit beamforming for a downlink MIMO system \cite{huang2018unsupervised}. The unsupervised learning approach outperforms the supervised learning method for various settings such as the SNR and number of transmit antennas. This interesting observation shows that (deep) unsupervised learning has the potential to deliver high performance, especially when truly labeled datasets are not available. 
The availability of domain knowledge and truly labeled examples (i.e., real data or data generated by global optimization algorithms) would render semisupervised learning a promising solution for this problem as it combines both the supervised and unsupervised approaches to exploit their respective advantages. 
Active learning provides other ways to solve the challenge of creating labeled instances. In particular, the central server (e.g., eNBs equipped with computing capabilities) may query end devices to provide labeled instances, thereby improving the learning performance using the collected labeled data. 
Applications of active learning for signal processing can be found in a few recent studies of the initial access problem in mmWave systems (for example, \cite{Chiu2019ActiveLearning}) and of joint power and resource block allocation in vehicular networks \cite{AbdelAziz2020uRLLC}.


\subsection{Standardization and Open-Source Activities} 
Although AI for radio signal processing has been a hot topic and has received much interest in the last few years, the successful integration of AI for practical applications and standardization activities is still in its infancy. This could be due to several reasons, including the lack of cooperation between academia and industry, competition among high-tech countries, the impact of the coronavirus pandemic, etc. However, certain bodies have already made initial efforts to integrate AI techniques into both existing and future networks. The Telecommunication Standardization Sector of the International Telecommunication Union (ITU) proposed a unified architecture for ML in future networks \cite{ITU2019Unified}. The unified architecture is composed of three main blocks, including 1) \textit{a management subsystem}, which allows operators to deploy on-top services without affecting the underlying infrastructure, 2) \textit{an ML pipeline}, which defines a set of logical entities to perform specific functions, and 3) \textit{a closed-loop subsystem}, which helps the ML pipeline adapt to network dynamics. In particular, the ITU architecture needs to satisfy four high-level requirements: use of multisource and correlated data; support for multiple technologies and network layers; support for multilevel, multidomain, and distributed services; and negligible effect on the underlying infrastructure. Very recently, the 3GPP started a study item focusing on traffic characteristics and performance requirements of AI models in 5G. Methods of overcoming particular challenges are specified \cite{3FPP_AI_ML}; for example, achieving AI inference on devices with limited computing and battery capacities, and designing on-board AI inference according to network dynamics. 

In addition to standardization, open-source activities play an important role in accelerating the adoption of AI-based signal processing schemes in real wireless communications. However, it is worth noting that the availability of many open-source platforms would limit industrial collaboration and reduce the verification of intelligent signal processing solutions \cite{Pham2020ASurvey_MEC}. 

\subsection{Support for New Services}
Although few metrics are used in most AI-enabled schemes to evaluate their performance, emerging applications and new services are expected to have more stringent requirements \cite{saad2020vision}. For example, smart railway station services could be considered as a type of mission-critical services, but they have additional considerations, which may include massive devices with different characteristics, heterogeneous environments, and use cases (e.g., energy charging, emergency, and public safety) \cite{3FPP_Railway}. 
Current AI-enabled signal processing frameworks are often unable to meet the requirements of these services. 
Embedding AI techniques in signal processing and signaling protocols is a challenging task as the specification is still under development. As a result, partners in academia and industry need to cooperate more closely to accelerate the standardization and specification phase. Moreover, more performance metrics and factors would have to be considered in the design of AI solutions, for example, the training and inference time, reliability, explainability, scalability, simplicity, and security. 
 


\section{Conclusion} 
\label{Sec:Conclusion}
In this paper, we presented a survey of state-of-the-art AI solutions for the intelligent processing of radio signals. We first presented a brief overview of AI, ML, FL, and well-known DL architectures such as DNN, CNN, RNN, and DRL. Then, we reviewed applications of AI techniques for three main themes of radio signal processing: modulation classification, signal detection, and channel estimation and beamforming. We discussed these themes according to various classifications and provided necessary background knowledge to enable readers with different types of related expertise to obtain a good understanding of the corresponding topic. We also emphasized a number of challenges that remain unsolved and also offered several suggestions and directions for future research. We hope that this paper can serve as an important reference for both academic and industrial audiences and drive further research and innovation in the domain.

\end{document}